\documentclass[a4paper]{article}
\usepackage{etex}
\usepackage{a4wide}
\usepackage{amsmath,amsfonts,amssymb,color,latexsym}
\usepackage{pictex}
\usepackage[breaklinks=true,bookmarks=true,colorlinks=true,linkcolor=blue,citecolor=blue,allcolors=blue]{hyperref}
\usepackage{graphicx}
\usepackage{epstopdf}
\def\drawline#1#2{\raisebox{2.5pt}{\rule{#1pt}{#2pt}}}
\def\spacce#1{\hskip #1pt}
\def\solid{\protect\rule[2.5pt]{24.pt}{.5pt}\hspace*{1ex}}
\def\bdash{\protect\hbox{\protect\drawline{4}{.5}\spacce{2}}}
\def\dashed{\bdash\bdash\bdash\bdash\nobreak\ }

\def\bdot{\protect\hbox{\protect\drawline{1}{.5}\spacce{2}}}
\def\dotted{\protect\hbox{\leaders\bdot\hskip 24pt}\nobreak\ }
\def\chndot{\protect\hbox
{\protect\drawline{9.5}{.5}\spacce{2}\protect\drawline{1}{.5}\spacce{2}\protect\drawline{9.5}{.5}}
    \nobreak\ }

\def\trian{\raise 1.25pt\hbox{$\scriptstyle\triangle$}\nobreak\ }
\def\squar{\raise 1.25pt\hbox{$\scriptstyle\Box$}\nobreak\ }
\def\dtrian{\raise 1.25pt\hbox
   {$\scriptscriptstyle\bigtriangledown$}\nobreak\ }

\newcommand{\ben} {\begin{equation}}
\newcommand{\een} {\end{equation}}
\newcommand{\be} [1] {\begin{equation} \label{#1}}
\newcommand{\ee} {\end{equation}}
\newcommand{\bse} [1] {\begin{subequations} \label{#1}}
\newcommand{\ese} {\end{subequations}}
\newcommand{\ban} {\begin{eqnarray*} }
\newcommand{\ean} {\end{eqnarray*} }
\newcommand{\bea} {\begin{eqnarray}}
\newcommand{\eea} {\end{eqnarray}}

\begin{document}
\epstopdfsetup{suffix=}
\title{An Immersed Boundary Method with Direct Forcing for the
  Simulation of Particulate Flows}
\author{Markus Uhlmann\footnote{Now at Karlsruhe Institute of
    Technology, 76131 Karlsruhe, Germany,
    \href{mailto:markus.uhlmann@kit.edu}{(markus.uhlmann@kit.edu)}}
  \\[1ex]
  {\small 
Departamento de Combustibles F\'osiles, CIEMAT, Avenida
  Complutense 22,}\\ 
{\small
  28040 Madrid, Spain}}
\date{\small (published in
  \href{http://dx.doi.org/10.1016/j.jcp.2005.03.017}{{\it
      J. Comput. Phys.}, 209(2):448--476, 2005})} 
\maketitle
\begin{abstract}
  We present  
  an improved 
  method for computing incompressible viscous flow
  around suspended rigid particles using a fixed 
  and uniform 
  computational
  grid. The main idea is to incorporate Peskin's regularized delta
  function approach [Acta Numerica 11 (2002) 1] into a direct
  formulation of the 
  fluid-solid interaction force in order to allow for a smooth
  transfer between Eulerian and Lagrangian representations while at
  the same time avoiding strong restrictions of the time step. This
  technique was implemented in a finite-difference and fractional-step
  context. A variety of two- and three-dimensional simulations are
  presented, ranging from the flow around a single cylinder to 
  the sedimentation of 1000 spherical particles. 
  The accuracy and efficiency of the current
  method are clearly demonstrated.
\end{abstract}
\section{Introduction}\label{sec-intro}
Fluid-particle systems are of considerable scientific and
technological interest in a wide range of disciplines. Some examples
are: chemical engineering (fluidized beds), medical sciences (blood
flow)
and civil engineering (sediment transport near river beds). Our
present understanding of the dynamics of these systems is far from
complete and complex phenomena such as the
formation of particle clusters under ``turbulent'' conditions are still
awaiting a definite explanation~\cite{sundaresan:00}. 

In the framework of single-phase turbulent flow the analysis of data
from direct numerical simulation (DNS) has proven particularly
fruitful~\cite{moin:98}. This strategy appears equally 
promising for the future of multi-phase flows, but the computational
challenge is only starting to become accessible. Indeed, in recent
years much effort has been devoted to the design of a feasible
method for DNS of the motion of rigid particles immersed in an
incompressible fluid
\cite{hu:01,hoefler:00,kim:01,kajishima:02,glowinski:02,zhang:03,feng:04}.
By ``feasible'' it is understood that the method should at the same 
time: (a) be efficient enough to allow for the treatment of a large
number of particles and at sufficiently high values of the relative
Reynolds number;
(b) provide adequate accuracy in representing the dynamics of the
fluid-solid flow.

One way of tackling the computation of suspended particles is to solve
the Navier-Stokes equations in the time-dependent fluid domain subject 
to the no-slip condition at the interfaces with the solid
objects. This, however, 
implies adapting the mesh to the varying positions of the particles
during the course of the simulation and leads to a substantial
computational cost. An example for such a technique is the arbitrary
Lagrangian-Eulerian particle mover of Hu et al.~\cite{hu:01} which has
been successfully applied to various sedimentation problems.

In order to avoid frequent re-meshing, the flow equations can instead
be solved on a fixed grid while the presence of the solid
bodies is imposed by means of adequately formulated source terms added
to the Navier-Stokes equations. This class of techniques is 
termed ``fictitious domain methods''.
One of the precursors, the immersed boundary (IB) method of Peskin, 
was originally conceived 
for flows around flexible membranes, specifically the flow in the
human heart~\cite{peskin:72}. The basic idea is to determine a
singular force distribution at arbitrary (Lagrangian) positions and to
apply it to the flow equations in the fixed reference frame via a
regularized Dirac delta function. At the same time, the membrane is
moving at the local flow velocity. The additional force term for this
problem is simply a function of the deformation of the membrane and of
its elastic properties. The careful design of Peskin's delta function
is vital to the efficiency of the method. 

The IB method was later extended to Stokes flow around suspended
particles~\cite{fogelson:88} and Navier-Stokes flow around fixed
cylinders~\cite{lai:00}. H\"ofler and Schwar\-zer used similar ideas to
compute many-particle systems, albeit at relatively low Reynolds
numbers~\cite{hoefler:00}. Recently, Feng and Michaelides coupled the
IB method with the lattice Boltzmann technique~\cite{feng:04}. In
references \cite{lai:00,hoefler:00,feng:04} as well as in related
studies~\cite{saiki:96,lee:03} the singular forces are obtained by
means of a feedback mechanism first proposed by Goldstein et
al.~\cite{goldstein:93} and termed ``virtual boundary
method''. Therein, a deviation from the local desired value of
velocity (or position) generates a force in the opposite direction
which tends to restore the target value. In other words, a system of
virtual springs and dampers is attached to the virtual boundary
points, locally forcing a predetermined behavior. An undesirable
feature of this indirect formulation of the fluid-solid interaction 
force is the introduction of additional free parameters. In practice,
values for the spring stiffness and damping constant need to be
determined in a problem-dependent fashion. Moreover, the
characteristic time scales of the oscillations of the spring-damper
systems need to be resolved, which can lead to severe restrictions on
the time step~\cite{lai:00,lee:03}.

In order to avoid the drawbacks of the virtual boundary force,
Fadlun et al.~\cite{fadlun:00} introduced a direct formulation of the 
force term. Roughly
speaking, the method consists of modifying the entries of the implicit
matrix of the discretized momentum equation such that the desired
velocity at the boundary points is obtained after each time step. 
The authors demonstrated that this scheme does not suffer from the
time step restrictions of the virtual boundary method. Kim
et al.~\cite{kim:01} later proposed an explicit variant of the above
direct forcing method which allows to maintain the original simple
matrix structure of a standard finite-difference method. In both 
references~\cite{fadlun:00,kim:01} the objective was the
efficient computation of flow in complex domains. Although Fadlun et
al.~\cite{fadlun:00} present an example of a flow involving moving
boundaries, the smoothness of the boundary force during the relative
motion was not demonstrated. It was later recognized that the
interpolation procedure relating values at fixed grid nodes and values
at arbitrarily located boundaries can lead to force oscillations which
are undesirable for the purpose of particulate flow
simulations~\cite{uhlmann:03}.

Kajishima and Takiguchi~\cite{kajishima:02} use an extremely simple
scheme for modelling the fluid-solid interaction. At the end of a 
time-step the velocity is explicitly set to the particle's rigid-body
velocity inside each solid sub-domain. At the interfaces, fluid and
solid velocities are smoothly connected by using the solid volume
fraction of each computational cell as a weight factor. The method is
quite efficient, allowing for the long-time integration of the
sedimentation of ${\mathcal O}(1000)$ particles at a value of 350 for
the particle Reynolds number. Although this strategy avoids spurious
oscillations of the hydrodynamic force acting on a
particle, the force still shows a strong grid
dependency~\cite{uhlmann:03}. 
Furthermore, it should be mentioned that the resulting
flow-field does not verify the divergence-free condition in the
vicinity of the particles.

A different approach was taken by Glowinski et al.~\cite{glowinski:99}
who impose rigid-body motion upon the region occupied by the particles
by means of a Lagrangian multiplier technique in a finite-element
context. In subsequent simulations Glowinski et
al.~\cite{glowinski:02} use a first-order accurate, four-step
operator splitting scheme for the temporal discretization, including
reduced local time steps when updating the particle positions. The
method was applied to various sedimentation
problems~\cite{glowinski:02} and to the fluidization of ${\mathcal
  O}(1000)$ spheres in a narrow gap~\cite{pan:02}. It should be kept
in mind, however, that the use of a grid system based on
tetrahedral elements can break inherent symmetries of the problem. 
Patankar et al.~\cite{patankar:00} proposed a related
Lagrangian multiplier technique where---instead of velocity---the
deformation-rate tensor was imposed in the particle sub-domains,
thereby simplifying the treatment of irregularly-shaped particles.
A further improvement was introduced by Patankar~\cite{patankar:01}
who showed how the need for an iterative procedure could be
eliminated when imposing the rigidity constraint. During the review
process it was brought to our attention that this scheme has 
meanwhile been implemented in a control volume
context~\cite{sharma:05} and successfully applied to DNS of
particulate flow.  

Finally, it should be mentioned that Zhang and
Prosperetti~\cite{zhang:03} have recently proposed a semi-analytic
method based upon local Stokesian dynamics in the
close vicinity of spherical particles. The matching with the outer
(Navier-Stokes) solution is performed iteratively. Up to the present
date, however, only computations of two-dimensional sedimentation
problems have been reported by these authors. 

The objective of the present work is to develop a fictitious domain
method in which the forcing term is not obtained by any kind of
feed-back mechanism and where oscillations due to the fixed grid are
suppressed as much as possible. 
We will present a strategy to 
combine the original IB method's ability to smoothly transfer
quantities between Lagrangian and Eulerian positions on the one hand
with the advantages of a direct and explicit formulation of the
fluid-solid interaction force on the other hand. Thereby, the present
method yields less oscillatory particle forces than existing direct
methods and a higher efficiency compared to indirect methods. 

The organization of the paper is as follows. First, we will briefly
state the flow problem in mathematical terms (\S~\ref{sec-form})
before presenting our technique for imposing the presence of solid
bodies upon the fluid in \S~\ref{sec-forcing}. The treatment of the
equations of motion of the particles is described in
\S~\ref{sec-particles}. Results from a number of test problems of
increasing complexity as well as an evaluation of the efficiency of
our method are presented in \S~\ref{sec-results}.

\section{Formulation of the problem}\label{sec-form}
\begin{figure}
  \begin{center}
    \input{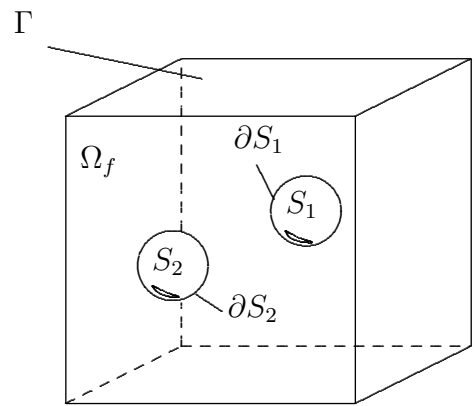}
  \end{center}
  \hfill
  \caption{Example of a configuration with two spherical
    objects $S_1$, $S_2$ and an interstitial fluid domain
    $\Omega_f$. The outer boundary is called $\Gamma$ and the
    interfaces between the fluid and the $i$th object
    $\partial S_i$. 
  }
  \label{fig-domain1}
\end{figure}

The Navier-Stokes equations for an incompressible fluid read:
\begin{equation}\label{equ-n-s}
  \left.
  \begin{array}{rcl}
    \partial_t\mathbf{u}+\left(\mathbf{u}\cdot\nabla\right)\mathbf{u}
    +\nabla p
    &=&\nu\nabla^2\mathbf{u}+\mathbf{f}\\
    \nabla\cdot\mathbf{u}&=&0
  \end{array}
  \quad\right]\quad
  \mathbf{x}\in\Omega
  \quad\mbox{}
\end{equation}
where $\mathbf{u}$ is the vector of fluid velocities, $p$ the pressure
normalized with the fluid density and $\mathbf{f}$ a volume force
term. These equations are enforced throughout the entire domain
$\Omega$, comprising the actual fluid domain $\Omega_f$ and the space
occupied by the $N_p$ suspended solid objects $\cup_{i=1}^{N_p}S_i$
(cf.\ figure~\ref{fig-domain1}). In \S~\ref{sec-forcing} below
the force term $\mathbf{f}$ will be formulated in such a way as to
represent the action of the solids upon the fluid.

In addition to providing appropriate initial conditions and 
conditions on the outer boundary $\Gamma$, 
we need to describe the motion of the suspended particles under the
action of gravity and hydrodynamic forces. This topic will be
discussed in \S~\ref{sec-particles}. 
\section{The action of the solids upon the fluid}\label{sec-forcing}
\subsection{Spatial discretization of Eulerian and Lagrangian
  variables}\label{sec-forcing-markers} 
We employ separate
discretizations for the Eulerian and Lagrangian quantities. 
First, we define a Cartesian, fixed grid $g_h$ consisting of uniformly
distributed nodes $\mathbf{x}_{ijk}=(i,j,k)\,h$ covering the domain
$\Omega$ (the constant $h$ is the mesh width, the integers $i,j,k$ are
the grid indices). 
A uniform grid is necessary in the present context in
order for essential identities of the
interpolation scheme to hold (cf.\ equations
\ref{equ-deltah-prop}-\ref{equ-deltah-prop-conseq} below).

Next, let us define for each embedded solid
object a number of $N_L$ points which are evenly
distributed over the fluid-solid interface and whose locations are
denoted by  
\begin{equation}\label{equ-def-force-point-loc}
  \mathbf{X}_l^{(i)}\in\partial S_i\qquad\forall\quad 1\leq l\leq N_L,
  \quad 1\leq i\leq N_p
\end{equation}
We will call these points {\it Lagrangian force points}. For the sake
of simplicity we will henceforth assume that all solid objects are of
equal shape and size. Therefore, the number of force points $N_L$ is
the same for each one of them.   

The locations $\mathbf{X}_l^{(i)}$ are used for interpolation purposes and
are constant in time with respect to a coordinate system attached to
the $i$th particle. This concept is related to the {\it
  Lagrangian marker points} used in the framework of the IB
method~\cite{peskin:02}. However, in the latter technique the marker
points are advected with the local fluid velocity whereas our
Lagrangian force points follow the 
rigid-body motion of the particles and, therefore, do not require
additional tracking, i.e.\ they do not constitute additional degrees
of freedom. Furthermore, we associate a discrete volume $\Delta
V_l^{(i)}$ with each force point such that the union of all these
volumes forms a thin shell (of thickness equal to one mesh width)
around each particle. 
Similar to references~\cite{fogelson:88,hoefler:00}, this allows us to
formulate a volume force at each Lagrangian force point as opposed to
the original IB method where a singular force is defined at the
Lagrangian marker points. 
Here we do not apply any forcing to the
interior of the particles for reasons of efficiency (cf.\ related
discussion in reference~\cite{hoefler:00}). 
From two-dimensional test computations of particle sedimentation we
could conclude that locating force points throughout the particle
volume does not lead to significantly different
results~\cite{uhlmann:04c}.

In appendix~\ref{app-marker} the geometrical definitions related to
the force point distribution for spherical particles and---in the case
of two dimensions---circular objects can be found. Although only these
two simple shapes are considered hereafter, the present method equally
applies to arbitrarily-shaped objects and even to rigid particle
surfaces which evolve in time (e.g.\ due to combustion processes). 
\subsection{Formulation of the volume force}\label{sec-forcing-force}
The forcing scheme has the purpose of imposing desired velocity
values at selected grid nodes. For a few geometrically simple and
stationary solid objects the grid nodes can be located on the
interface and forcing reduces to directly modifying the respective
matrix entries such that the local velocity vanishes. However, real 
particles have interfaces with arbitrary, time-dependent locations
w.r.t.\ the grid. Therefore, interpolation steps between Eulerian and
Lagrangian positions are necessary.  

For the purpose of discussion of the general concepts, let us write
the time-discretized momentum equation in the following form:  
\begin{equation}\label{equ-n-s-discr}
  \frac{\mathbf{u}^{n+1}-\mathbf{u}^{n}}{\Delta t}
  =\mathbf{rhs}^{n+1/2}+\mathbf{f}^{n+1/2}\qquad\forall\quad\mathbf{x}\in
  g_h\,, 
\end{equation}
where $\mathbf{rhs}^{n+1/2}$ regroups the convective, pressure and viscous
terms at some intermediate time level between $t^n$ and $t^{n+1}$. The force 
term which yields the desired velocity
$\mathbf{u}^{(d)}$ is then simply~\cite{fadlun:00}:
\begin{equation}\label{equ-n-s-discr-force}
  \mathbf{f}^{n+1/2}=\frac{\mathbf{u}^{(d)}-\mathbf{u}^{n}}{\Delta t}
  -\mathbf{rhs}^{n+1/2}
\end{equation}
at some selected grid nodes (and zero elsewhere). 
Kim {\it et al.}~\cite{kim:01} use grid nodes 
which are located inside the immersed object and adjacent to its 
interface, evaluating the desired velocity
$\mathbf{u}^{(d)}$ by means of a linear interpolation
procedure. Fadlun {\it et al.}~\cite{fadlun:00} discuss several
related interpolation techniques. Our personal experience is that in 
the case of
arbitrarily moving objects these procedures can lead to strong
oscillations of the hydrodynamical forces due to insufficient
smoothing~\cite{uhlmann:03}.  

Instead, we propose to evaluate the force term at the Lagrangian force
points $\mathbf{X}_l^{(i)}$, viz.
\begin{equation}\label{equ-n-s-discr-force-lag}
  \mathbf{F}^{n+1/2}=\frac{\mathbf{U}^{(d)}-\mathbf{U}^{n}}{\Delta t}
  -\mathbf{RHS}^{n+1/2}\qquad\forall\quad\mathbf{X}_l^{(i)}\,,
\end{equation}
In (\ref{equ-n-s-discr-force-lag}) and henceforth we use upper-case
letters for quantities evaluated at the locations of the Lagrangian
force points $\mathbf{X}_l^{(i)}$. The desired
velocity at a location on the interface between fluid and
solid is simply given by the rigid-body motion of the solid object:
\begin{equation}\label{equ-desired-vel}
  \mathbf{U}^{(d)}(\mathbf{X}_l^{(i)})=
  \mathbf{u}_c^{(i)}+\boldsymbol{\omega}_c^{(i)}\times
  (\mathbf{X}_l^{(i)}-\mathbf{x}_c^{(i)})\,,
\end{equation}
where $\mathbf{u}_c^{(i)}$, $\boldsymbol{\omega}_c^{(i)}$,
$\mathbf{x}_c^{(i)}$ are the translational and rotational velocity and
center coordinates of the $i$th solid, respectively. 

The two remaining terms on the right hand side of 
(\ref{equ-n-s-discr-force-lag}) can be collected as
\begin{equation}\label{equ-n-s-pre-predicted-lag}
  \tilde{\mathbf{U}}=\mathbf{U}^{n}+\mathbf{RHS}^{n+1/2}{\Delta t}
\end{equation}
which corresponds to a preliminary velocity obtained without applying
a force term. Its Eulerian counterpart,
\begin{equation}\label{equ-n-s-pre-predicted}
  \tilde{\mathbf{u}}=\mathbf{u}^{n}
  +\mathbf{rhs}^{n+1/2}{\Delta t}
  \qquad\forall\quad\mathbf{x}\in g_h
\end{equation}
is available explicitly in our scheme (cf.\
equation~\ref{equ-hybrid-discr-algo-lag-rhs} below).  
In order to complete the evaluation of the forcing term in 
(\ref{equ-n-s-discr}), we still need to provide a mechanism for
transferring the preliminary velocity
($\tilde{\mathbf{u}},\tilde{\mathbf{U}}$) and the force itself
($\mathbf{F}^{n+1/2},\mathbf{f}^{n+1/2}$) back and forth between
Lagrangian and Eulerian locations.  
\subsection{Transfer of quantities between Lagrangian and Eulerian
  locations} 
\label{sec-forcing-delta}
Here we use the class of regularized delta functions introduced by
Peskin~\cite{peskin:72,peskin:02} as kernels in the transfer steps
between Lagrangian and Eulerian locations. Dropping the temporal
superscripts for convenience, we write:
\bse{equ-transfer}
\bea\label{equ-transfer-eul-lag}
\tilde{\mathbf{U}}(\mathbf{X}_l^{(m)})&=&
\sum_{\mathbf{x}\in g_h}\tilde{\mathbf{u}}(\mathbf{x})\,
\delta_h(\mathbf{x}-\mathbf{X}_l^{(m)})\,h^3
\qquad\forall\,1\le m\le N_p;\,1\le l\le N_L
\\\label{equ-transfer-lag-eul}
\mathbf{f}(\mathbf{x})&=&
\sum_{m=1}^{N_p}
\sum_{l=1}^{N_L}\mathbf{F}(\mathbf{X}_l^{(m)})\,
\delta_h(\mathbf{x}-\mathbf{X}_l^{(m)})\,\Delta V_l^{(m)}
\qquad\forall\,\mathbf{x}\in g_h
\eea
\ese
The salient properties of the kernels $\delta_h$ are the following:
\begin{itemize}
\item %
  $\delta_h$ is a continuously differentiable function and
  therefore yields a smoother transfer than e.g.\ linear
  interpolation.  
\item %
  Interpolation using the kernels $\delta_h$ is
  second-order accurate for smooth fields (cf.\
  \S~\ref{sec-results-1way-tg}). 
\item %
  The support of the regularized delta function is
  small, which makes the evaluation of the sums in
  (\ref{equ-transfer}) relatively cheap. In particular, we use the
  expression for $\delta_h$ defined by Roma {\it et
  al.}~\cite{roma:99}, involving only three grid points in each
  coordinate direction.
\item %
  For all real shifts $\mathbf{X}$ we have:
  \bse{equ-deltah-prop}
  \bea
  \sum_{\mathbf{x}\in g_h}\delta_h(\mathbf{x}-\mathbf{X})\,h^3&=&1\\ 
  \sum_{\mathbf{x}\in g_h}(\mathbf{x}-\mathbf{X})
  \,\delta_h(\mathbf{x}-\mathbf{X})\,h^3&=&0
  \eea
  \ese
  which are discrete analogues of basic properties of the Dirac delta
  function. As a consequence it can be shown~\cite{peskin:02} that the
  total amount of force and torque added to the fluid is not changed
  by the transfer step in (\ref{equ-transfer-lag-eul}), i.e.
  \bse{equ-deltah-prop-conseq}
  \bea\label{equ-deltah-prop-conseq-force}
  \sum_{\mathbf{x}\in g_h}\mathbf{f}(\mathbf{x})\,h^3&=&
  \sum_{m=1}^{N_p}\sum_{l=1}^{N_L}\mathbf{F}(\mathbf{X}_l^{(m)})\,
  \Delta V_l^{(m)}
  \\\label{equ-deltah-prop-conseq-torque}
  \sum_{\mathbf{x}\in g_h}\mathbf{x}\times\mathbf{f}(\mathbf{x})\,h^3&=&
  \sum_{m=1}^{N_p}\sum_{l=1}^{N_L}\mathbf{X}_l^{(m)}\times
  \mathbf{F}(\mathbf{X}_l^{(m)})\, 
  \Delta V_l^{(m)}
  \eea
  \ese
\end{itemize}
\subsection{The flow solver} 
\label{sec-forcing-flow-solver}
Our Navier-Stokes solver is based upon a conventional fractional-step
method for enforcing continuity. A three-step Runge-Kutta scheme  
is used for the convective terms while the viscous terms are treated
by the Crank-Nicholson method, leading to overall formal second-order
temporal accuracy. 

The spatial derivatives are evaluated by means of second-order,
central finite-difference operators on a staggered 
grid. Staggering implies that each component of velocity $u_\beta$ is
defined at its own Eulerian grid locations, say $\mathbf{x}\in
g_h^{(\beta)}$. Therefore, the transfer between Eulerian and
Lagrangian locations in (\ref{equ-transfer}) needs to be carried out
for each component individually. 

The discretized flow equations, including the fluid-solid coupling
term, for the $k$th Runge-Kutta step are the following:
\bse{equ-hybrid-discr-algo}
\bea\nonumber
\tilde{\mathbf{u}}&=&
\mathbf{u}^{k-1}
+\Delta t\left(
2\alpha_k\nu\nabla^2\mathbf{u}^{k-1}
-2\alpha_k\nabla p^{k-1}\right.
\\\label{equ-hybrid-discr-algo-lag-rhs}
&&\left.
-\gamma_k\left[(\mathbf{u}\cdot\nabla)\mathbf{u}\right]^{k-1}
-\zeta_k\left[(\mathbf{u}\cdot\nabla)\mathbf{u}\right]^{k-2}
\right)
\\\label{equ-hybrid-discr-algo-lag-interpol}
\tilde{{U}}_\beta(\mathbf{X}_l^{(m)})&=&
\sum_{\mathbf{x}\in g_h^{(\beta)}}\tilde{{u}_\beta}(\mathbf{x})\,
\delta_h(\mathbf{x}-\mathbf{X}_l^{(m)})\,h^3
\quad\forall\,l;\,m;\,1\le \beta\le 3
\\\label{equ-hybrid-discr-algo-lag-force}
\mathbf{F}(\mathbf{X}_l^{(m)})&=&
\frac{\mathbf{U}^{(d)}(\mathbf{X}_l)
-\tilde{\mathbf{U}}(\mathbf{X}_l)}{\Delta t}
\quad\qquad\qquad\forall\,l;\,m
\\\nonumber
{f}_\beta(\mathbf{x})&=&
\sum_{m=1}^{N_p}
\sum_{l=1}^{N_L}{F}_\beta(\mathbf{X}_l^{(m)})\,
\delta_h(\mathbf{x}-\mathbf{X}_l^{(m)})\,\Delta V_l^{(m)}
\quad\forall\,\mathbf{x}\in g_h^{(\beta)}
\\\label{equ-hybrid-discr-algo-eul-force}
&&\quad\qquad\qquad\qquad\qquad\qquad\qquad\qquad\qquad 1\le \beta\le 3
\\\label{equ-hybrid-discr-algo-predict}
\nabla^2\mathbf{u}^\ast-\frac{\mathbf{u}^\ast}{\alpha_k\nu\Delta t}&=&
-\frac{1}{\nu\alpha_k}\left(\frac{\tilde{\mathbf{u}}}{\Delta
      t}+\mathbf{f}^{k}\right) 
  +\nabla^2\mathbf{u}^{k-1}
\\\label{equ-hybrid-discr-algo-poisson}
\nabla^2\phi^{k}&=&\frac{\nabla\cdot\mathbf{u}^\ast}{2\alpha_k\Delta t}\,,
\\\label{equ-hybrid-discr-algo-update-vel}
\mathbf{u}^{k}&=&\mathbf{u}^\ast-2\alpha_k\Delta
t\nabla\phi^k\,,
\\\label{equ-hybrid-discr-algo-update-press}
p^{k}&=&p^{k-1}+\phi^k-\alpha_k\Delta t\,\nu\nabla^2\phi^k
\eea
\ese
where the set of coefficients $\alpha_k$, $\gamma_k$, $\zeta_k$ ($1\le
k\le 3$) is given in \cite{rai:91}. The intermediate variable $\phi$
is the so-called ``pseudo-pressure'' and has no physical
meaning. 
Equations
(\ref{equ-hybrid-discr-algo-predict})-
(\ref{equ-hybrid-discr-algo-update-press}) with $\mathbf{f}$ set to
zero correspond to the basic fractional-step
method~\cite{verzicco:96}.

In the case of periodic boundary conditions, the 
spatial average of the force term,
$\int_\Omega\mathbf{f}\mbox{d}\mathbf{x}/||\Omega||$, needs to be
subtracted from the momentum equation for compatibility
reasons~\cite{fogelson:88,hoefler:00}. 

It should be pointed out that the resulting velocity field
$\mathbf{u}^k$ is divergence-free in the sense of the discrete
operators. As in previous studies on explicit formulations of the
coupling force~\cite{fadlun:00}, there are no additional restrictions
of the time step stemming from the fluid-solid coupling. This means
that stable integration is possible with values of the CFL number
close to the theoretical limit of $\sqrt{3}$ imposed by the basic
Runge-Kutta scheme. 

In practice, the solution of the Helmholtz
(\ref{equ-hybrid-discr-algo-predict}) and Poisson 
(\ref{equ-hybrid-discr-algo-poisson}) 
problems is performed as follows. In two space dimensions, a direct
solution method based on cyclic reduction~\cite{schumann:76} is used
for solving both types of implicit problems. For reasons of efficient
implementation on multi-processor machines in the case of three space
dimensions, the Helmholtz problems are simplified by
second-order-accurate approximate factorization and the Poisson
problem is solved by a multi-grid technique. 

\section{The motion of the solid particles}\label{sec-particles}
The motion of the particles is governed by Newton's equations for
linear and angular momentum of a rigid body. 
Evaluating the hydrodynamic forces acting upon a particle by means of
a momentum balance over the corresponding fluid domain we can write 
(cf.~appendix~\ref{app-newton}):
\bse{equ-particles-newton-2-present}
\bea\nonumber
V_c^{(m)}\,\left(\rho_p^{(m)}-\rho_f\right)\,\dot{\mathbf{u}}_c^{(m)}
&=&
-\rho_f\underbrace{\sum_l\mathbf{F}(\mathbf{X}_l^{(m)})\,\Delta V_l^{(m)}}_{
  \displaystyle{\mathcal F}^{(m)}}
+(\rho_p^{(m)}-\rho_f)\,V_c^{(m)}\,\mathbf{g}
\\\label{equ-particles-newton-2-present-translation}
\\\nonumber
I_c^{(m)}\,\dot{\boldsymbol{\omega}}_c^{(m)}
&=&
-\rho_f\underbrace{\sum_l\left(\mathbf{X}_l^{(m)}-\mathbf{x}_c^{(m)}\right)
  \times\mathbf{F}(\mathbf{X}_l^{(m)})\,\Delta V_l^{(m)}}_{\displaystyle{\mathcal T}^{(m)}}
\\\label{equ-particles-newton-2-present-rotation}
&&+\rho_f\frac{\mbox{d}}{\mbox{d}t}\underbrace{\int_{\mathcal
    S^{(m)}}\left((\mathbf{x}-\mathbf{x}_c^{(m)})\times\mathbf{u}\right)\,\mbox{d}\mathbf{x}}_{
  \displaystyle{\mathcal I}^{(m)}}
\eea
\ese
where $V_c^{(m)}$, $I_c^{(m)}$, $\rho_p^{(m)}$ are the volume, moment
of inertia and density of the $m$th particle; $\rho_f$ the fluid
density; $\mathbf{g}$ the vector of gravitational acceleration.

The second term on the r.h.s.\ of
(\ref{equ-particles-newton-2-present-rotation})   
represents the rate of change of angular momentum of the fluid
occupying the domain of the $m$th solid. Its contribution is due to
the fact that applying the fluid-solid coupling force
only to the surface of each particle
causes a residual non-rigid motion of fluid inside the particle domain 
${\mathcal S}^{(m)}$. 
In practice the integral ${\mathcal I}^{(m)}$ was evaluated as a sum
over each grid cell with the cell's volumetric solid fraction as a
weight. In our three-dimensional applications the rate-of-change term
was approximated by supposing rigid-body motion inside the solid
volume, i.e.\ using equation (\ref{equ-rate-of-change-torque}). This
was done for reasons of efficiency and a justification is given in
\S~\ref{sec-results-2d-wake}. 

The equations of motion (\ref{equ-particles-newton-2-present}) are
discretized in time by the same Runge-Kutta procedure as the 
the fluid equations:
\bse{equ-particles-newton-2-present-translation-discrete}
\bea\label{equ-particles-newton-2-present-translation-discrete-u}
\frac{\mathbf{u}_c^k-\mathbf{u}_c^{k-1}}{\Delta t}
&=&
-\frac{\rho_f}{V_c(\rho_p-\rho_f)}\,{\mathcal F}^k
+2\alpha_k\mathbf{g}
\\\label{equ-particles-newton-2-present-translation-discrete-x}
\frac{\mathbf{x}_c^k-\mathbf{x}_c^{k-1}}{\Delta t}
&=&
\alpha_k\left(
  \mathbf{u}_c^k+\mathbf{u}_c^{k-1}
\right)
\\\label{equ-particles-newton-2-present-translation-discrete-om}
\frac{\boldsymbol{\omega}_c^k-\boldsymbol{\omega}_c^{k-1}}{\Delta t}
&=&
-\frac{\rho_f}{I_c}\,{\mathcal T}^k
+\frac{\rho_f}{I_c}
\frac{\left({\mathcal I}^{k}-{\mathcal I}^{k-1}\right)}{\Delta t}
\eea
\ese
where we have dropped the superscript for the particle index ${}^{(m)}$
in favor of the Runge-Kutta sub-step index. 
Also note that the angular position is not needed for advancing the
equations. 
In the case of evaluating the rate-of-change term from a full
rigidity approximation, equation
(\ref{equ-particles-newton-2-present-translation-discrete-om}) is
replaced by:  
\begin{equation}\label{equ-particles-newton-2-present-translation-discrete-om-approx}
  \frac{\boldsymbol{\omega}_c^k-\boldsymbol{\omega}_c^{k-1}}{\Delta t}
  =
  -\frac{\rho_f}{\rho_p-\rho_f}\frac{1}{(I_c/\rho_p)}\,{\mathcal T}^k
\end{equation}
\subsection{Weak coupling of fluid and particle equations}
\label{sec-particles-weak}
Our scheme consists in first solving the fluid equations
(\ref{equ-hybrid-discr-algo}) with the particle positions and
velocities known from the previous Runge-Kutta level and then solving
the particle equations
(\ref{equ-particles-newton-2-present-translation-discrete}) as
indicated, using the most recent flow field. In order to simplify the
notation, consider the following model system 
where each sub-system contains only one variable, flow velocity
$\mathbf{u}$ and particle center velocity $\mathbf{u}_c$, respectively: 
\bse{equ-numa-newton-coupling-model}
\bea\label{equ-numa-newton-coupling-model-u}
\mathbf{u}^k&=&\mathbf{u}^{k-1}+\Psi_1\left(
\mathbf{u}^k,\mathbf{u}^{k-1},\mathbf{u}^{k-2},\mathbf{u}_c^{k-1}
\right)
\\\label{equ-numa-newton-coupling-model-uc}
\mathbf{u}_c^k&=&\mathbf{u}_c^{k-1}+\Psi_2\left(
\mathbf{u}_c^{k-1},\mathbf{u}^{k-1},\mathbf{u}^{k-2}
\right)
\eea
\ese
and the functions $\Psi_1$, $\Psi_2$ represent the time advancement of
each subsystem. This model is representative of our full system
inasmuch as it is implicit in the former case and explicit in the
latter. The coupling between both sub-systems is explicit, also called
``weak coupling''. 

It has been noted in the past that the treatment of very light
particles presents a problem for methods where the fluid equations
are weakly coupled to the equations of motion for the rigid
particles. Hu {et al.}~\cite{hu:01} show how growing oscillations of
the particle velocity can arise depending on the added mass in the
case of a particle accelerating from rest due to gravity while using
fully explicit coupling. 
In practice we have found that there is a lower limit of the density
ratio for stable weakly-coupled integration of the fluid-particle
system for the present method: 
$\rho_p/\rho_f\gtrsim 1.05$ for circular disks,
$\rho_p/\rho_f\gtrsim 1.2$ for spherical particles. We have observed
that the limiting value does not depend significantly upon the
chosen time step. 
Incidentally, the explicitly-coupled scheme of Kajishima and
Takiguchi~\cite{kajishima:02} allows for density ratios
$\rho_p/\rho_f\gtrsim 2$ (circular disks) according to our
experience.  
In cases where the weakly coupled procedure is unstable, 
Gauss-Seidel-like sub-iterations for each Runge-Kutta
step can be performed~\cite{uhlmann:04c}. In order to avoid the
additional overhead associated with iterative coupling, fully
implicit coupling---like the method proposed in
reference~\cite{patankar:01}---is in principle preferable. This aspect
is left as a future extension of our scheme. In the following examples
we have chosen density ratios above the indicated threshold. 
\section{Results}\label{sec-results}
\subsection{Test cases with one-way coupling}\label{sec-results-1way}
First we consider configurations where the equations for the motion
of particles
(\ref{equ-particles-newton-2-present-translation-discrete}) need not
be solved because Lagrangian velocity and position data is explicitly 
known. Thereby, the principle features of our new force formulation
can be validated in a separate way. Furthermore, we will initially
focus on two-dimensional flows for simplicity.   
\subsubsection{Taylor-Green vortices}\label{sec-results-1way-tg}
In order to establish the influence of the relative position of the
immersed boundary with respect to the fixed grid, we consider the case
of an array of decaying vortices with analytical solution 
\bse{equ-taylor-green}
\begin{eqnarray}\label{equ-taylor-green-u}
  u(x,y,t)&=&\sin(k_xx)\cos(k_yy)e^{-(k_x^2+k_y^2)\nu t}
  \\\label{equ-taylor-green-v}
  v(x,y,t)&=&-\frac{k_x}{k_y}\sin(k_yy)\cos(k_xx)e^{-(k_x^2+k_y^2)\nu t}
  \\\label{equ-taylor-green-pressure}
  p(x,y,t)&=&\frac{1}{2}\left(\cos^2(k_yy)\frac{k_x^2}{k_y^2}-\sin^2(k_xx)\right) 
  e^{-2(k_x^2+k_y^2)\nu t}
\end{eqnarray}
\ese
where $k_x=k_y=\pi$. This case is simulated in an embedded circular
domain with radius unity and centered at the origin of the
computational domain $\Omega=[-1.5,1.5]\times[-1.5,1.5]$. 
This flow has been computed in reference~\cite{kim:01} in a
quadrilateral embedded domain. 
The viscosity is set to $\nu=0.2$
and the equations are advanced for $0\le t\le 0.3$ using
a time step of $\Delta t=0.001$. The exact solution
(\ref{equ-taylor-green}) provides: (a) the initial field at $t=0$; (b) the
time-dependent boundary conditions at the domain boundary $\Gamma$;
(c) the time-dependent desired velocity values $\mathbf{U}^{(d)}$ at
the circumference of the embedded circle.  

\begin{figure}
  \begin{center}
    \begin{minipage}{1cm}
      \rotatebox{90}{
        max.\ error}
    \end{minipage}
    \begin{minipage}{10cm}
      \includegraphics*[width=\linewidth]{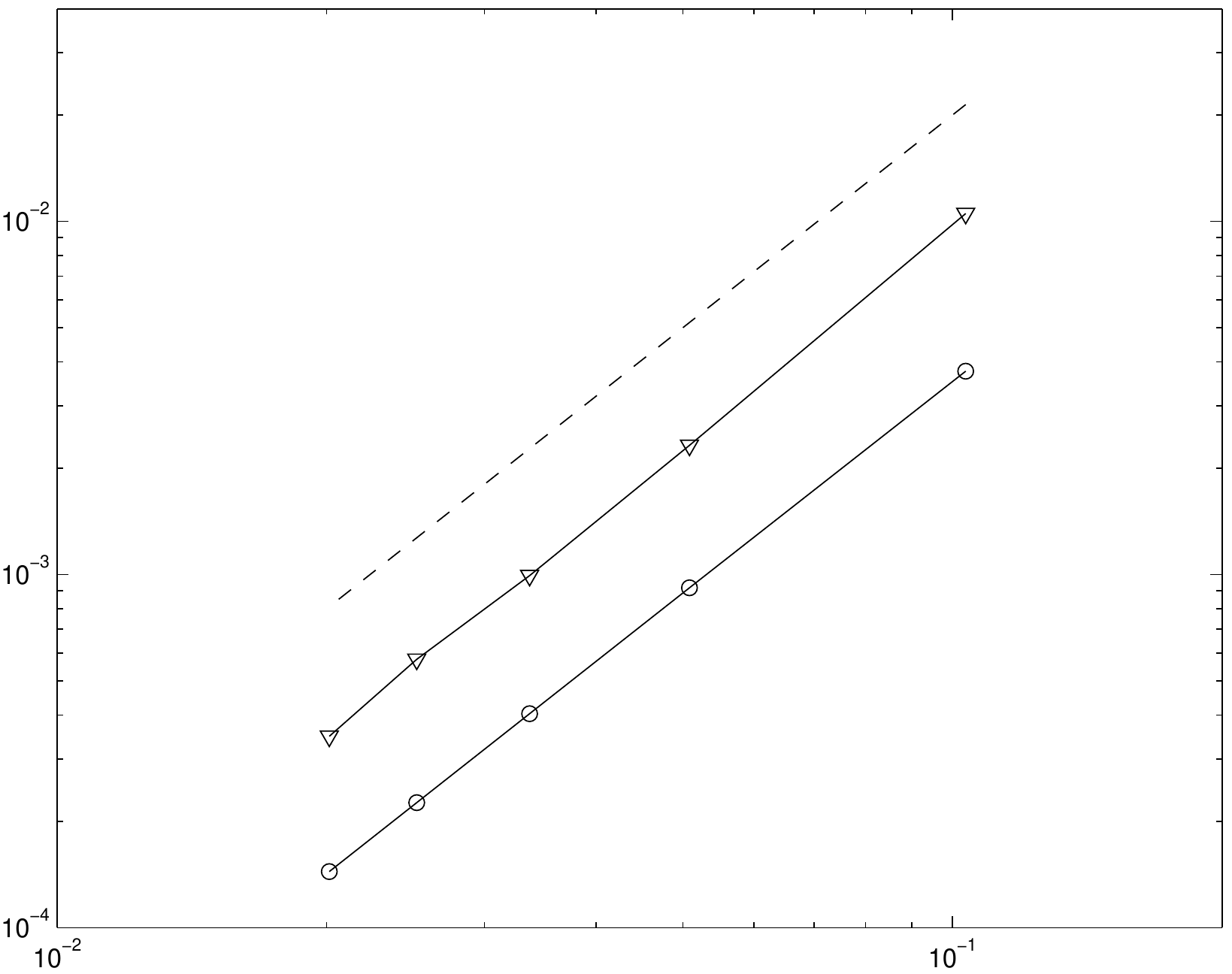}
    \end{minipage}
    \\
    $h$
  \end{center}
  \caption{Maximum error of the velocity field in the case of
    Taylor-Green vortices. The error is shown as a function of the
    mesh width $h$ without embedded boundary ($\circ$) and with
    circular embedded boundary ($\triangle$); the dashed line is
    proportional to $h^2$.}
  \label{fig-tg-conv}
\end{figure}

\begin{figure}
  \begin{center}
    \begin{minipage}{1cm}
      \rotatebox{90}{
        max.\ error}
    \end{minipage}
    \begin{minipage}{10cm}
      \includegraphics*[width=\linewidth]{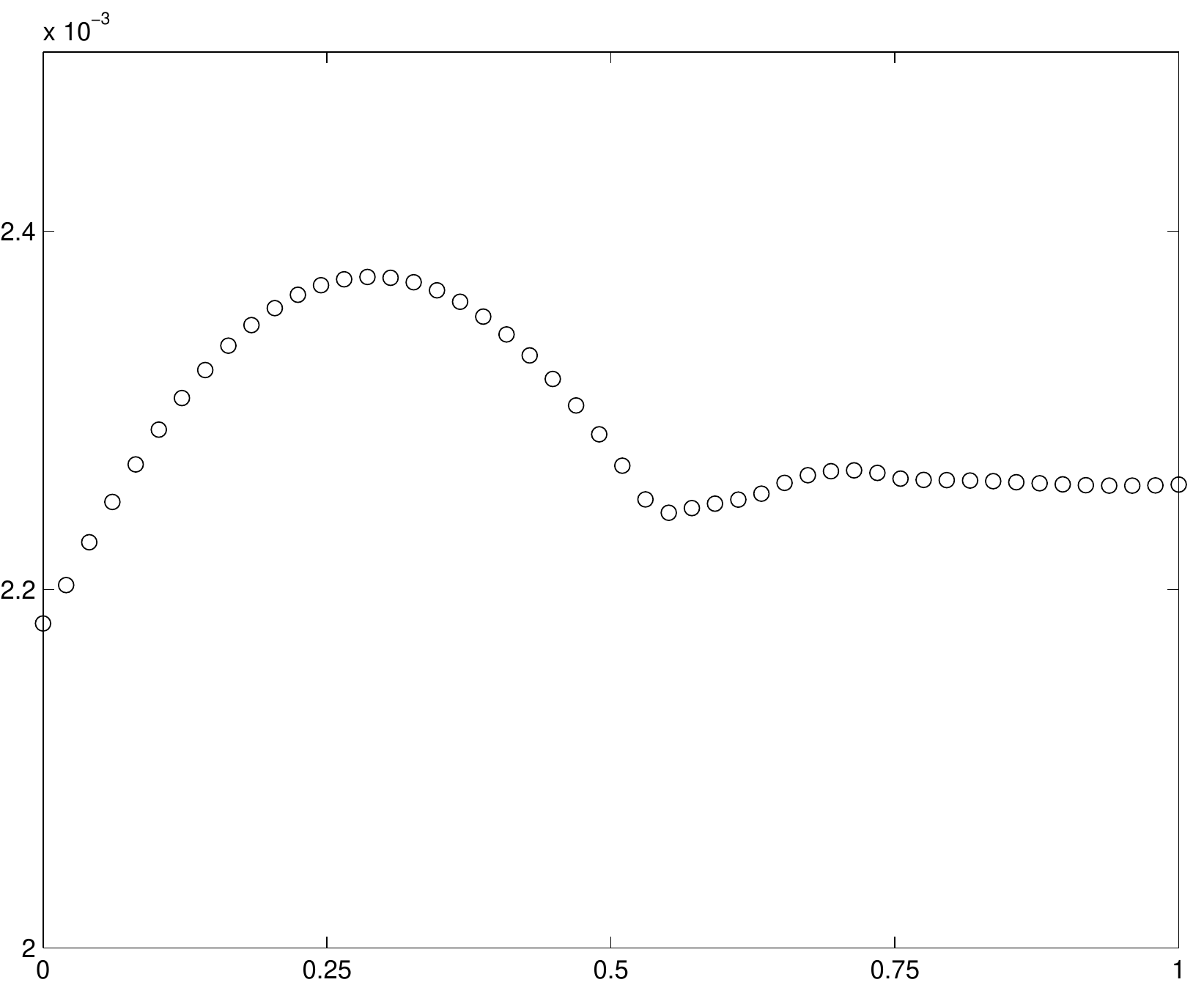}
    \end{minipage}
    \\
    $x_c/h$
  \end{center}
  \caption{Maximum error of the velocity field in the case of
    Taylor-Green vortices. The error is shown as a function of
    the horizontal position of the embedded circle, for fixed
    $h=0.05$.}
  \label{fig-tg-shift}
\end{figure}

Figure~\ref{fig-tg-conv} shows the maximum error of velocity for grid
nodes located inside the embedded domain, plotted as a function of the
mesh size $h$. 
Second order convergence is observed, which confirms 
the accuracy of the interpolation with the regularized delta function
in the case of smooth fields. 
The important result of this case is that the error 
is not very sensitive to the position of the immersed boundary
relative to the grid. This feature can be demonstrated by fixing the
resolution ($h=0.05$) and shifting the circular sub-domain
horizontally by fractions of the mesh-width.
Figure~\ref{fig-tg-shift} shows that the error varies indeed very
smoothly as a function of the shift.  
\subsubsection{A cylinder in uniform cross-flow}
\label{sec-results-1way-cyl} 
We place a cylinder with radius $r_c=0.15$ at the origin of the domain
$\Omega_1=[-1.85,6.15]\times[-4,4]$. 
At the three boundary segments
$x=-1.85$, $y=-4$ and $y=4$ we impose a uniform free-stream velocity
$\mathbf{u}=(1,0)$. The boundary at $x=6.85$ is treated by a
convective outflow condition. A homogeneous Neumann condition is used
at all four boundaries for the Poisson equation of 
pseudo-pressure (\ref{equ-hybrid-discr-algo-poisson}). 

The uniform grid has $1024\times 1024$ nodes, i.e.\ the ratio
of particle diameter to mesh size is $D/h=38.4$. This corresponds to
the finest grid used in reference~\cite{lai:00}. The
Reynolds number $Re_D=\frac{u_\infty D}{\nu}$ is set to 100. 
The time step is $\Delta t=0.003$, leading to a maximum CFL number of
approximately 0.6. 

Table~\ref{tab-cylinder-stat} shows the resulting drag and lift
coefficients $C_D$, $C_L$ as well as the Strouhal number $St$ defined
from the oscillation frequency of the lift force. 
It should be
mentioned that drag and lift forces were evaluated as sums over the
fluid-solid coupling terms (\ref{equ-hybrid-discr-algo-lag-force}) and
summing contributions from the three Runge-Kutta sub-steps
(cf.~\cite{kim:01} and the discussion on methods 
for determining drag/lift forces in~\cite{lai:00}).
The agreement with reference values from the
literature~\cite{liu:98} is generally good. 
In particular, the Strouhal number is predicted
within 4\% error, the amplitude of the lift and drag fluctuations with
errors of 3\% and 8\% respectively. 
However, the mean drag is overpredicted by 11\%; this is also true for
the following case of an oscillating cylinder where an overprediction
of approximately 10\% is obtained (cf.\
table~\ref{tab-cylinder-osc}). Using the IB method, Lai and
Peskin~\cite{lai:00} obtained a 
similar overprediction of the mean drag in this case and attributed 
the discrepancy to the confinement effect due to the
finite distance of the lateral boundaries which are treated as slip
walls. 
The importance of the domain size was previously 
demonstrated by Behr {\it et al.}~\cite{behr:95}. 
More recently, Linnick and Fasel~\cite{linnick:05} reported
an irregular drag coefficient for a computational domain measuring
approximately $18D$ in the cross-stream direction; using $43D$
corrected the problem in their case. 
For the purpose of verification, we have repeated our simulation in
a larger domain $\Omega_2=1.5\Omega_1$ (i.e.\ $40D\times 40D$) while
maintaining the same spatial resolution ($1536\times 1536$ nodes)
and time step. As can be seen in table~\ref{tab-cylinder-stat} the
effect is a decrease of the mean drag, leading to a reduced error of
less than 8\%. Furthermore, the amplitude of the lift fluctuations
now matches with the reference values from~\cite{liu:98} and
the prediction of the Strouhal number is further improved.

\begin{figure}
  \begin{center}
    \begin{minipage}{.5cm}
      $C_p$
    \end{minipage}
    \begin{minipage}{.43\linewidth}
      \includegraphics*[width=\linewidth]{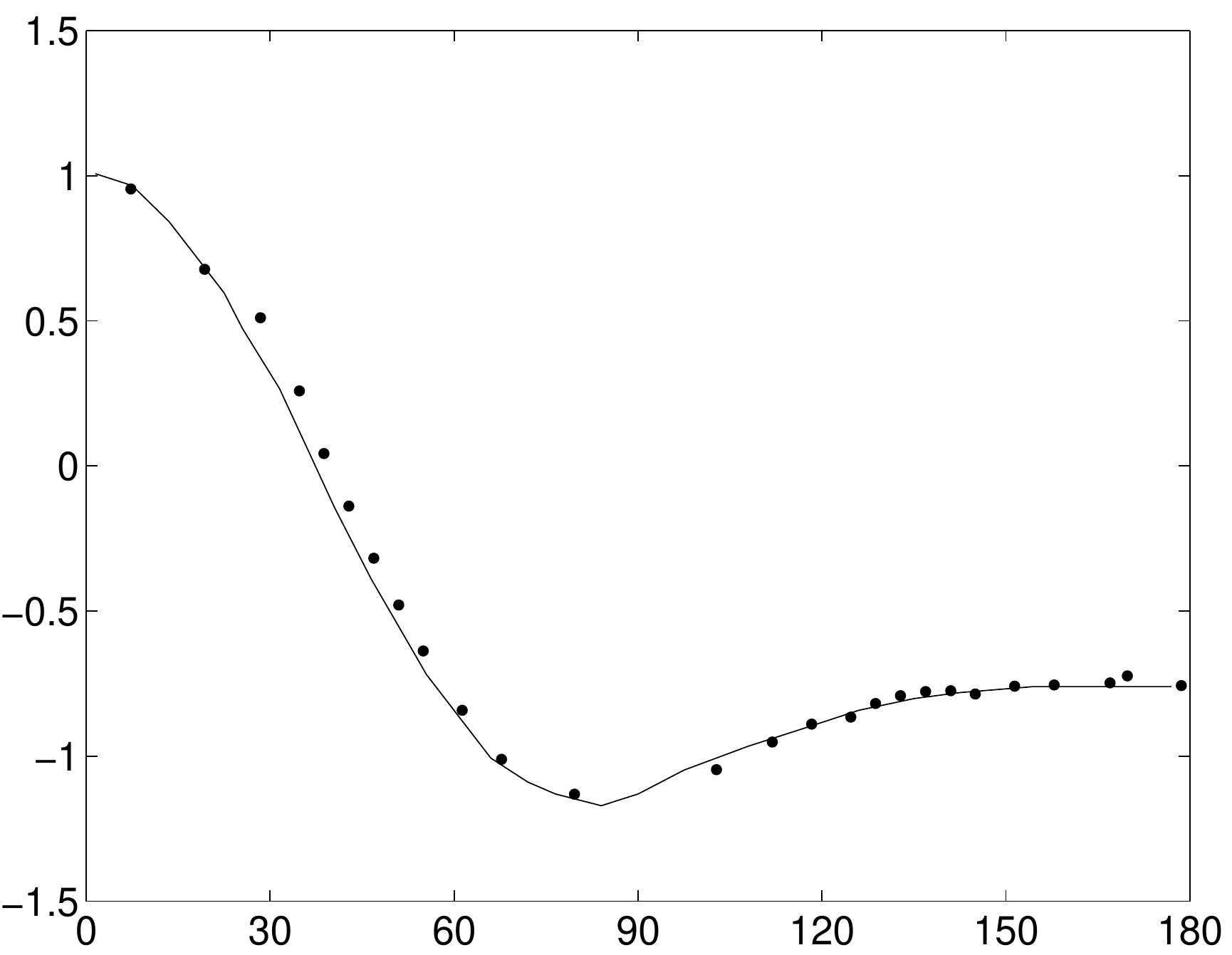}
    \end{minipage}\\
    $\theta$ (degrees)
  \end{center}
  \caption{Surface pressure coefficient $C_p$ as a
      function of the angle $\theta$ ($\theta=0^\circ$ corresponds to the
      stagnation point, $\theta=180^\circ$ to the base point) for the
      case of a stationary cylinder in uniform flow at
      $Re_D=100$. $\bullet$, present results; \solid, data from Park
      et al.~\cite{park:98}.}
  \label{fig-stat-cyl-cp}
\end{figure}

Figure~\ref{fig-stat-cyl-cp} shows the time-averaged
pressure coefficient $C_p=(p-p_\infty)/u_\infty^2$ along the
cylinder surface. The data is plotted at the nearest pressure nodes
outside the cylinder. A very good agreement with the
well-established results of Park et al.~\cite{park:98} is obtained,
including the stagnation and base region.

We now set the cylinder in time-periodic motion in the direction which
is perpendicular to the cross-flow, i.e.:
\begin{equation}\label{equ-results-osc-cyl-motion}
  y_c(t)=A\,\sin(2\pi\,f_f\,t)\,,
\end{equation}
with the amplitude set to $A=0.2D$ and the frequency $f_f$ to $0.8$
times the natural shedding frequency, i.e.\ $f_f=0.52$. The maximum
velocity of the cylinder motion is approximately $0.2u_\infty$. The
value for the Reynolds number is set to $Re_D=185$ in order to match
the conditions of reference~\cite{lu:96}. 
All other parameters remain the same as in the corresponding
stationary case above.
In particular, the smaller domain $\Omega_1$ was used if
not otherwise stated.

\begin{figure}
  \begin{minipage}{.5cm}
    $C_D$
  \end{minipage}
  \begin{minipage}{.43\linewidth}
    \begin{center}
      present, using 3-point $\delta_h$ of~\cite{roma:99}\\
      \includegraphics*[width=\linewidth]{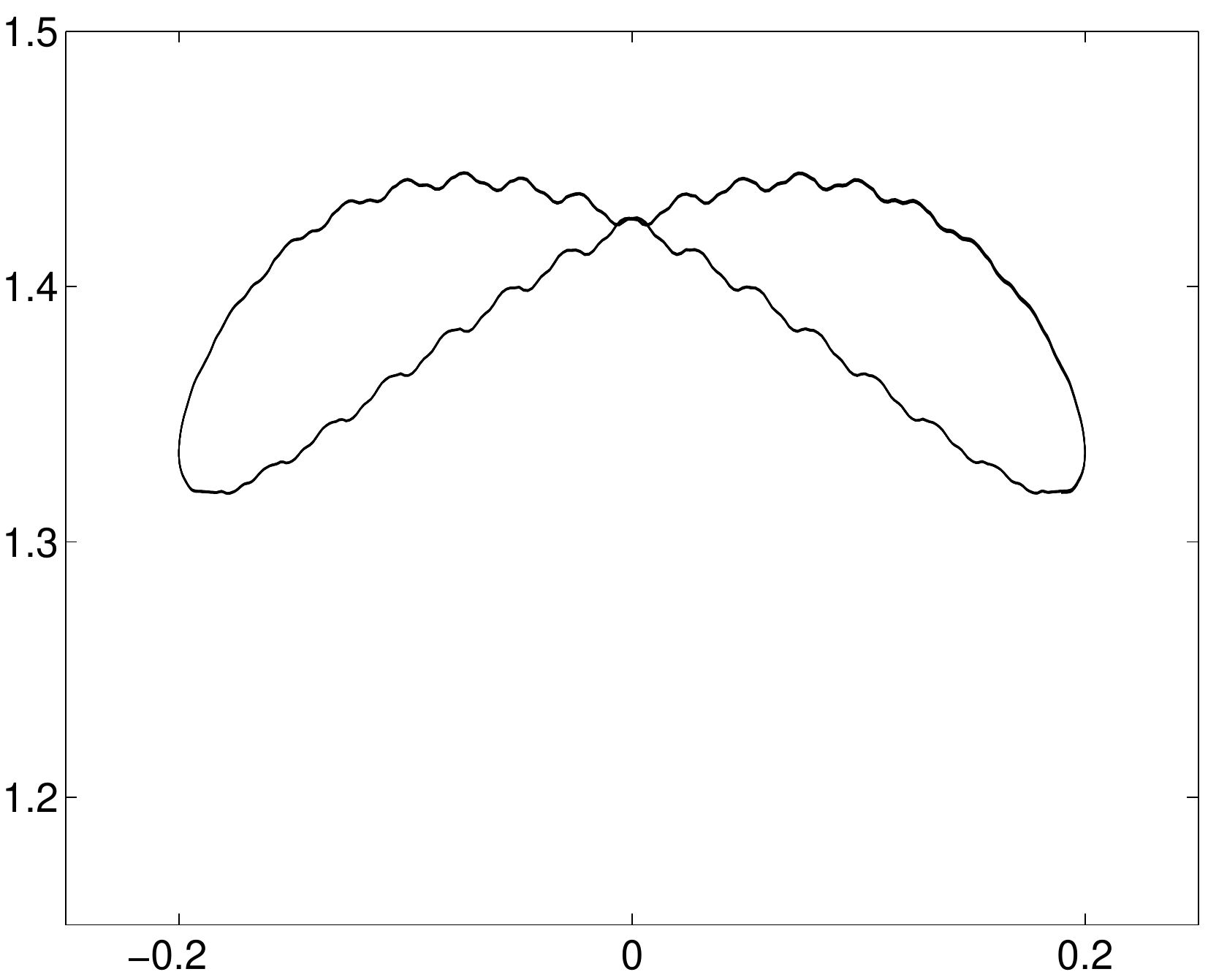}\\[-2.ex]
      $y_c(t)/D$
    \end{center}
  \end{minipage}
  \hfill
  \begin{minipage}{.5cm}
    $C_D$
  \end{minipage}
  \begin{minipage}{.43\linewidth}
    \begin{center}
      method of~\cite{kajishima:02}\\
      \includegraphics*[width=\linewidth]{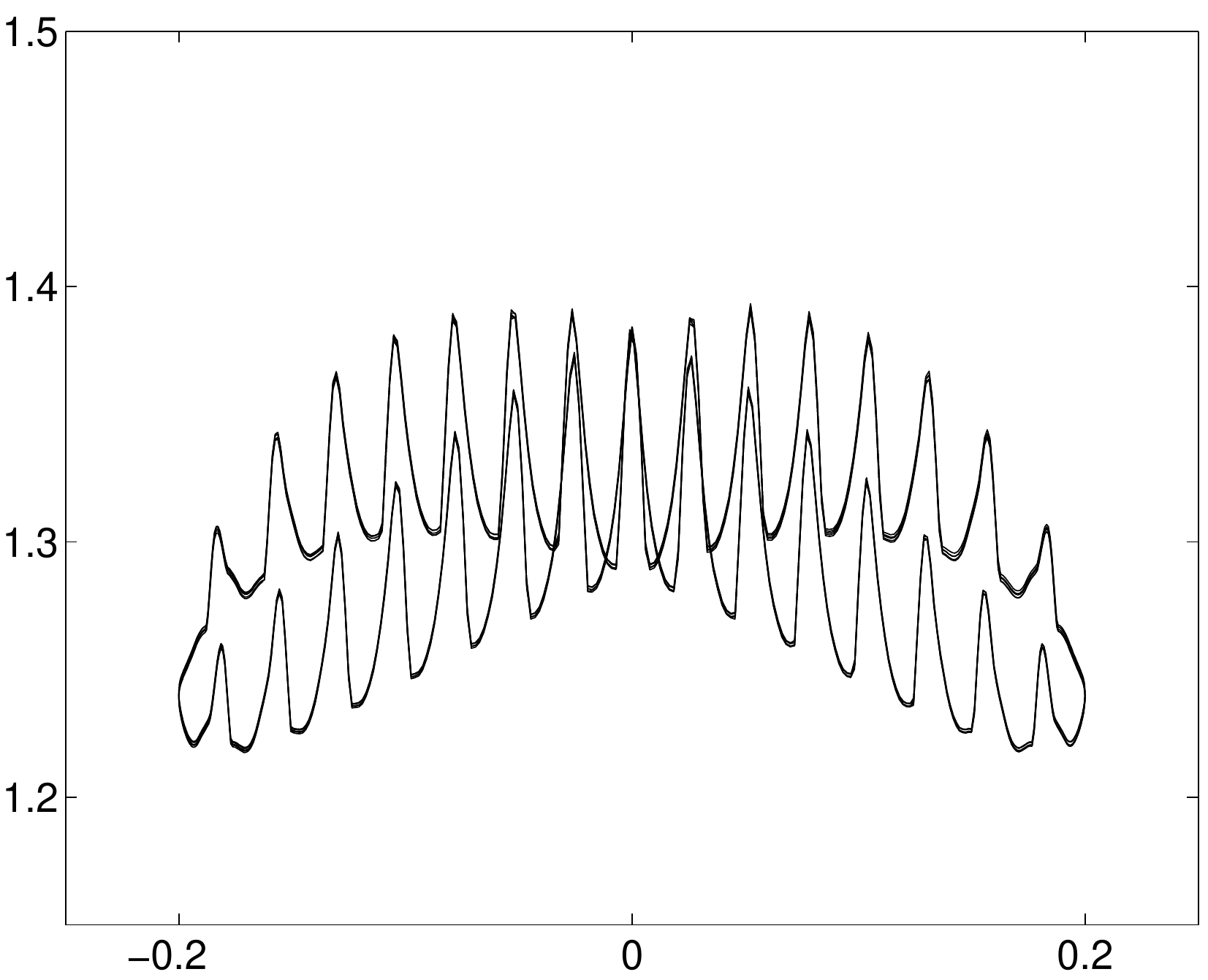}\\[-2.ex]
      $y_c(t)/D$
    \end{center}
  \end{minipage}\\[3ex]
  \begin{minipage}{.5cm}
    $C_D$
  \end{minipage}
  \begin{minipage}{.43\linewidth}
    \begin{center}
      present, using 4-point $\delta_h$ of~\cite{peskin:02}\\
      \includegraphics*[width=\linewidth]{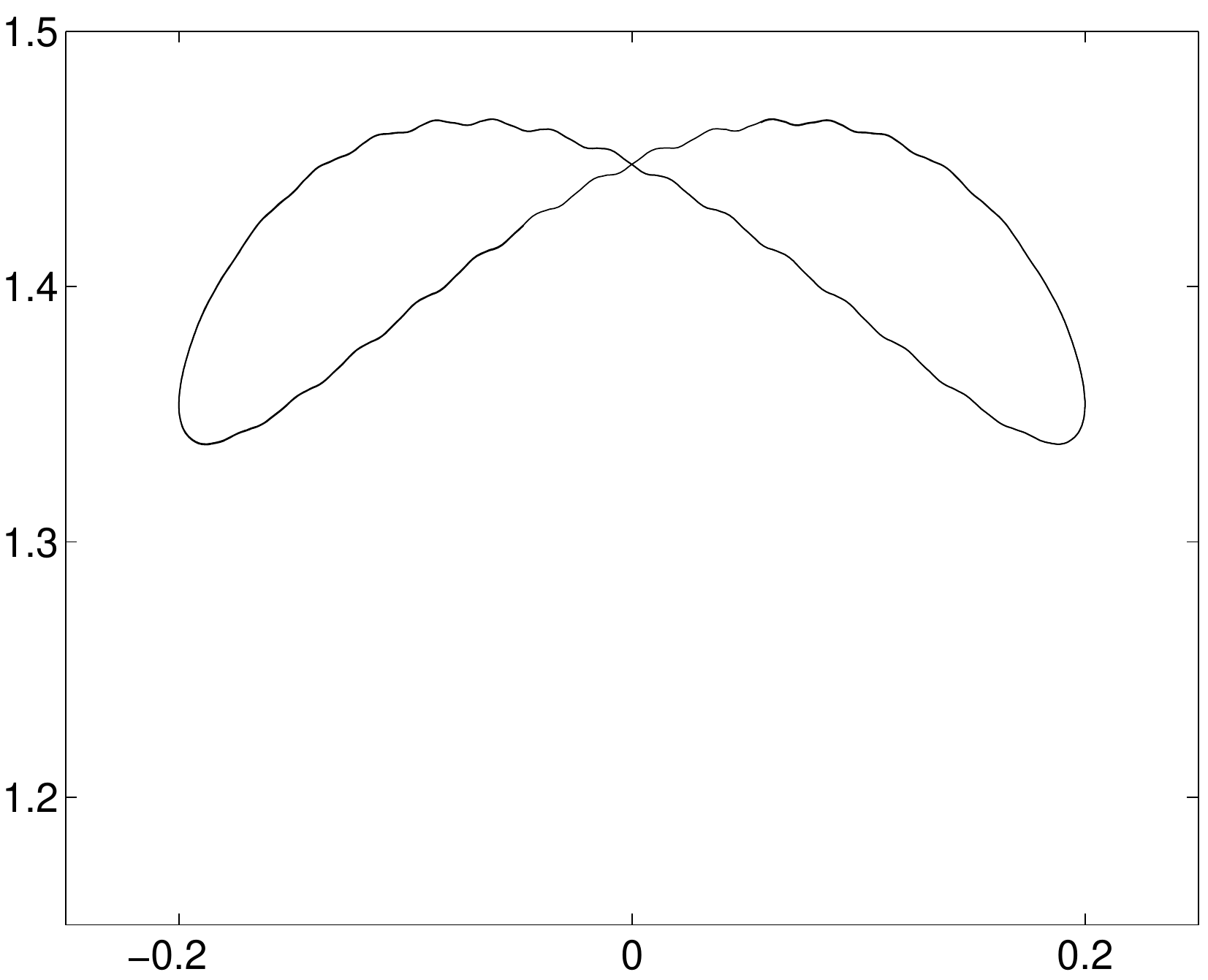}\\[-2.ex]
      $y_c(t)/D$
    \end{center}
  \end{minipage}
  \caption{Time-periodic variation of the drag coefficient
      in the case of a translationally oscillating cylinder in  
      uniform cross-flow at $Re_D=185$ with $D/h=38.4$ and $CFL\approx
      0.6$. Left column: present method, using two regularized delta
      functions with different support. Right graph: method of
      Kajishima and Takiguchi~\cite{kajishima:02}, implemented into
      the present solver as described in~\cite{uhlmann:03}.}
  \label{fig-oscillating-cyl-case006-185}
\end{figure}

Figure~\ref{fig-oscillating-cyl-case006-185} shows the
periodic variation of the drag coefficient as a function of the
cylinder's position. The important observation here is that the curve
is reasonably smooth, which demonstrates our present scheme's ability
to handle arbitrary motion w.r.t.\ the fixed grid. 
Using the regularized delta function of
Peskin~\cite{peskin:02} with wider support of 4 grid points reduces
the remaining mild oscillations even further
(figure~\ref{fig-oscillating-cyl-case006-185}). However, in the
latter case the cost of evaluating the interpolation sums is
significantly higher (more than twice in three dimensions).
For the purpose of comparison we have included in
figure~\ref{fig-oscillating-cyl-case006-185} the corresponding
result obtained by means of the forcing method of Kajishima and
Takiguchi~\cite{kajishima:02}, implemented into the present solver as
described in detail in~\cite{uhlmann:03}. Strong oscillations on the
scale of the mesh-width are evident, indicating that the
solid-volume-fraction-weighting used therein is a less efficient
smoothing mechanism. 
It can be seen from table~\ref{tab-cylinder-osc} that the
present method yields a higher drag than our computations using the 
method of reference~\cite{kajishima:02}; the use of the 4-point
delta function increases the mean drag even further. It should be
noted that the smoothing scheme proposed by Kajishima and
Takiguchi~\cite{kajishima:02} corresponds to the most compact
stencil among the three methods discussed here
(solid-volume-fraction-weighting, 3-point delta function, 4-point
delta function). Consequently, our results
indicate that the smoother the representation of the interface, the
higher the value of the mean drag. Finally, as in the case of the
stationary cylinder, it was verified that our over-prediction of the
drag diminishes with the domain size: the error of $\bar{C}_D$ decreases
to approximately $8\%$ for domain $\Omega_2$
(table~\ref{tab-cylinder-osc}). The predicted value for the
r.m.s.\ lift coefficient also diminishes with the domain size,
yielding an error of $8\%$ with respect to reference~\cite{lu:96}
for $\Omega_2$.
\subsection{Sedimentation of circular discs}\label{sec-results-2d}
The following two cases treat the sedimentation of circular discs in
an ambient container. At $t=0$ all particles are at rest. 
The initial velocity field is $\mathbf{u}(t=0)=0$
$\forall\,\mathbf{x}\in\Omega$ and no-slip conditions apply at the
boundaries, 
$\mathbf{u}=0$ $\forall\,\mathbf{x}\in\Gamma$. Homogeneous Neumann
boundary 
conditions are used for the pseudo-pressure. The particle Reynolds
number is defined from the particle velocity,
$Re_D=|\mathbf{u}_c|D/\nu$.  
\subsubsection{Drafting-kissing-tumbling case}
\label{sec-results-2d-dkt}
Two particles with identical density and radius are accelerating from
rest due to the action of gravity. Initially, they have the same horizontal
position, but some vertical offset. The trailing particle catches up
with the leading one due to the reduced drag in the former particle's
wake. This case has frequently been considered in the
literature~\cite{hu:01,glowinski:02,zhang:03,feng:04}. At a later
stage the present case involves direct particle-particle 
interaction, i.e.\ the particles approach each other closely, albeit
probably not closely enough for collision/\-film rupture to take
place. However, very thin liquid inter-particle films cannot be
resolved by a typical grid and therefore the correct build-up of
repulsive pressure is not captured which in turn can lead to possible 
partial ``overlap'' of the particle positions in the numerical
computation. In practice, various authors use artificial repulsion
potentials which prevent such non-physical
situations~\cite{hoefler:00,hu:01,glowinski:02}. 
In order to allow for comparison with available data, we apply the
collision strategy of Glowinski et al.~\cite{glowinski:99}, relying
upon a short-range repulsion force (stiffness $\varepsilon_p=5\cdot
10^{-7}$ and force range $\rho=3h$, in the terminology of
reference~\cite{glowinski:99}).

\begin{figure}
  \begin{minipage}{.5cm}
    $x_c$
  \end{minipage}
  \begin{minipage}{.44\linewidth}
    \includegraphics*[width=\linewidth]{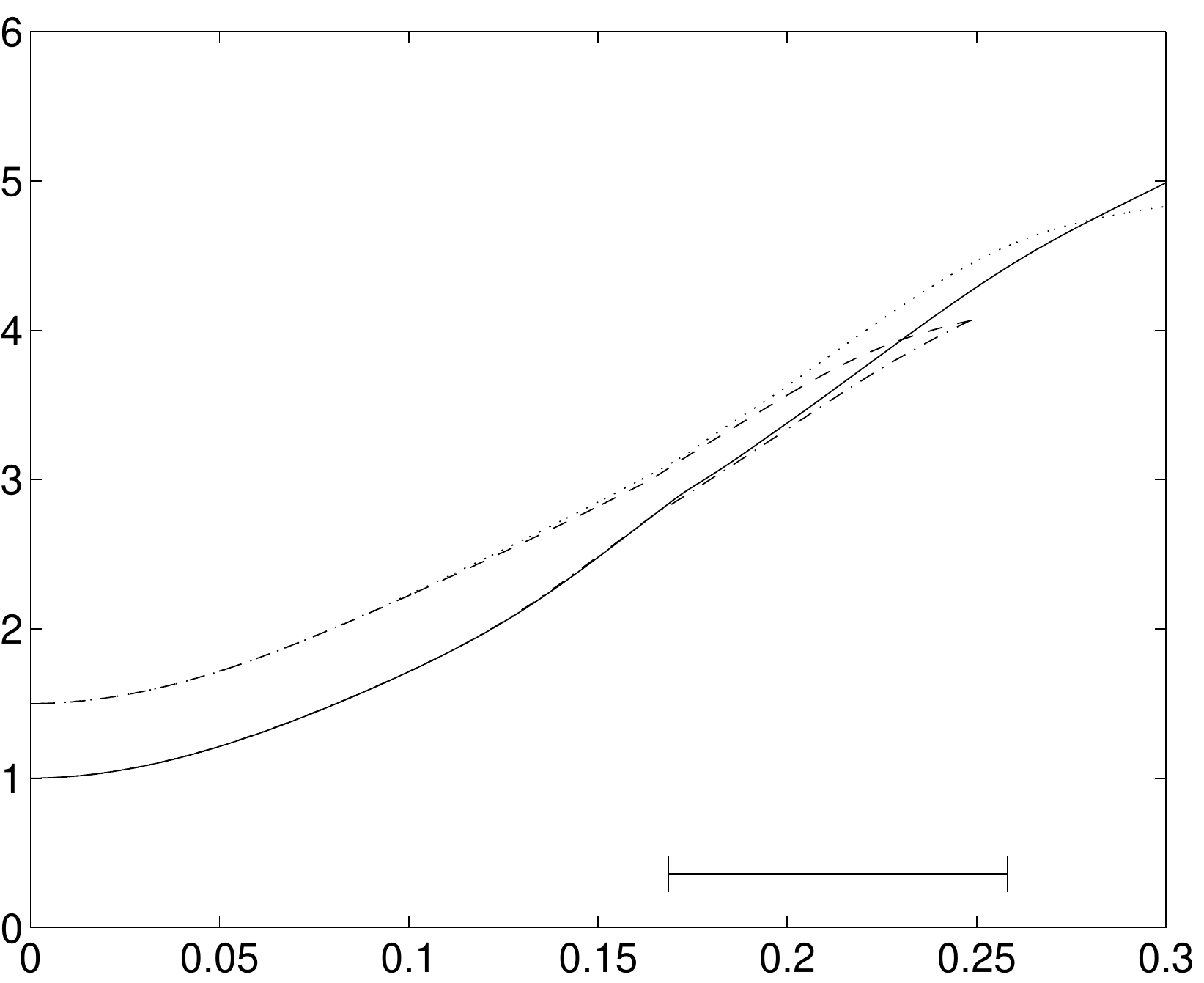}
  \end{minipage}
  \hfill
  \begin{minipage}{.5cm}
    $u_c$
  \end{minipage}
  \begin{minipage}{.44\linewidth}
    \includegraphics*[width=\linewidth]{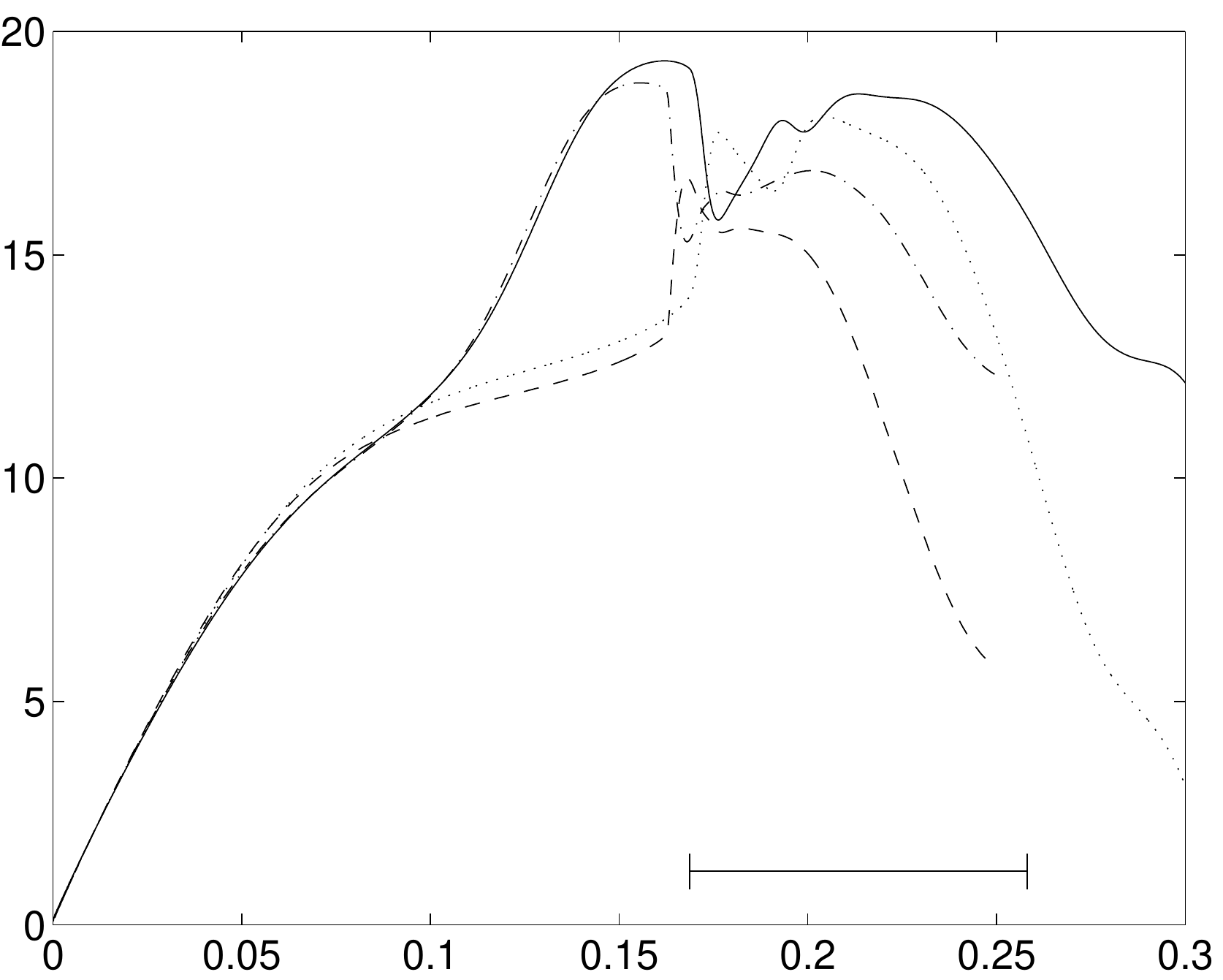}
  \end{minipage}
  \centerline{$t$\hspace*{.4\linewidth}$t$}
  \caption{
    Drafting-kissing-tumbling of two sedimenting discs with density 
    $\rho_p/\rho_f=1.5$ which are initially aligned vertically. 
    Results obtained with the present method, $\Delta t=0.0001$,
    $\Delta x=1/256$: {\solid trailing}, {\dotted leading}. Results
    provided by T.-W.\ Pan: {\chndot
      trailing}, {\dashed leading}. Vertical position (left), vertical
    velocity (right). The interval during which the repulsion force
    takes finite values is indicated near the abscissa. 
  }
  \label{fig-dkt-1.5-a}
\end{figure}

\begin{figure}
  \begin{minipage}{.5cm}
    $y_c$
  \end{minipage}
  \begin{minipage}{.44\linewidth}
    \includegraphics*[width=\linewidth]{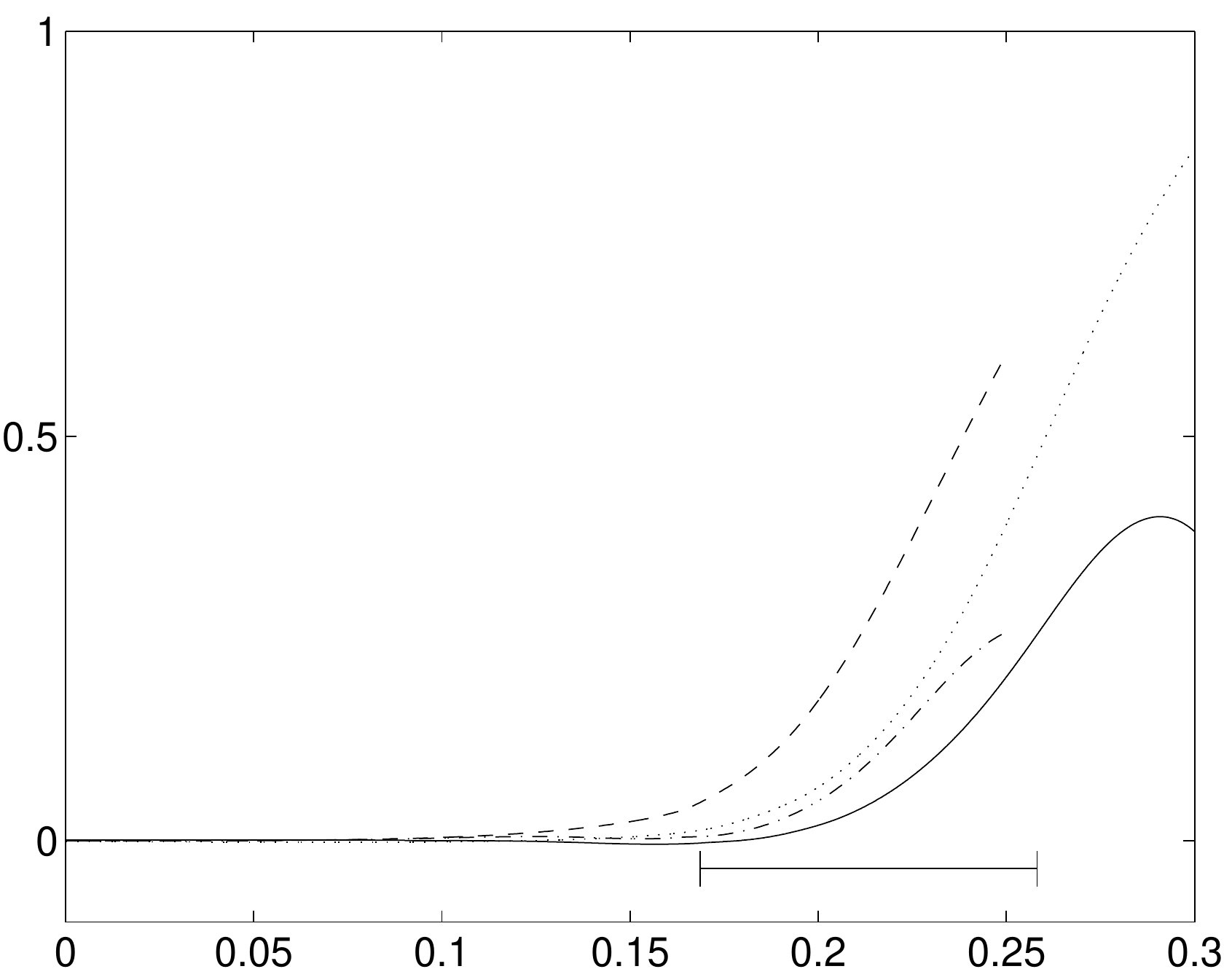}
  \end{minipage}
  \hfill
  \begin{minipage}{.5cm}
    $v_c$
  \end{minipage}
  \begin{minipage}{.44\linewidth}
    \includegraphics*[width=\linewidth]{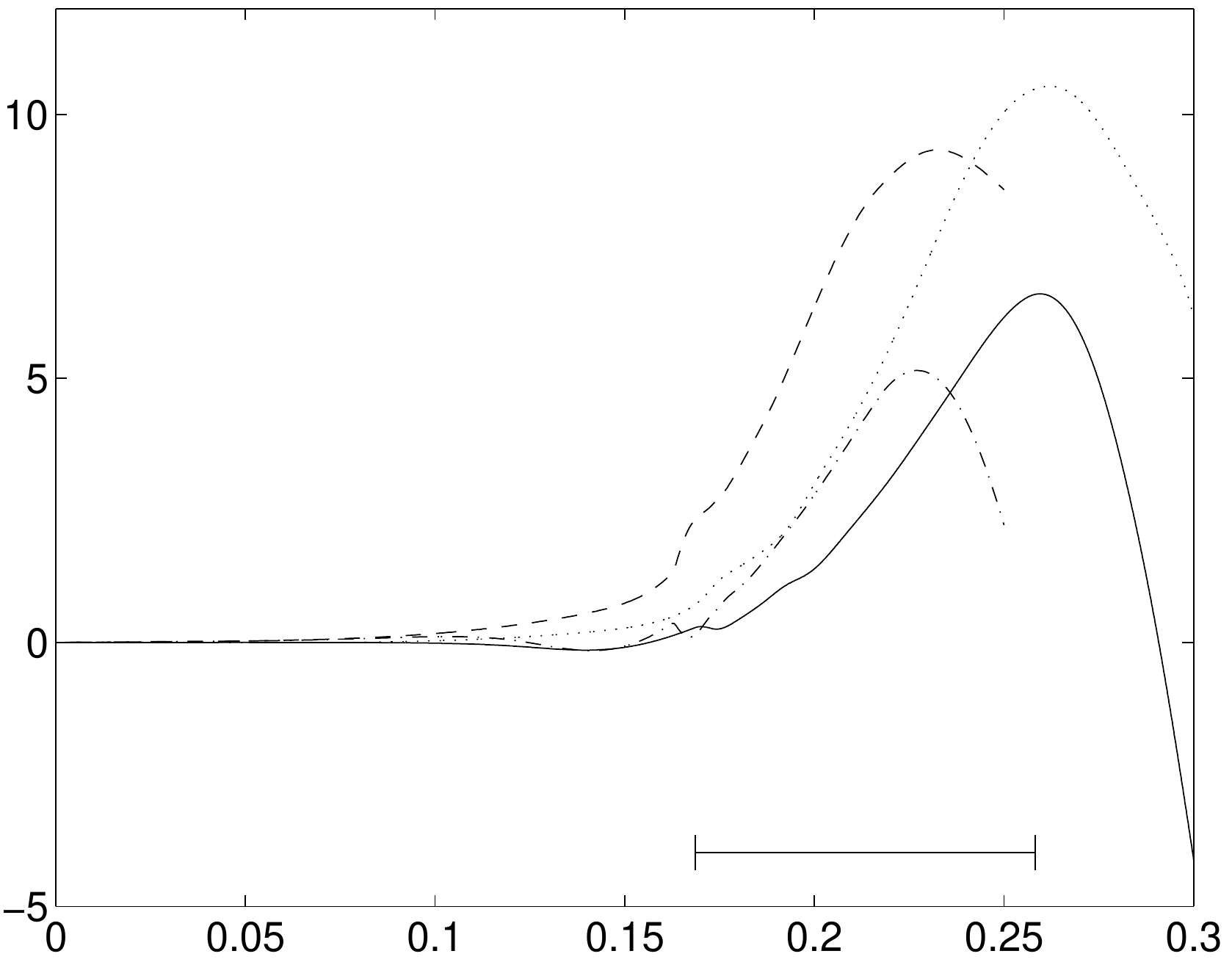}
  \end{minipage}\\
  \centerline{$t$\hspace*{.4\linewidth}$t$}
  \caption{
    Drafting-kissing-tumbling of two sedimenting discs with density 
    $\rho_p/\rho_f=1.5$ which are initially aligned vertically. 
    Results obtained with the present method, $\Delta t=0.0001$,
    $\Delta x=1/256$: {\solid trailing}, {\dotted leading}. Results
    provided by T.-W.\ Pan: {\chndot
      trailing}, {\dashed leading}. Horizontal position (left),
    horizontal velocity (right).
  }
  \label{fig-dkt-1.5-b}
\end{figure}

\begin{figure}
  \begin{center}
    \begin{minipage}{.5cm}
      $\omega_c$
    \end{minipage}
    \begin{minipage}{.44\linewidth}
      \includegraphics*[width=\linewidth]{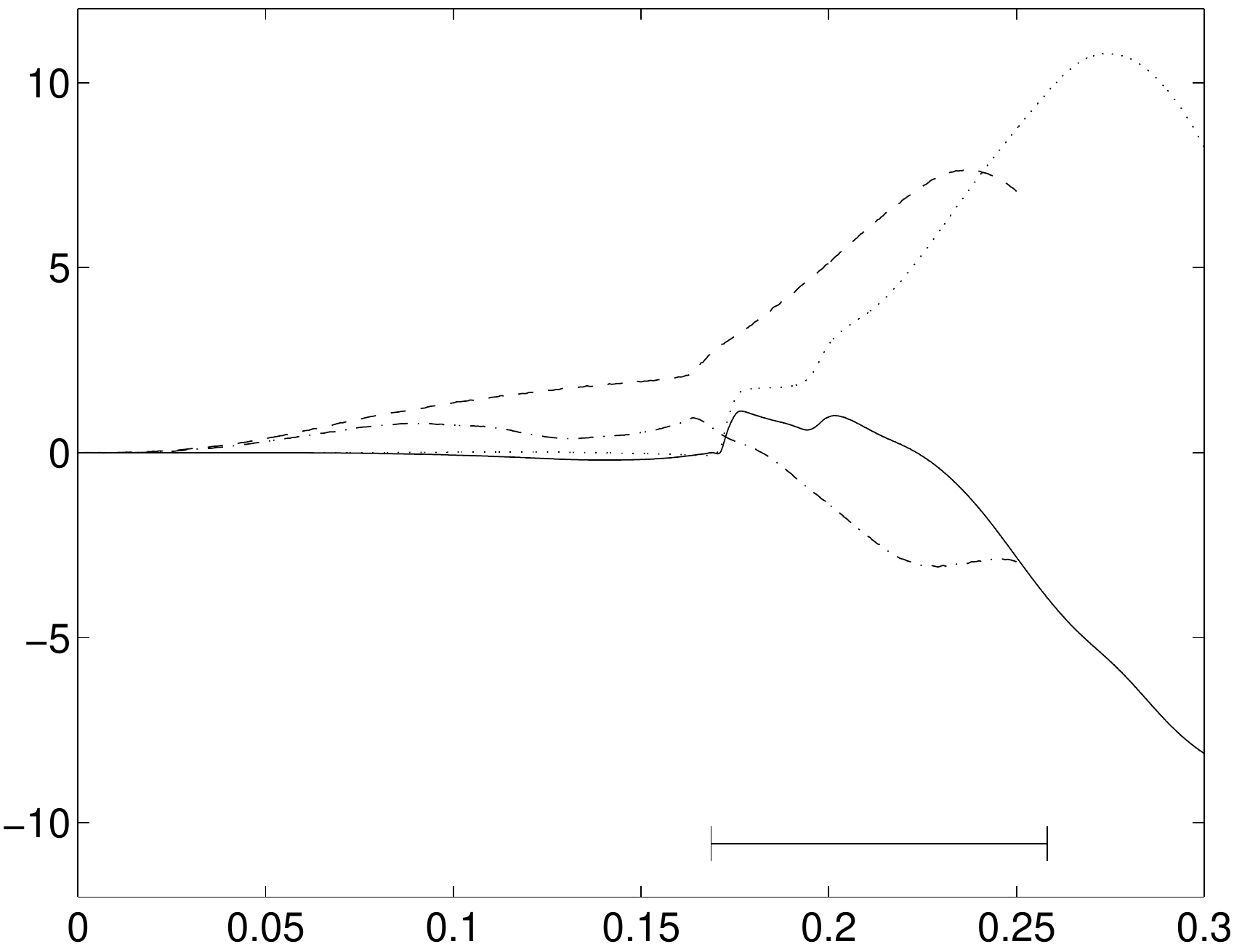}
    \end{minipage}\\
    $t$
  \end{center}
  \caption{
    Rotational velocity vs.\ time during the interaction of two
    sedimenting discs with density  
    $\rho_p/\rho_f=1.5$ which are initially aligned vertically. 
    Results obtained with the present method, $\Delta t=0.0001$,
    $\Delta x=1/256$: {\solid
      trailing}, {\dotted leading}; results 
    provided by T.-W.\ Pan: {\chndot trailing}, {\dashed leading}. 
  }
  \label{fig-dkt-1.5-d}
\end{figure}

This case corresponds to the one computed
in~\cite[\S~8.4]{glowinski:02}. The physical parameters of the
problem are the following:
\begin{itemize}
\renewcommand{\itemsep}{-.0cm}
\item domain size $\Omega=[0,6]\times[-1,1]$;
\item disc radius $r_c^{(1)}=r_c^{(2)}=0.125$;
\item initial location of the discs $\mathbf{x}_c^{(1)}(t=0)=(1,+0.001)$,
    $\mathbf{x}_c^{(2)}(t=0)=(1.5,-0.001)$;
\item density ratio $\rho_p^{(1)}/\rho_f=\rho_p^{(2)}/\rho_f=1.5$;
\item fluid viscosity $\nu=0.01$;
\item gravitational acceleration $\mathbf{g}=(981,0)$.
\end{itemize}
This leads to maximum values for the particle Reynolds number of
approximately 480 and 430, respectively. The numerical para\-meters
were: 
\begin{itemize}
\renewcommand{\itemsep}{-.0cm}
\item mesh width $h=1/256$, i.e.\ $D/h=64$;
\item time step $\Delta t=0.0001$, which leads to a maximum $CFL$
  number around 0.85. 
\end{itemize}
Figures~\ref{fig-dkt-1.5-a}-\ref{fig-dkt-1.5-d} show our present
results as well as the ones kindly re-computed and provided by T.-W.\
Pan using the method of reference~\cite{glowinski:02}. The latter
results are, therefore, not exactly equivalent to those presented
in~\cite{glowinski:02}. 
As marked in the figures, the artificial repulsion force
is non-zero during the following interval: $0.1687\leq t\leq
0.2582$. 

For the vertical position and velocity, we observe a very close
agreement between both results---up to the time of direct particle
interaction (``kissing''). During the ``tumbling'' stage, which is
the manifestation of a strong instability, we cannot expect more
than a qualitative accord among simulations performed with quite
different numerical methods. It is noteworthy that the leading and
trailing particle reverse their roles (i.e.\ the vertical position
curves cross-over) in both results, albeit at different times. 
The results for the horizontal position and velocity, on the other
hand, differ considerably during the ``drafting'' stage.
Particularly, a much more pronounced lateral motion
is manifest in the data-set of Pan. 
In our computations, a lateral motion of the particle during
``drafting'' is only observed if the initial position is chosen
non-symmetric w.r.t.\ the grid since our spatial scheme fully
preserves the symmetry and perturbations due to finite-precision
arithmetic do not grow fast enough for these short times. 
With the present small lateral offset of the initial particle
position, lateral motion sets in quickly, albeit to a much lesser
extent than exhibited by the results provided by Pan. We believe
that the anisotropic triangular grid used therein is responsible for
the larger lateral motion as well as for a higher angular velocity
(cf.\ related discussion in reference~\cite{zhang:03}).
During ``kissing'' and ``tumbling'', however, the lateral motion and
rotation obtained by our method show a similar behavior as Pan's
results. It should be noted that our results for these later 
stages are sensitive to the choice of the initial horizontal offset.  
\subsubsection{Pure wake interaction}
\label{sec-results-2d-wake}
Two particles with a vertical and horizontal offset are released at
$t$=$0$. The ``trailing'' particle has a higher density and 
therefore passes the leading particle, subjecting it to 
perturbations in its wake. The computation is stopped before the
heavier particle reaches the bottom boundary of the computational
domain. 
There are two reasons for discussing this test case:
\begin{itemize}
\item No direct particle-particle interactions are observed. Therefore, no
  numerical collision model is needed, making this case attractive as
  a possible future ``benchmark'' for testing the basic fluid-solid
  interaction method.  
\item Since the angular motion of the particles is non-negligible
  here, we can gauge the importance of the representation of the
  rate-of-change term ${\mathcal I}$ (cf.\ equation
  \ref{equ-particles-newton-2-present-rotation}). Specifically, we can
  deduce that using the approximate formulation given in equation 
  (\ref{equ-particles-newton-2-present-translation-discrete-om-approx})  
  is an acceptable simplification.
\end{itemize}

The physical parameters of the problem are the following:
\begin{itemize}
\renewcommand{\itemsep}{-.0cm}
\item domain size $\Omega=[0,10]\times[-1,1]$;
\item disc radius $r_c^{(1)}=r_c^{(2)}=0.1$;
\item initial location of the discs
  $\mathbf{x}_c^{(1)}(t=0)=(0.8,-0.13)$,\\
    $\mathbf{x}_c^{(2)}(t=0)=(1.2,+0.13)$; 
\item density ratio $\rho_p^{(1)}/\rho_f=1.5$,
  $\rho_p^{(2)}/\rho_f=1.25$; 
\item fluid viscosity $\nu=0.0008$;
\item gravitational acceleration $\mathbf{g}=(9.81,0)$.
\end{itemize}
This yields maximum particle Reynolds numbers of $280$ and $230$,
respectively. The numerical para\-meters are set to the following
values: 
\begin{itemize}
\renewcommand{\itemsep}{-.0cm}
\item mesh width $h=1/200$, i.e.\ $D/h=40$;
\item time step $\Delta t=0.001$, which leads to a maximum $CFL$
  number around 0.5.
\end{itemize}
The final time shown below is $t_{fin}=8.629$, corresponding to the
center of the heavy particle being located at $2D$ above the bottom
boundary. 

\begin{figure}
  \begin{center}
    \begin{minipage}{.5cm}
      $y$
    \end{minipage}
    \begin{minipage}{.95\linewidth}
      \includegraphics*[width=\linewidth]{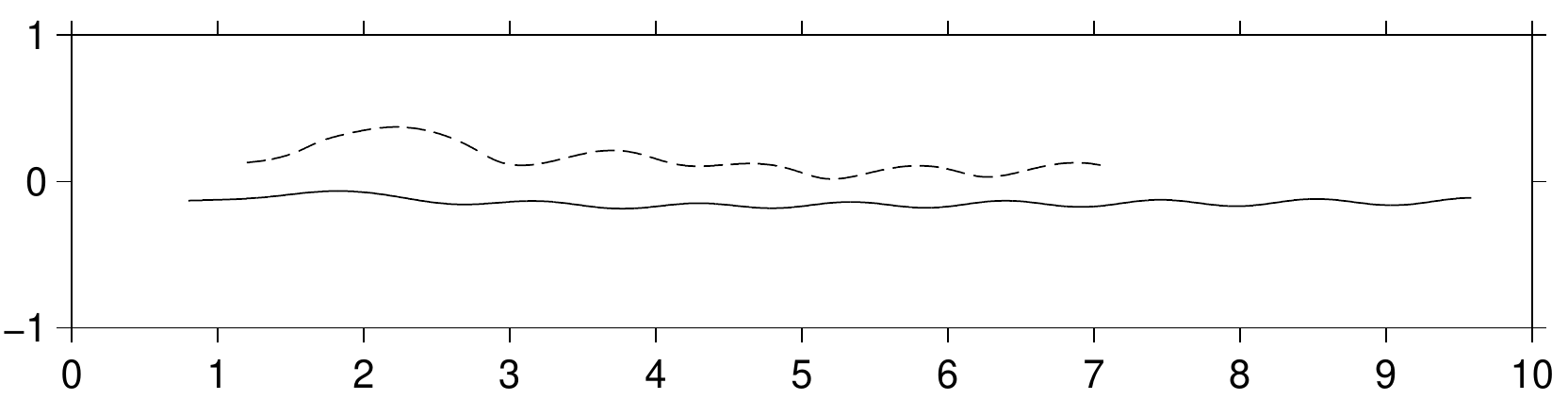}
    \end{minipage}
    \\$x$
  \end{center}
  \caption{Wake interaction of two sedimenting particles with density 
    $\rho_p^{(1)}/\rho_f=1.5$, $\rho_p^{(2)}/\rho_f=1.25$ and initial
    vertical and lateral offset. The gravity acts in the positive
    $x$-direction. Results obtained with $\Delta t=0.001$,
    $h=1/200$. The graph shows the particle trajectories with the line
    stlyes corresponding to: \solid~{heavy particle}, \dashed~light
    particle.} 
  \label{fig-022-traj}
\end{figure}

\begin{figure}
  \begin{center}
    \begin{minipage}{.5cm}
      $y$
    \end{minipage}
    \begin{minipage}{.95\linewidth}
      \includegraphics*[width=\linewidth]{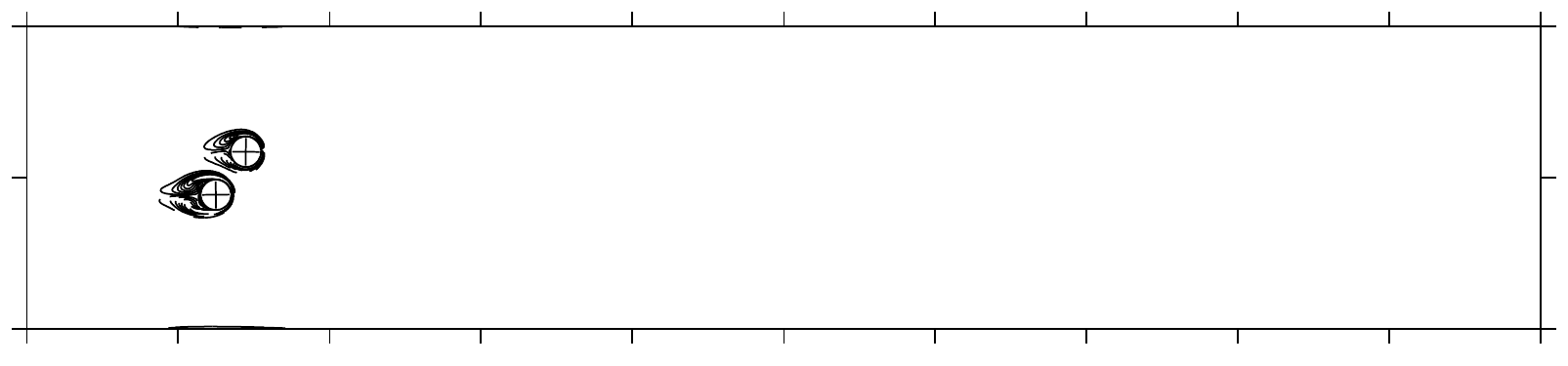}
    \end{minipage}\\
    \begin{minipage}{.5cm}
      $y$
    \end{minipage}
    \begin{minipage}{.95\linewidth}
      \includegraphics*[width=\linewidth]{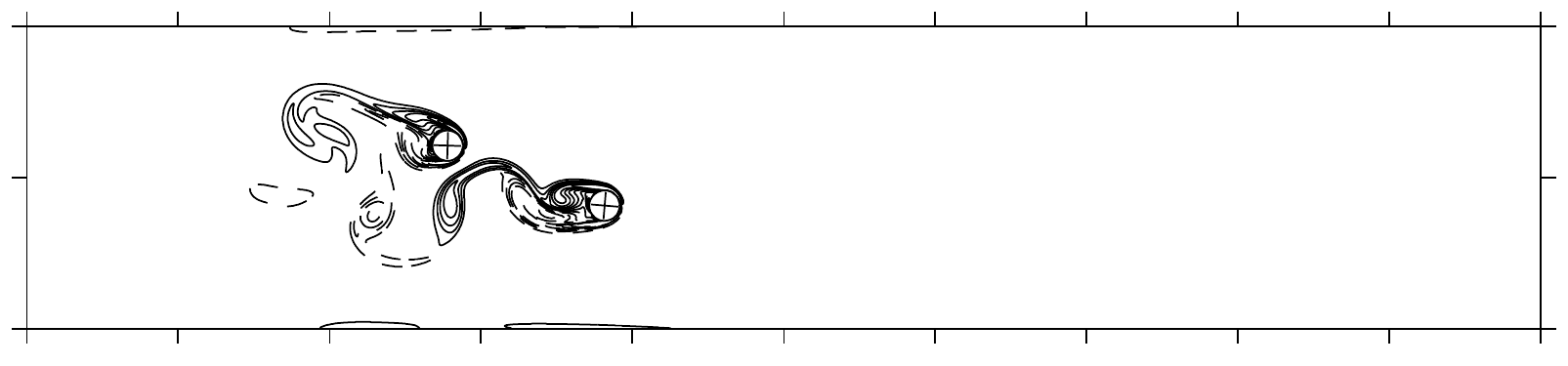}
    \end{minipage}\\
    \begin{minipage}{.5cm}
      $y$
    \end{minipage}
    \begin{minipage}{.95\linewidth}
      \includegraphics*[width=\linewidth]{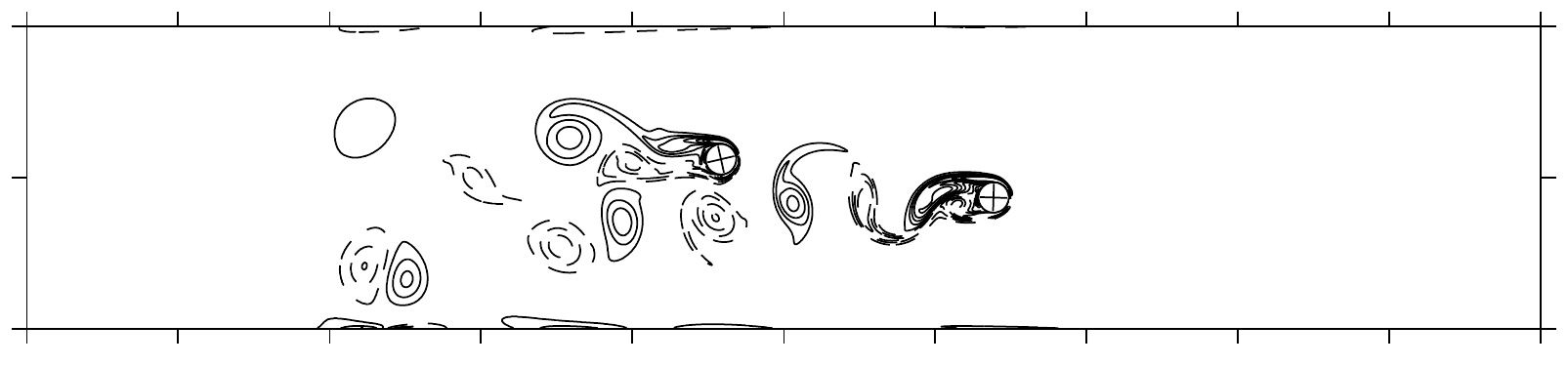}
    \end{minipage}\\
    \begin{minipage}{.5cm}
      $y$
    \end{minipage}
    \begin{minipage}{.95\linewidth}
      \includegraphics*[width=\linewidth]{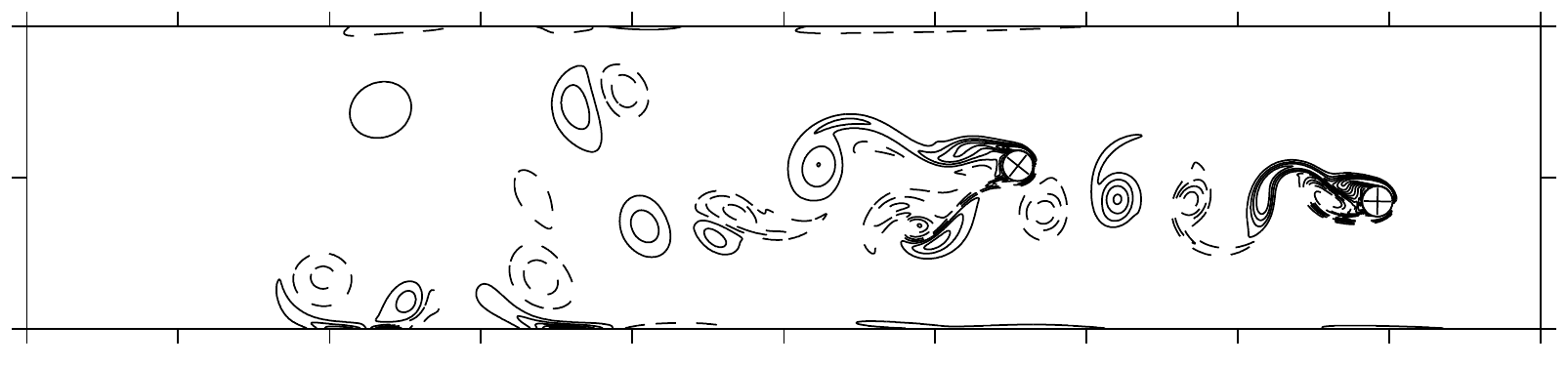}
    \end{minipage}\\$x$
  \end{center}
  \caption{Wake interaction case of
    figure~\ref{fig-022-traj}. Instantaneous contours of vorticity
    (values at 
    -30:4.6:30, negative values corresponding to dashed lines) and
    particle positions at times $t=0.8$, $t=3.2$, $t=5.6$, $t=8.0$
    (from top to bottom) are shown. The crosses inside the circles
    indicate the angular position of the particles.}  
  \label{fig-022-fields}
\end{figure}

\begin{figure}
  \begin{minipage}{.5cm}
    $x_c$
  \end{minipage}
  \begin{minipage}{.44\linewidth}
    \includegraphics*[width=\linewidth]{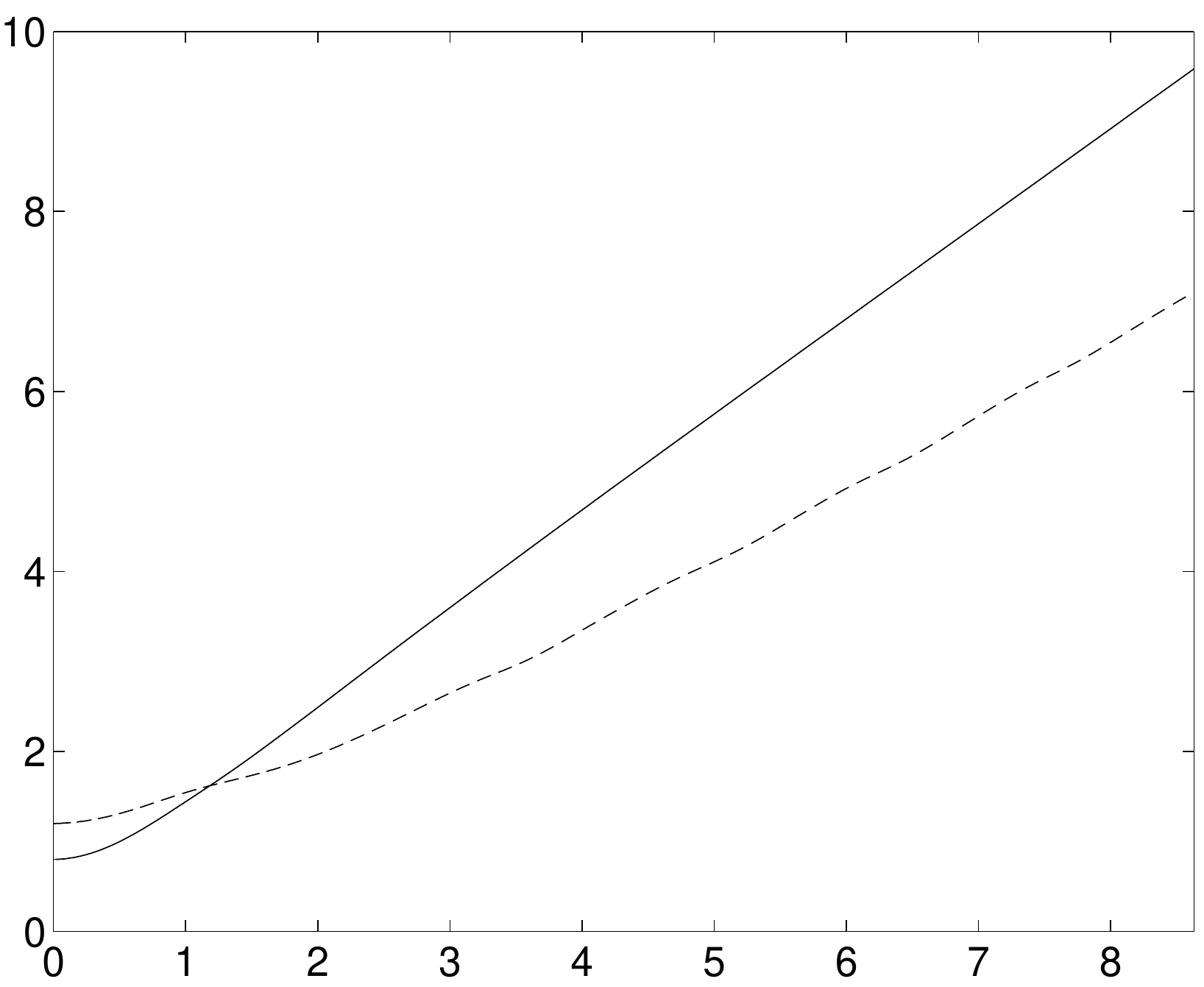}
  \end{minipage}
  \hfill
  \begin{minipage}{.5cm}
    $u_c$
  \end{minipage}
  \begin{minipage}{.44\linewidth}
    \includegraphics*[width=\linewidth]{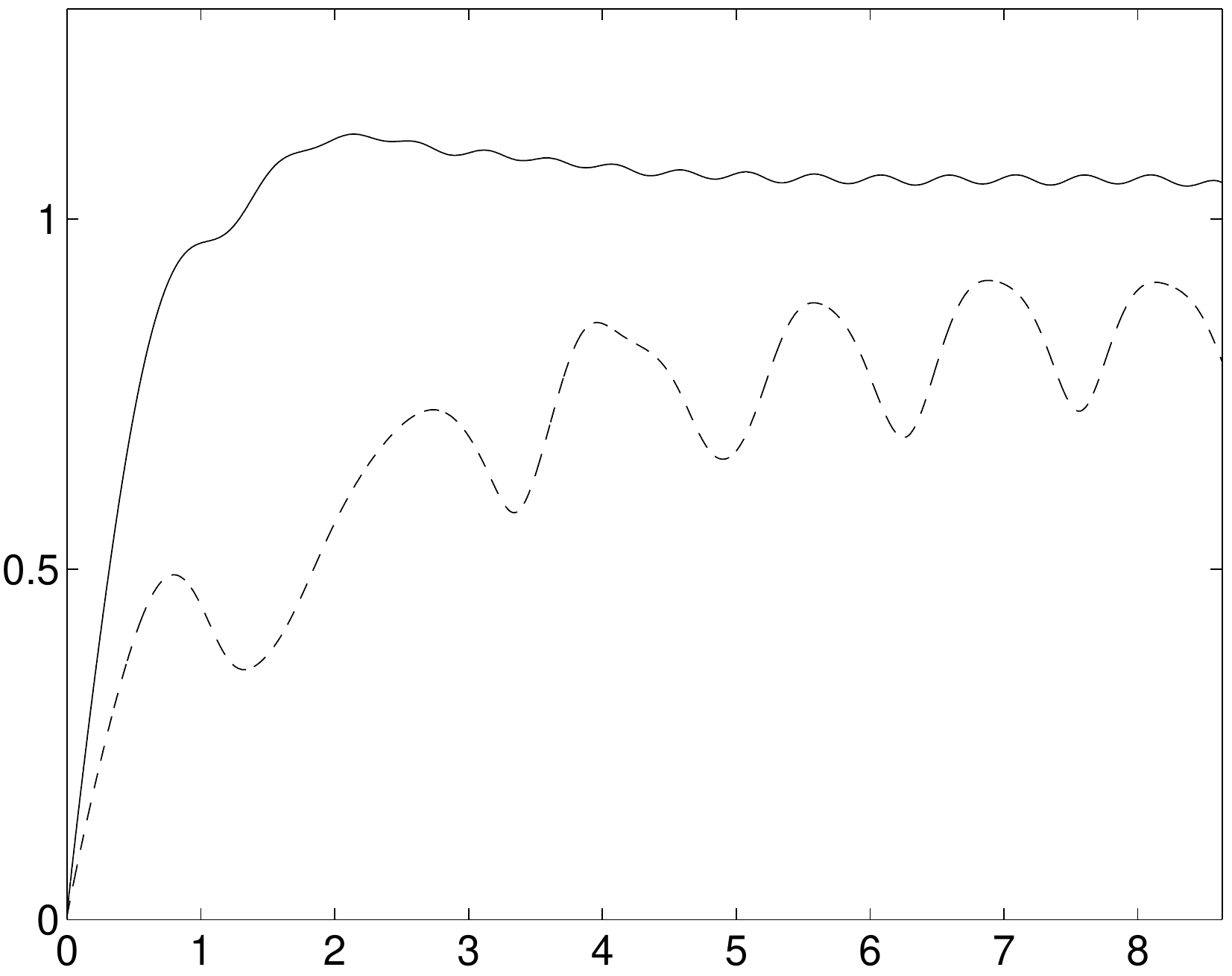}
  \end{minipage}
  \centerline{$t$\hspace*{.4\linewidth}$t$}
  \caption{Wake interaction case of figure~\ref{fig-022-traj}. The
    graphs show: vertical particle positions (left), vertical particle
    velocities (right).}  
  \label{fig-022-pos-vel-x}
\end{figure}

\begin{figure}
  \begin{minipage}{.5cm}
    $y_c$
  \end{minipage}
  \begin{minipage}{.44\linewidth}
    \includegraphics*[width=\linewidth]{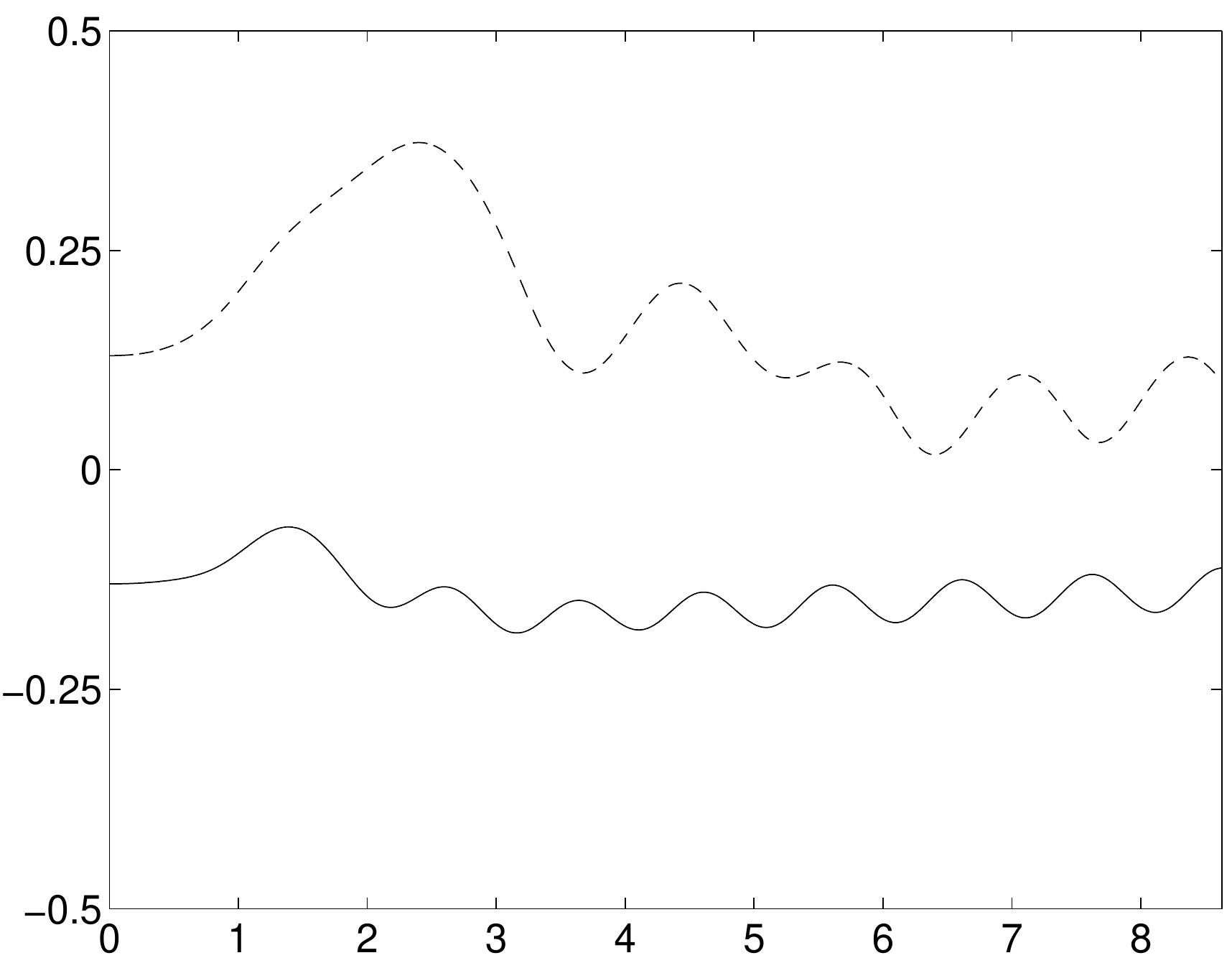}
  \end{minipage}
  \hfill
  \begin{minipage}{.5cm}
    $v_c$
  \end{minipage}
  \begin{minipage}{.44\linewidth}
    \includegraphics*[width=\linewidth]{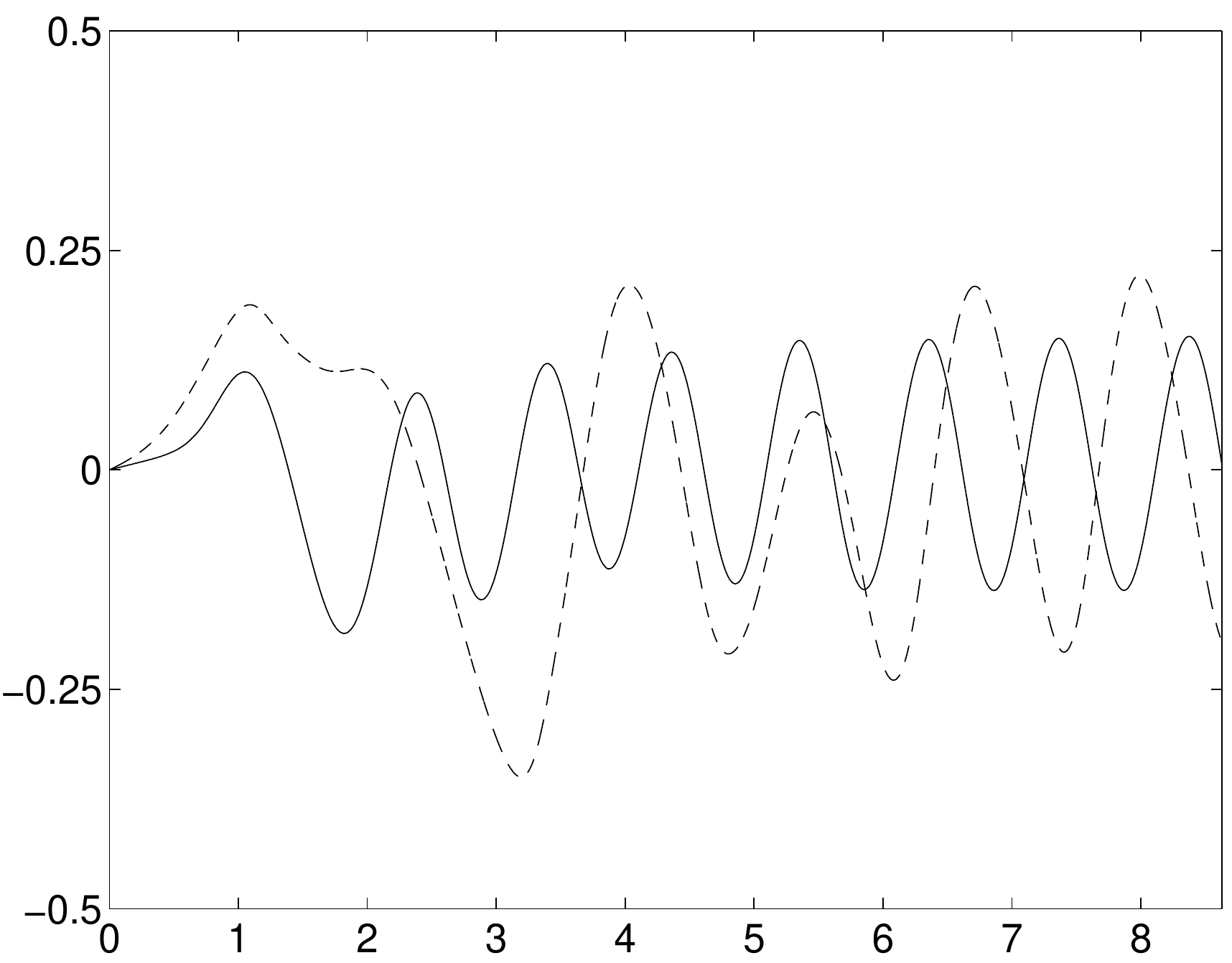}
  \end{minipage}
  \centerline{$t$\hspace*{.4\linewidth}$t$}
  \caption{Wake interaction case of figure~\ref{fig-022-traj}. The
    graphs show: horizontal particle positions (left), horizontal particle
    velocities (right).}
  \label{fig-022-pos-vel-y}
\end{figure}

Figure~\ref{fig-022-traj} shows the trajectories of the two particles
and figure~\ref{fig-022-fields} successive snapshots of the vorticity
field. It can be observed that the heavier particle follows a slightly
undulating path due to the oscillating lift force induced by
its own vortex shedding. The deviation of the lighter particle's path
from a vertical one is more pronounced, partially due to the
interaction with the preceding
vortices. The time-evolution of particle positions and translational
velocities is given in figures~\ref{fig-022-pos-vel-x} and
\ref{fig-022-pos-vel-y}. It is noteworthy that the heavy particle's
vertical velocity reaches its maximum value and then slightly
decelerates when vortex shedding has reached a periodic state.  

\begin{figure}
  \begin{minipage}{.5cm}
    $\theta_c$
  \end{minipage}
  \begin{minipage}{.44\linewidth}
    \includegraphics*[width=\linewidth]{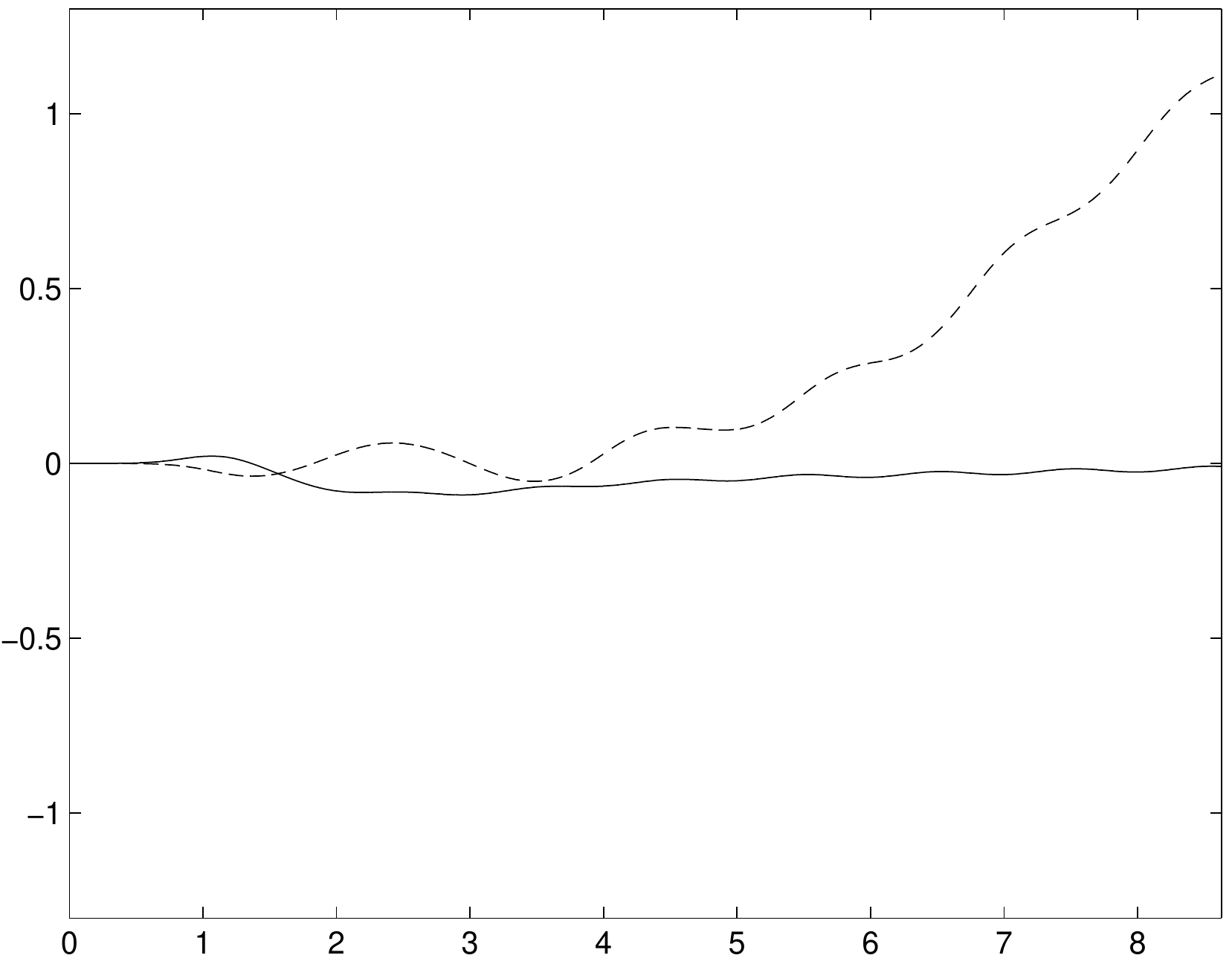}
  \end{minipage}
  \hfill
  \begin{minipage}{.5cm}
    $\omega_c$
  \end{minipage}
  \begin{minipage}{.44\linewidth}
    \includegraphics*[width=\linewidth]{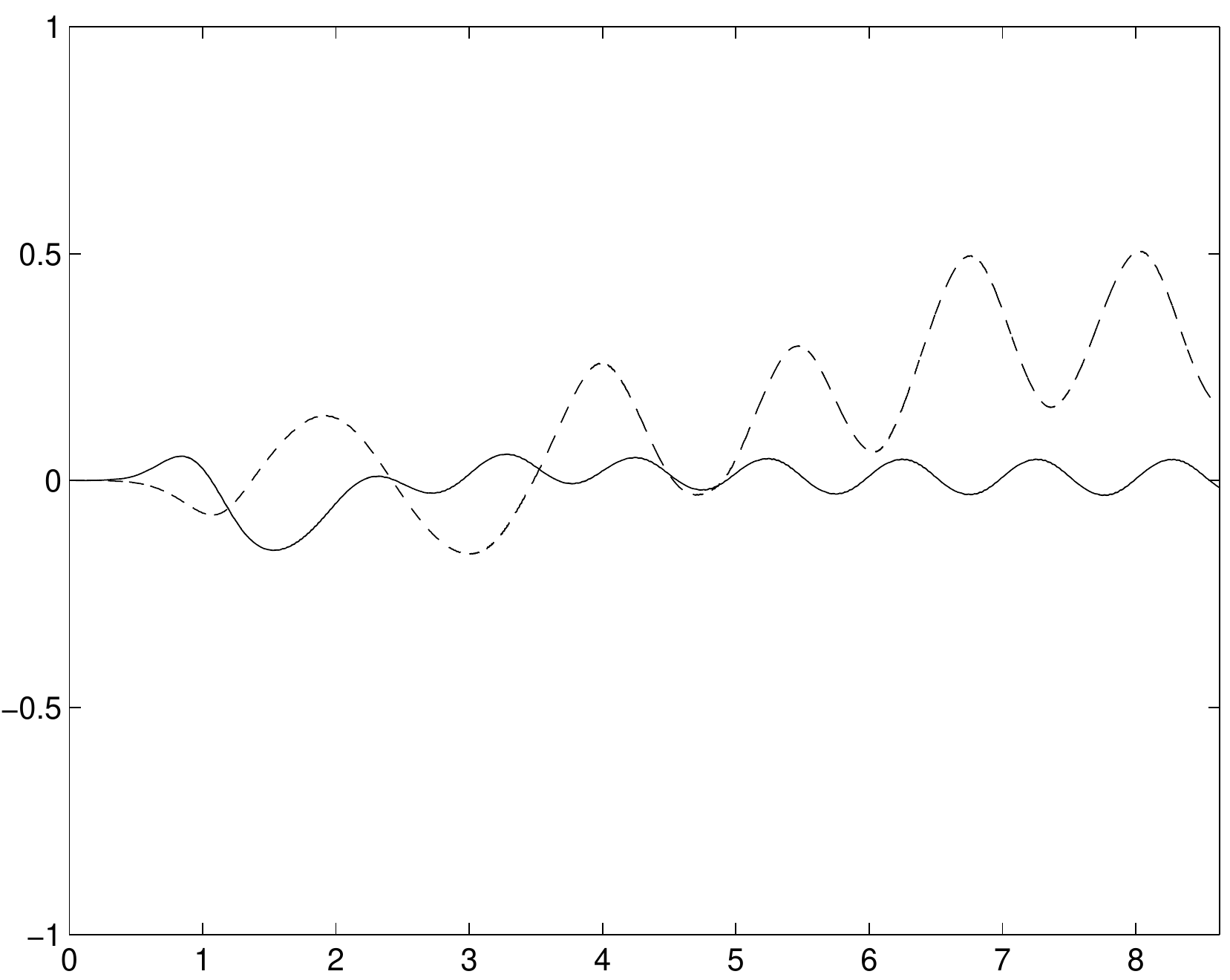}
  \end{minipage}
  \centerline{$t$\hspace*{.4\linewidth}$t$}
  \caption{Wake interaction case of figure~\ref{fig-022-traj}. The
    graphs show: angular particle positions (left),
    angular particle velocities (right).} 
  \label{fig-022-angle-angvel}
\end{figure}

\begin{figure}
  \begin{minipage}{.5cm}
    $\theta_c$
  \end{minipage}
  \begin{minipage}{.44\linewidth}
    \includegraphics*[width=\linewidth]{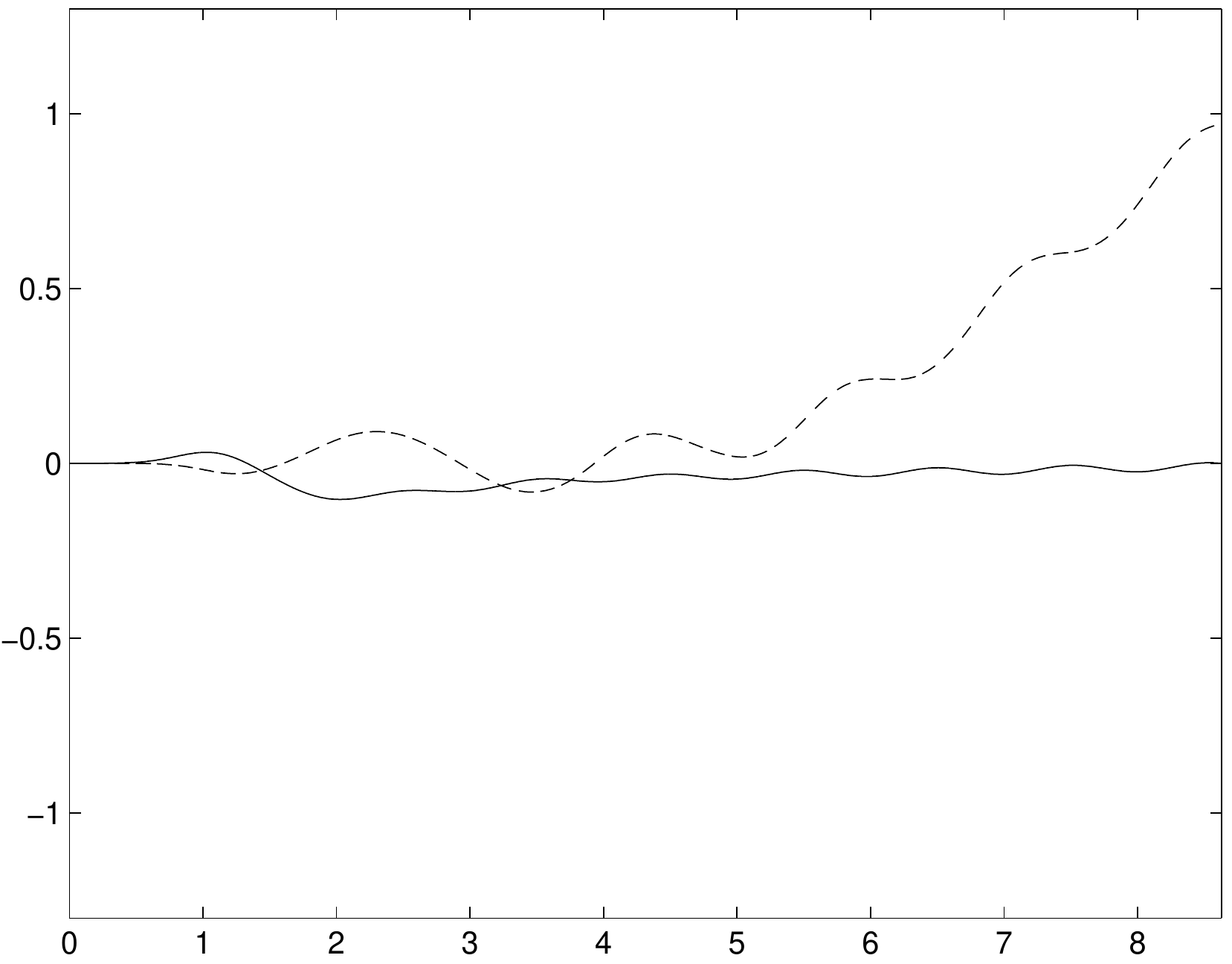}
  \end{minipage}
  \hfill
  \begin{minipage}{.5cm}
    $\omega_c$
  \end{minipage}
  \begin{minipage}{.44\linewidth}
    \includegraphics*[width=\linewidth]{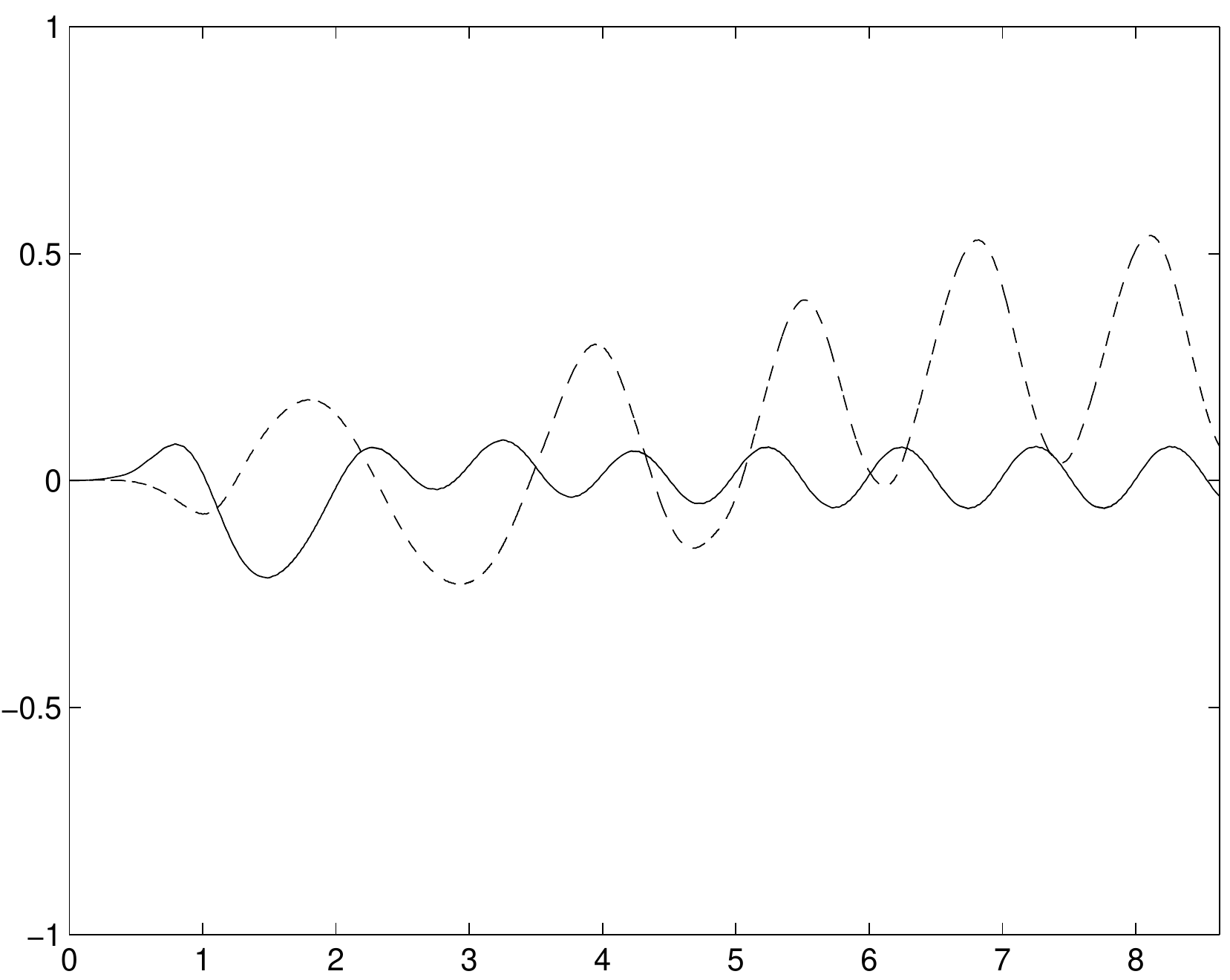}
  \end{minipage}
  \centerline{$t$\hspace*{.4\linewidth}$t$}
  \caption{The same as figure~\ref{fig-022-angle-angvel}, but the
    results were obtained by using the approximation of full rigidity
    (equation
    \ref{equ-particles-newton-2-present-translation-discrete-om-approx}) 
    for determining the rate-of-change term of angular particle momentum.} 
  \label{fig-022-angle-angvel-no-solid}
\end{figure}

Figure~\ref{fig-022-angle-angvel} shows the particles' angular
position and velocity; figure~\ref{fig-022-angle-angvel-no-solid} does
the same for the results obtained with the approximate formulation for
the angular momentum balance given in equation 
(\ref{equ-particles-newton-2-present-translation-discrete-om-approx}).
It can be seen that both variants yield qualitatively very similar
results. The effect of the simplification is an increase of the
amplitude of the oscillations of angular velocity. The
root-mean-square value of the difference amounts to approximately 13\%
(18\%) of the maximum angular velocity for the heavy (light)
particle. At the same time, the particle trajectories coincide to
within $0.08D$ and $0.54D$ for the heavy and light particle,
respectively. As a conclusion we consider it acceptable to simplify
the particles' angular momentum balance by using equation
(\ref{equ-particles-newton-2-present-translation-discrete-om-approx})
in the subsequent three-dimensional cases.
\subsection{Motion of spherical particles}\label{sec-results-3d}
Here we present simulations of the motion of three-dimensional
spherically-shaped particles. In all cases the flow-field and particle
positions are treated as triply-periodic. Again, initially the fluid
and the particles are at rest. In the following the particle-related
quantities will be normalized with the reference values
$u_{ref}=\sqrt{|\mathbf{g}|D}$, $t_{ref}=\sqrt{D/|\mathbf{g}|}$,
$l_{ref}=D$ for velocity, time and length, respectively. 
\subsubsection{A single sedimenting sphere}
\label{sec-results-3d-mordant}
We consider a single sphere which is released at $t=0$. The parameters
are chosen in order to match cases 1,2,4 of the experiment of Mordant
and Pinton~\cite{mordant:00}, where the motion of spherical beads in
water was investigated, while their material and diameter were
varied from case to case. 
This case has also been considered as a reference for the
computations in~\cite{sharma:05}. 
The experiment takes place in a large
container, justifying the use of periodic conditions in the
simulation. We have selected the following parameters by
similarity with the experiment, keeping the values for the density 
ratio, Froude number and particle Reynolds number constant:  
\begin{itemize}
  \renewcommand{\itemsep}{-.0cm}
\item domain size $\Omega=[0,1.25]\times[0,1.25]\times[0,L_z]$;
\item particle radius $r_c=1/12$;
\item initial particle location 
  $\mathbf{x}_c(t=0)=(0.625,0.625,9.5)$; 
\item gravitation vector $\mathbf{g}=(0,0,-9.81)$;
\item and:\\
  \begin{tabular}{*{4}{c}}
    case&1&2&4\\
    density ratio $\rho_p/\rho_f$&$2.56$&$2.56$&$7.71$
    \\
    domain length $L_z$&$10$&$10$&$15$\\
    fluid viscosity $\nu$&$0.005416368$&$0.00104238$&$0.00267626$
  \end{tabular}
\end{itemize}
The values for the numerical para\-meters are:
\begin{itemize}
\renewcommand{\itemsep}{-.0cm}
\item mesh width $h=1/76.8$, i.e.\ $D/h=12.8$;
\item time step $\Delta t=0.0025$, i.e.\ yielding a maximum $CFL$
  number of $0.3$, $0.75$, $0.5$, respectively.
\end{itemize}

\begin{figure}
  \begin{center}
    \begin{minipage}{.5cm}
      $\frac{w}{u_{ref}}$
    \end{minipage}
    \begin{minipage}{.43\linewidth}
      \includegraphics*[width=\linewidth]{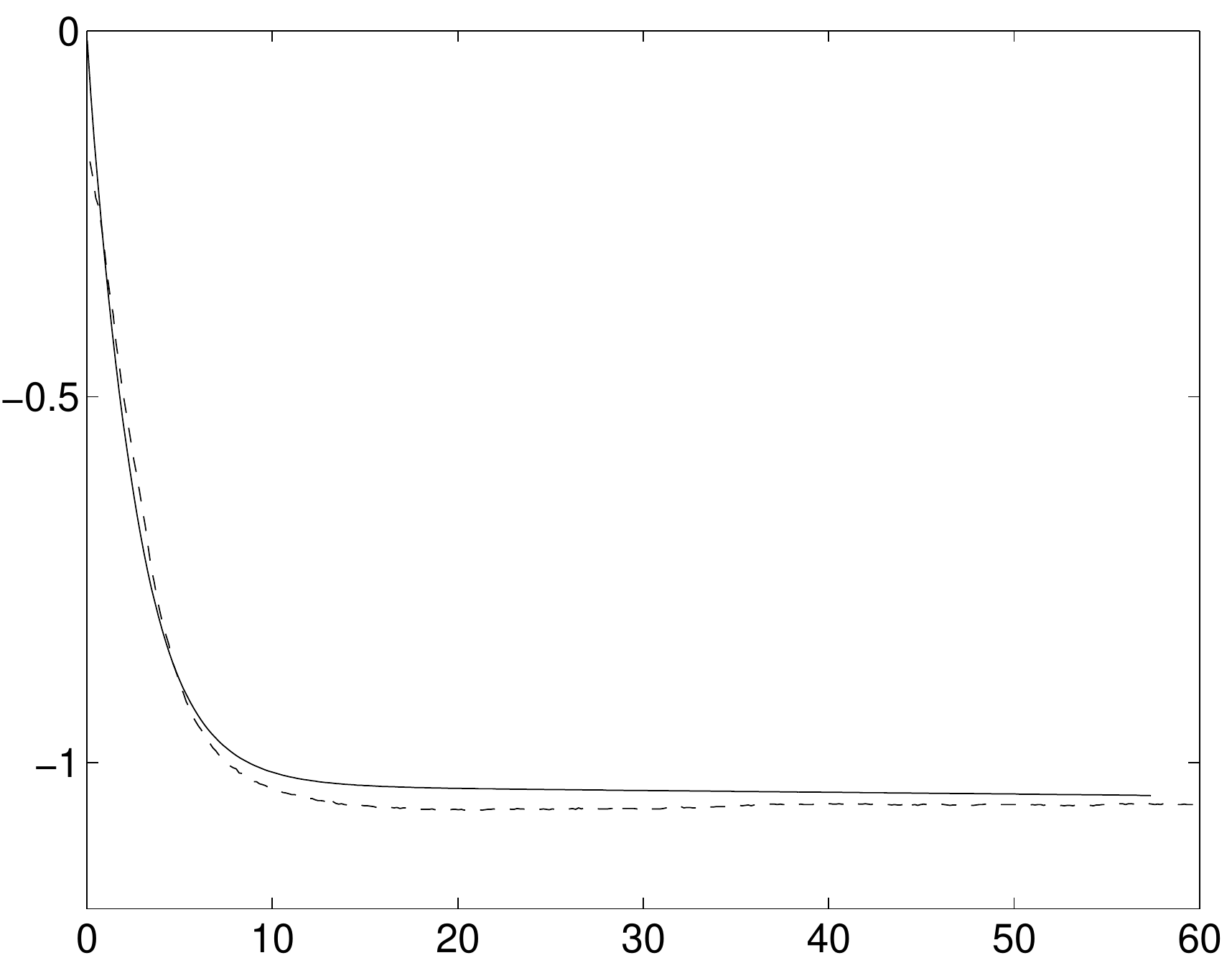}
    \end{minipage}
    \\$t/t_{ref}$
  \end{center}
  \caption{Sedimentation of a single sphere; case 1 of
    reference~\cite{mordant:00}. Vertical velocity:
    \solid present, \dashed experimental data.}
  \label{fig-mordant1-w}
\end{figure}

\begin{figure}
  \begin{center}
    \begin{minipage}{.5cm}
      $\frac{w}{u_{ref}}$
    \end{minipage}
    \begin{minipage}{.43\linewidth}
      \includegraphics*[width=\linewidth]{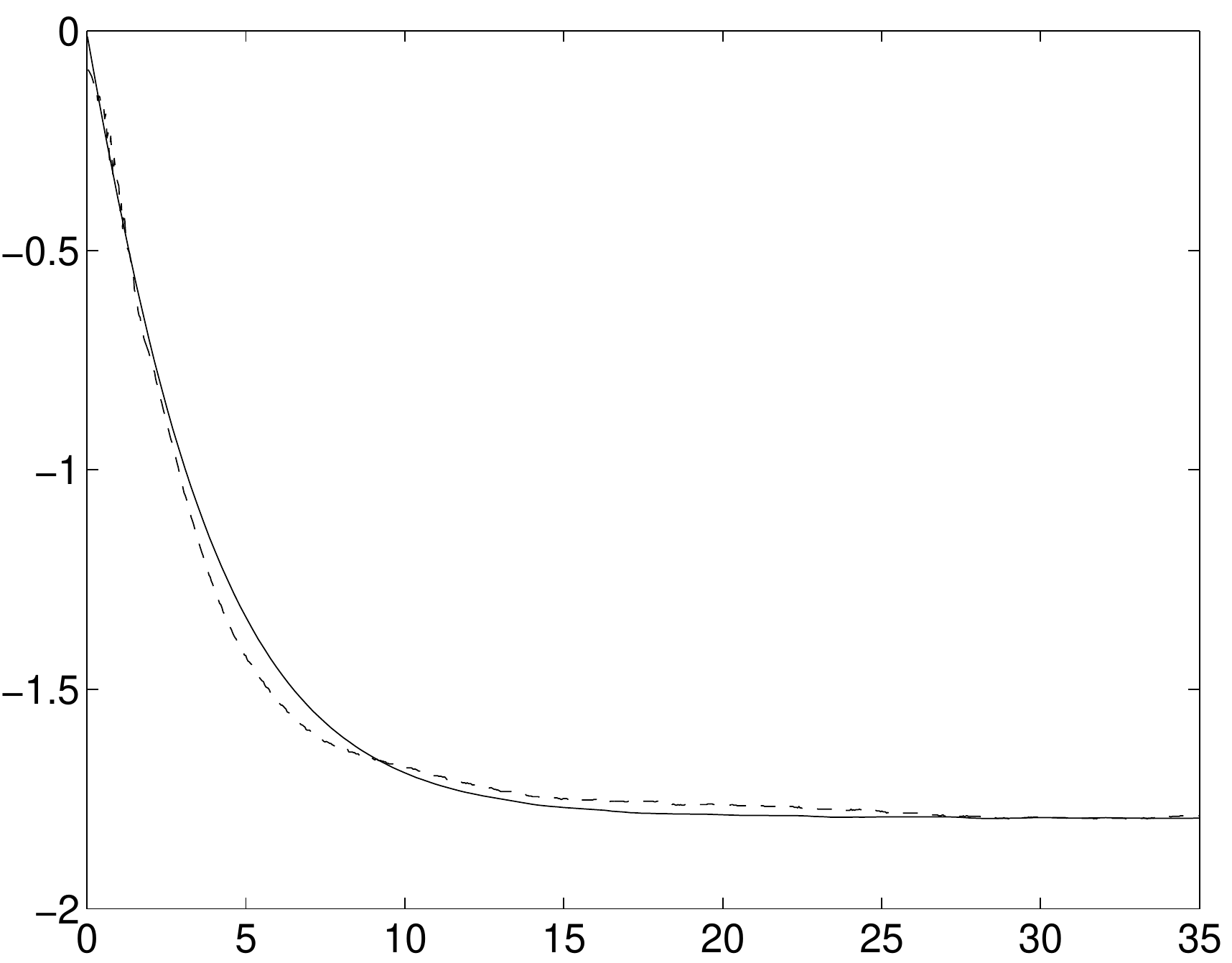}
    \end{minipage}
    \\$t/t_{ref}$
  \end{center}
  \caption{Sedimentation of a single sphere; case 2 of
    reference~\cite{mordant:00}. Vertical velocity:
    \solid present, \dashed experimental data.}
  \label{fig-mordant2-w}
\end{figure}

\begin{figure}
  \begin{center}
    \begin{minipage}{.5cm}
      $\frac{w}{u_{ref}}$
    \end{minipage}
    \begin{minipage}{.43\linewidth}
      \includegraphics*[width=\linewidth]{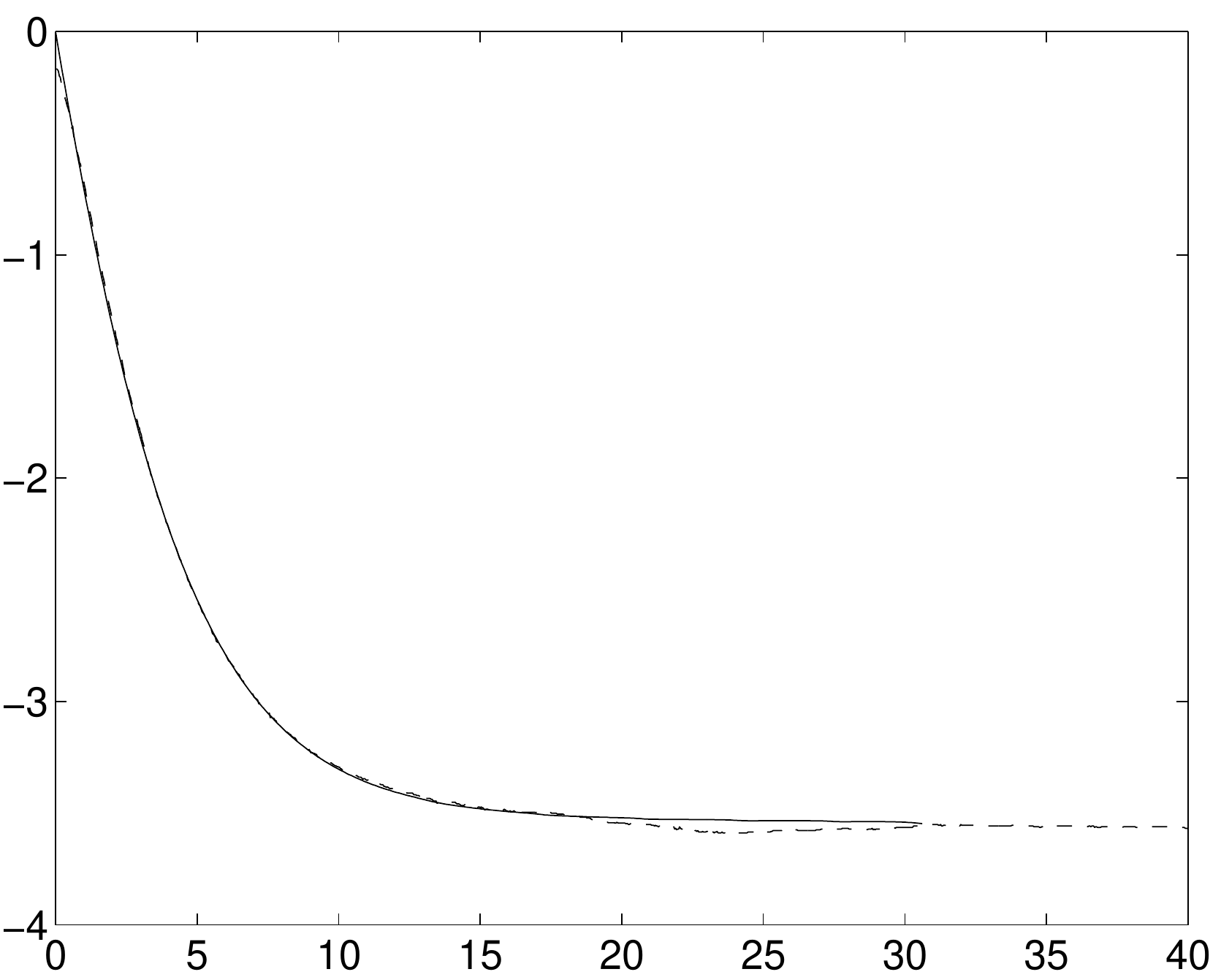}
    \end{minipage}
    \\$t/t_{ref}$
  \end{center}
  \caption{Sedimentation of a single sphere; case 4 of
    reference~\cite{mordant:00}. Vertical velocity:
    \solid present, \dashed experimental data.}
  \label{fig-mordant4-w}
\end{figure}

\begin{figure}
  \begin{minipage}{.5cm}
    $\frac{u}{u_{ref}}$
  \end{minipage}
  \begin{minipage}{.43\linewidth}
    \includegraphics*[width=\linewidth]{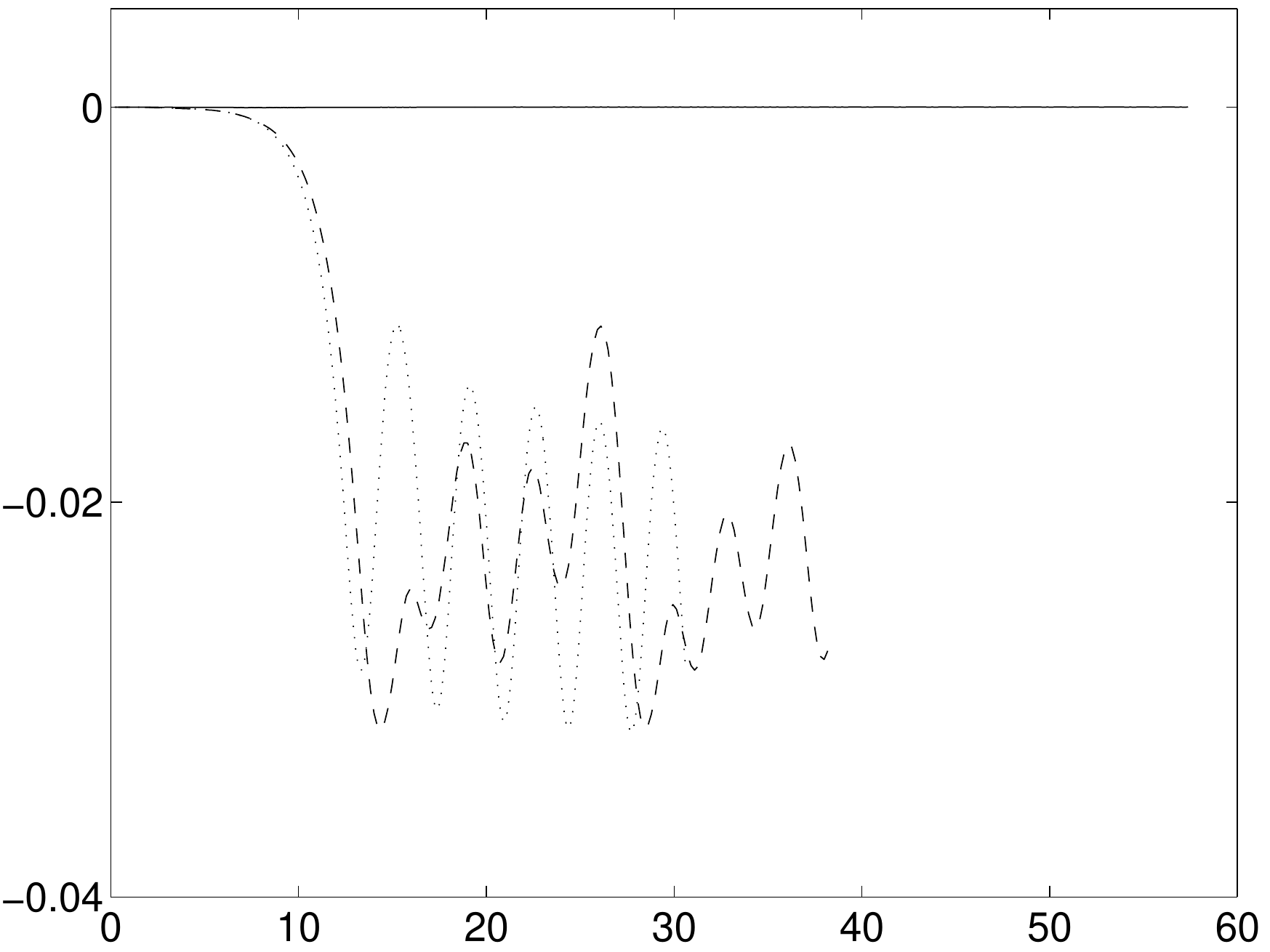}
  \end{minipage}
  \hfill
  \begin{minipage}{.5cm}
    $\frac{v}{u_{ref}}$
  \end{minipage}
  \begin{minipage}{.43\linewidth}
    \includegraphics*[width=\linewidth]{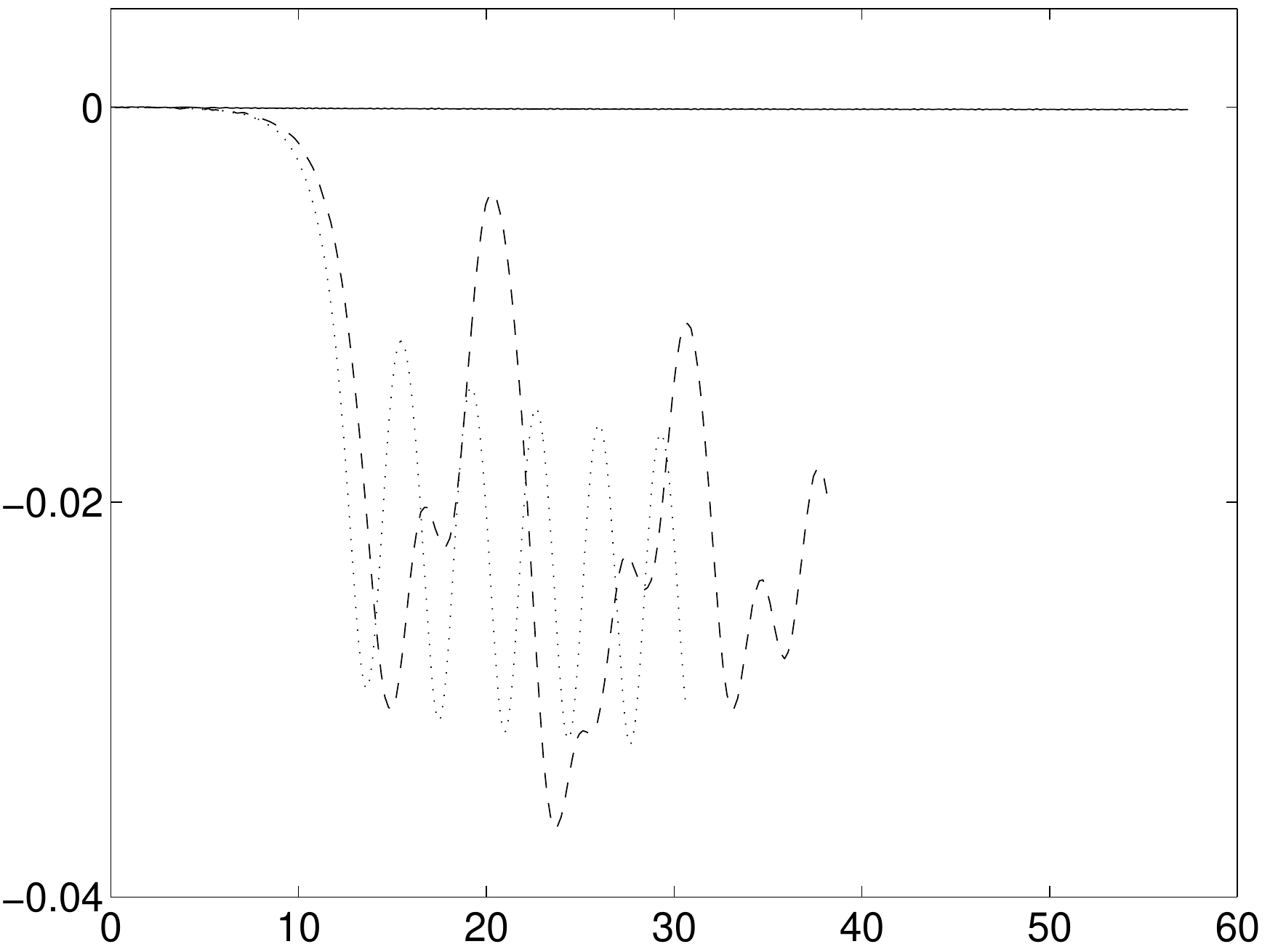}
  \end{minipage}
  \\
  \centerline{$t/t_{ref}$\hspace*{.5\linewidth}$t/t_{ref}$}
  \caption{Sedimentation of a single sphere. Horizontal velocities $u$
  (left) and $v$ (right). \solid~case~1, \dashed~case~2,
  \dotted~case~4.} 
  \label{fig-mordant-all-uv}
\end{figure}

Figures~\ref{fig-mordant1-w}-\ref{fig-mordant4-w} show the
vertical velocity as a function of the elapsed time. 
The computational results are shown for times before the particle
motion in the periodic domain is affected by the remnants of its own
wake.  
A very good agreement with the experimental measurements can be observed
in all three cases. Table~\ref{tab-mordant-re} shows the terminal
value of the Reynolds number whose maximum error is below 2\% (case 2).   
Finally, we have reported the variation of the two horizontal velocity
components in figure~\ref{fig-mordant-all-uv}. The lateral motion
is negligible in the low-Reynolds number case 1 due to the absence of
asymmetric vortex shedding. In the other two cases vortex shedding
induces horizontal velocities two orders of magnitude smaller than
the vertical velocity. 

The importance of this test case should be underlined: it
confirms---for a considerable range of Reynolds numbers---the present
method's ability to reproduce the dynamic behavior of a
three-dimensional, suspended particle under the action of gravity,
using reliable experimental data as a reference.  
\subsubsection{Many-particle systems}
\label{sec-results-3d-many}
Here we consider the collective behavior of a number of identical
particles under the action of gravity. The physical parameters are: 
\begin{itemize}
  \renewcommand{\itemsep}{-.0cm}
\item particle radius $r_c=1/12$;
\item density ratio $\rho_p/\rho_f=2.56$
\item gravitational acceleration $\mathbf{g}=(0,0,-9.81)$
\item fluid viscosity $\nu=0.001$
\end{itemize}
which corresponds to a terminal Reynolds number of approximately
$400$. The sedimentation process of many-particle systems in periodic
boxes has been simulated in reference~\cite{hoefler:00} for very low
Reynolds numbers and in reference~\cite{kajishima:02} for similar
values of the Reynolds number.  

The values for the numerical para\-meters are:
\begin{itemize}
\renewcommand{\itemsep}{-.0cm}
\item mesh width $h=1/76.8$, i.e.\ $D/h=12.8$;
\item time step $\Delta t=0.002$, i.e.\ yielding a maximum $CFL$
  number of approximately $0.5$ after the initial transient.
\end{itemize}
We have studied three different configurations with 1, 63 and 1000
particles and different domain sizes. All relevant definitions are
given in table~\ref{tab-many-p-config}. The initial particle locations  
consist of uniform and symmetric arrays (sizes indicated in the table)
with a small perturbation of the order of a few percent of the radius
in order to speed up the transient.
No collision model was used and the simulations were stopped when
unphysical overlapping of particle domains was detected.

\begin{figure}
  \begin{center}
    \begin{minipage}{.5cm}
      $\frac{\bar{w}_c}{u_{ref}}$
    \end{minipage}
    \begin{minipage}{.43\linewidth}
      \includegraphics*[width=\linewidth]{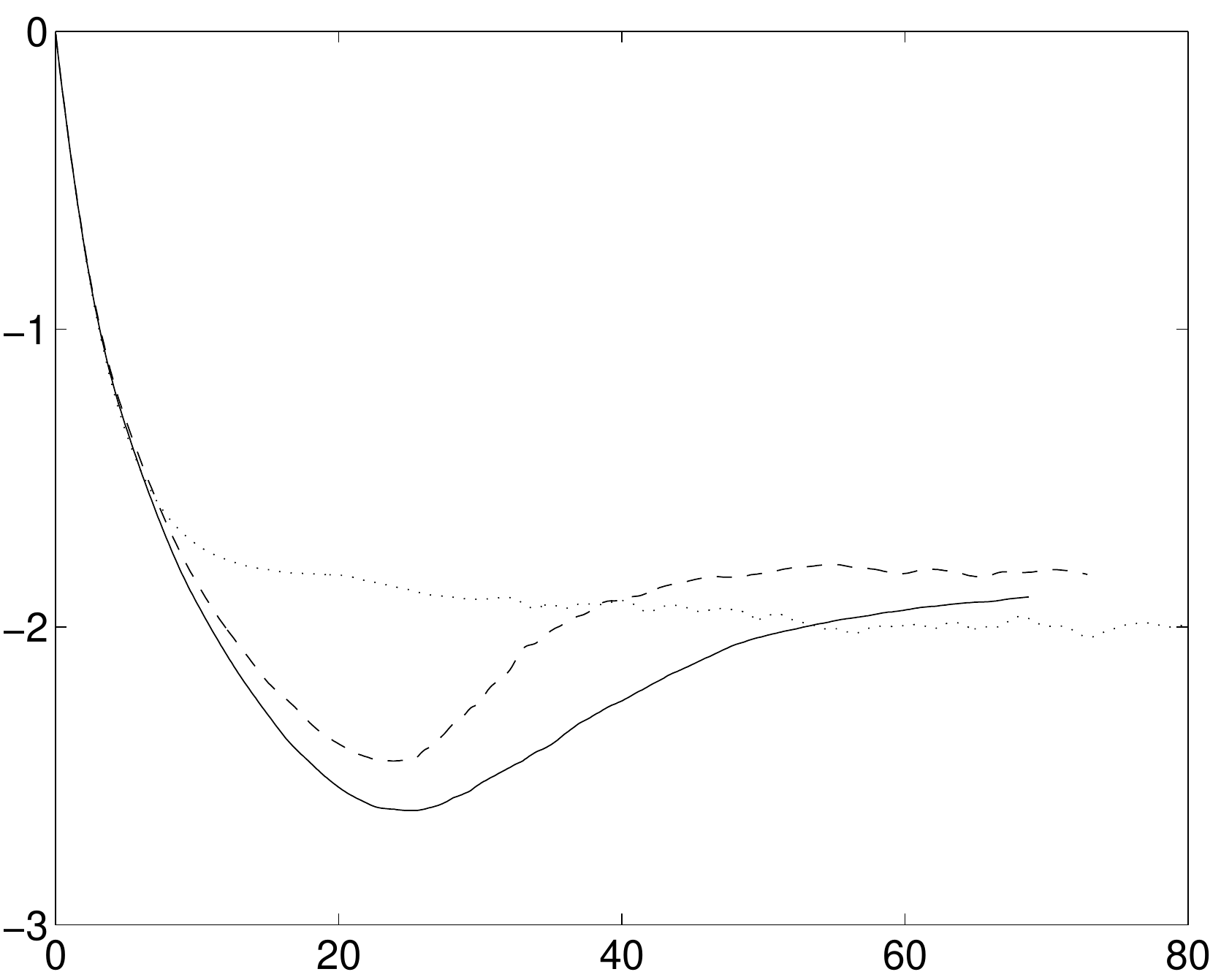}
    \end{minipage}\\
    $t/t_{ref}$
  \end{center}
  \caption{Sedimentation of an array of $N_p$ identical spheres with
    density ratio $\rho_p/\rho_f=2.56$ and terminal Reynolds number
    approximately $400$. Mean vertical particle velocity vs.\ time
    for: \dotted~case~1~($N_p$=$1$), \dashed~case~2~($N_p$=$63$), 
    \solid~case~3~($N_p$=$1000$).}  
  \label{fig-manyp-velzmean}
\end{figure}

Figure~\ref{fig-manyp-velzmean} shows the time-evolution of the
average vertical particle velocity,
$\bar{w}_c=\sum_i^{N_p}w_c^{(i)}/N_p$. In both many-particle cases
this quantity initially reaches very high values before 
levelling off to values which are similar to the single-particle
case~1. This transient behavior is due to the particles' initial
vertical alignment which leads to a relatively low drag because of
wake-sheltering. As soon as the configuration is perturbed through the
onset of asymmetric vortex-shedding the drag increases again leading
to the observed reduction of the settling velocity. 

\begin{figure}
  \begin{center}
    \begin{minipage}{.5cm}
      $\frac{\bar{d}_{min}}{D}$
    \end{minipage}
    \begin{minipage}{.43\linewidth}
      \includegraphics*[width=\linewidth]{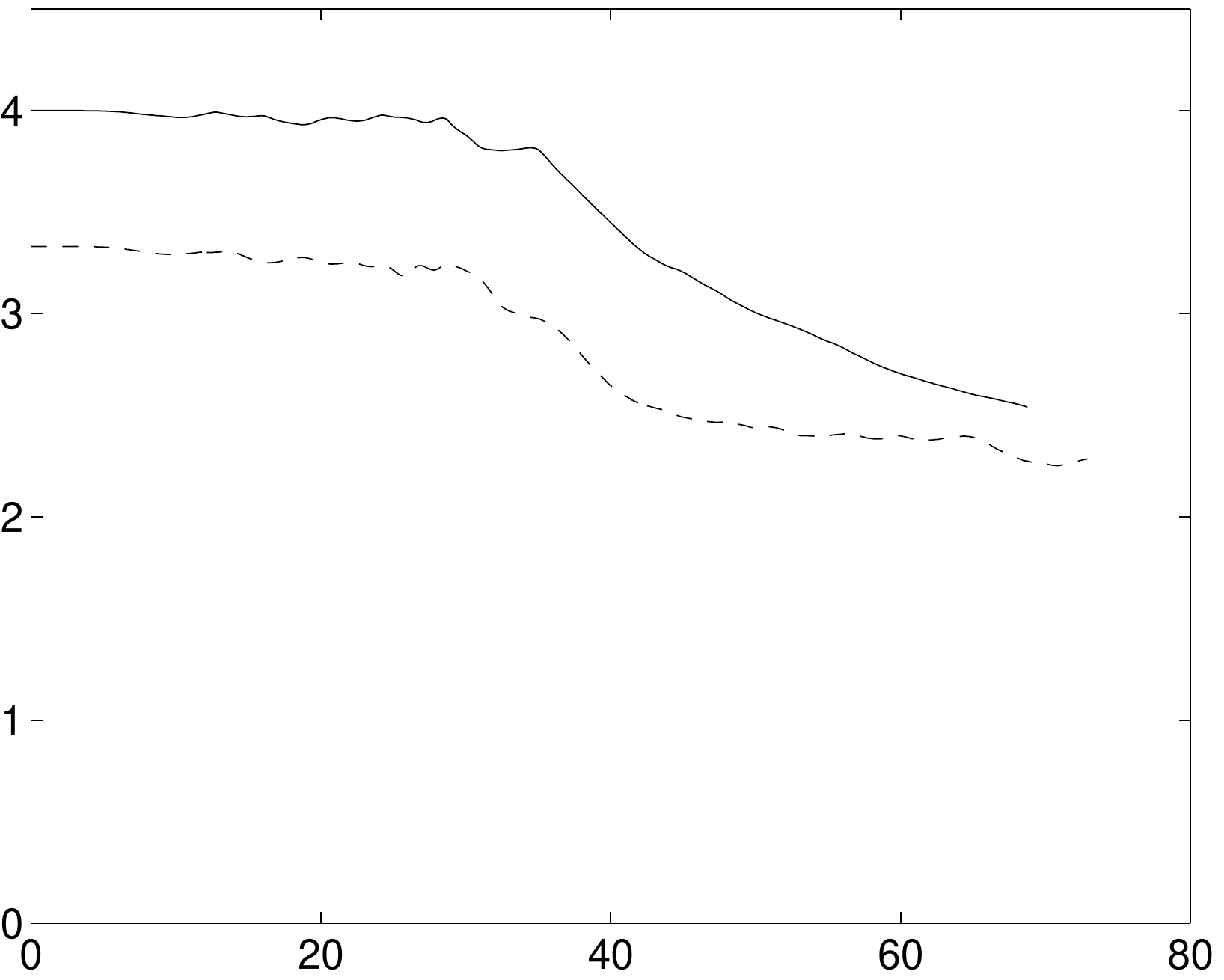}
    \end{minipage}\\
    $t/t_{ref}$
  \end{center}
  \caption{Sedimentation of an array of $n$ identical spheres with
    density ratio $\rho_p/\rho_f=2.56$ and terminal Reynolds number
    approximately $400$. Mean distance to the nearest particle vs.\
    time for: \dashed~case~2~($N_p$=$63$), \solid~case~3~($N_p$=$1000$).}   
  \label{fig-manyp-distmean}
\end{figure}

In figure~\ref{fig-manyp-distmean} we have plotted the 
time-evolution of the average distance to the nearest particle,
$\bar{d}_{min}=\sum_{i}^{N_p}\min_j(|\mathbf{x}_c^{(i)}-\mathbf{x}_c^{(j)}|)$,  
which is an indicator for the re-organization of the relative particle 
positions. It can be observed that $\bar{d}_{min}$ gradually decreases
in cases 2 and 3, showing that indeed there is a tendency for
particles to attract each other. 

\begin{figure}
  \begin{center}
    \begin{minipage}{.7\linewidth}
      \includegraphics*[width=\linewidth]{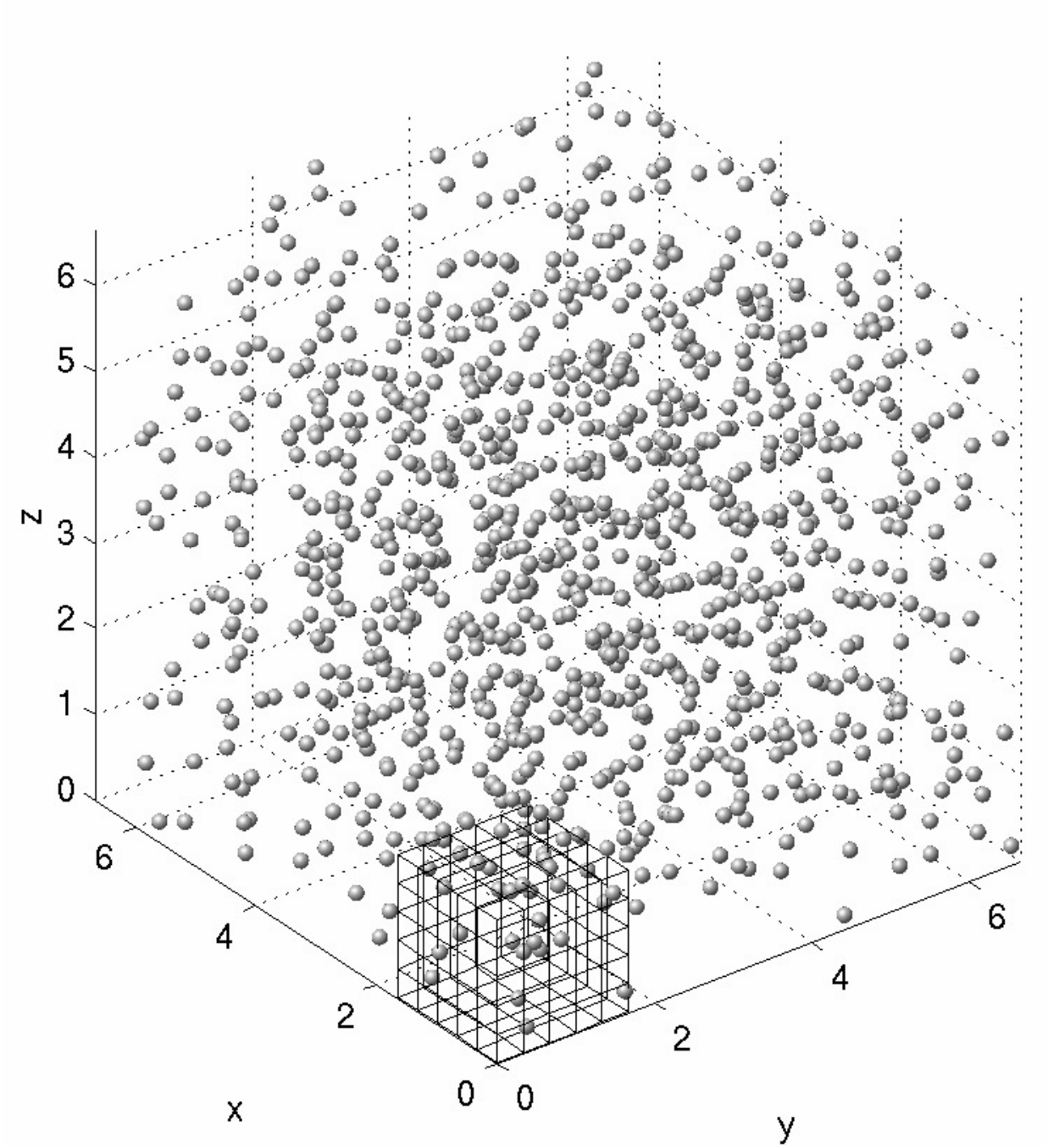}
    \end{minipage}
  \end{center}
  \caption{Instantaneous particle positions in the sedimentation
    case~3~($N_p$=$1000$) at $t/t_{ref}=64.445$. The small wire-frame cube
    indicates the sub-volume visualized in
    figure~\ref{fig-manyp-laplp-stream}.}    
  \label{fig-manyp-circpos}
\end{figure}

\begin{figure}
  \begin{minipage}{.45\linewidth}
    \includegraphics*[width=\linewidth]{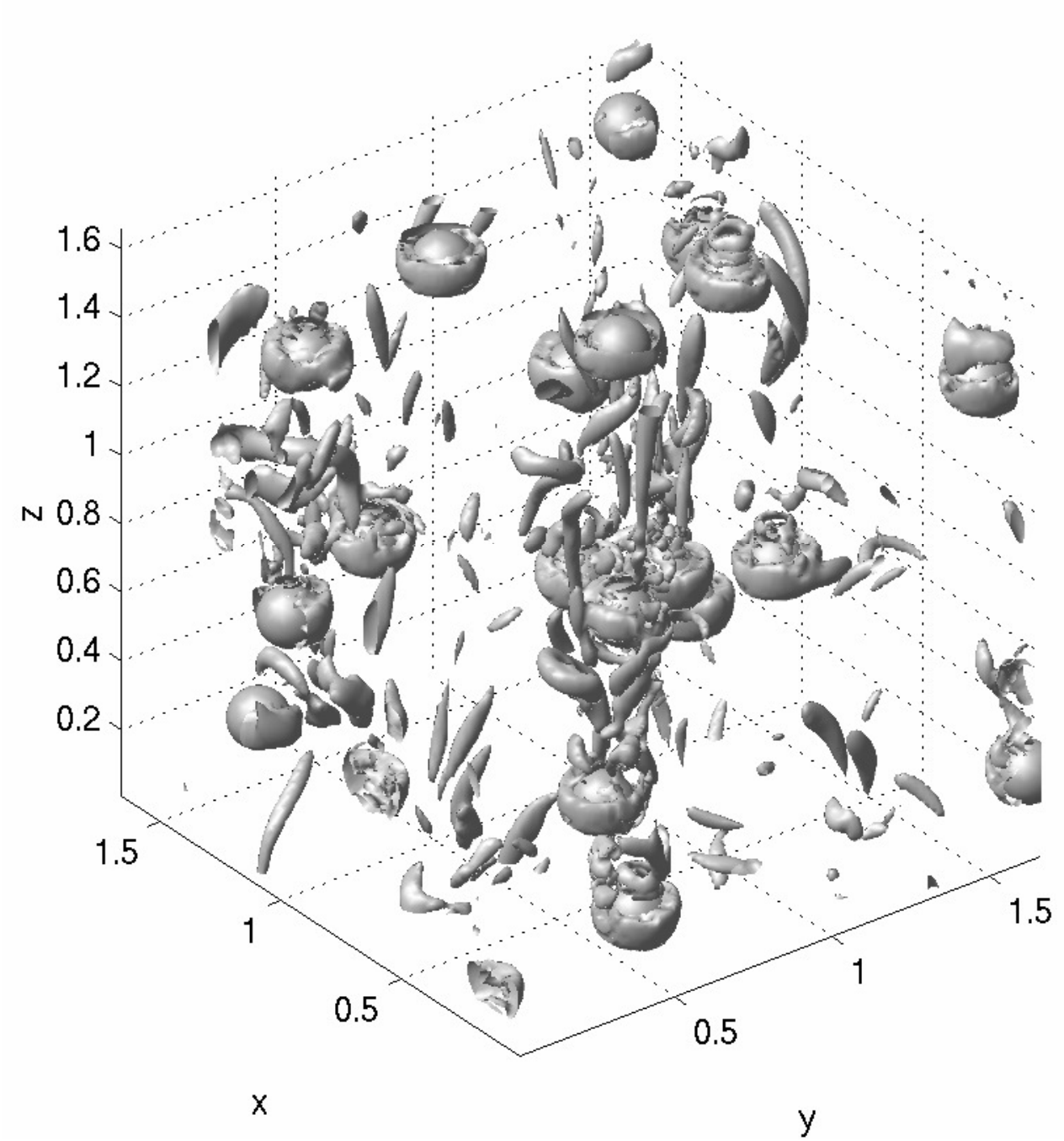}
  \end{minipage}
  \hfill
  \begin{minipage}{.45\linewidth}
    \includegraphics*[width=\linewidth]{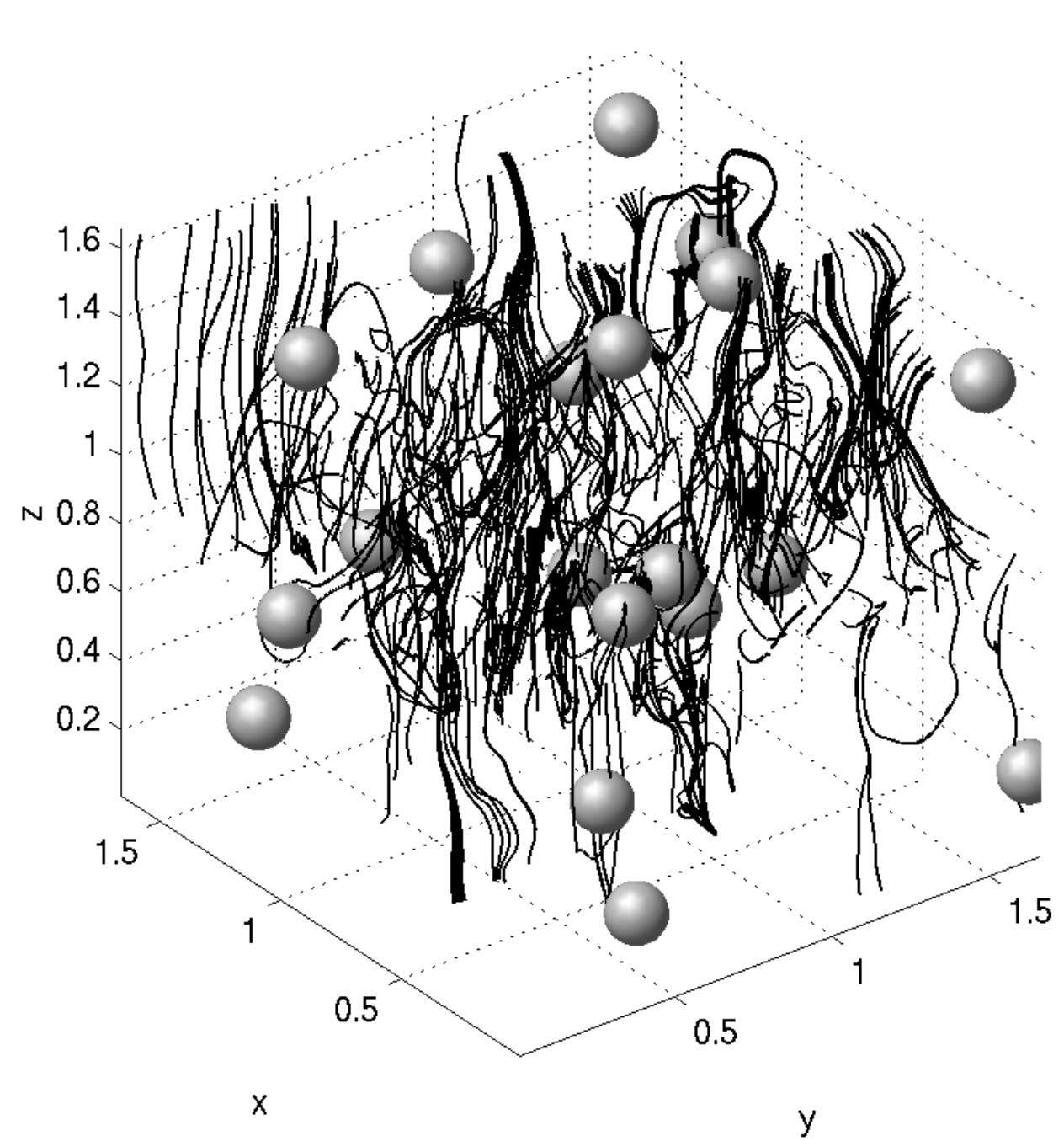}
  \end{minipage}
  \caption{Visualization of the instantaneous flow field and particle
    positions in a sub-volume indicated in
    figure~\ref{fig-manyp-circpos} of the sedimentation case~3~($N_p$=$1000$)
    at $t/t_{ref}=64.445$. The left graph shows iso-surfaces of positive
    values of the Laplacian of pressure, indicating vortex cores. The
    right graph shows streamlines released from the horizontal
    mid-plane.}  
  \label{fig-manyp-laplp-stream}
\end{figure}

Finally, figures~\ref{fig-manyp-circpos} and
\ref{fig-manyp-laplp-stream} show visualizations of the instantaneous
particle positions and the flow field in case 3 at
$t/t_{ref}=64.445$. It is evident that the initially uniform and
symmetric particle configuration has broken up and given way to a
seemingly disordered state by this time. The vortex tubes reaching
from particle to particle and the entangled streamlines show how
neighboring particles are indirectly interacting by way of the fluid. 
\subsection{A note on efficiency}\label{sec-conclusions-efficiency}
The main work in the pure fluid part of our method stems from the
multi-grid solution of the Poisson problem
(\ref{equ-hybrid-discr-algo-poisson}) and the factorized solution of
the Helmholtz problems
(\ref{equ-hybrid-discr-algo-predict}). Therefore, the
overall operation count for the fluid scales as ${\mathcal
  O}(N^3)$, where $N$ is the number of Eulerian grid nodes in one
spatial dimension. 
On the other hand, the solution of the Newton equations
(\ref{equ-particles-newton-2-present}) for $N_p$ particles requires 
simply ${\mathcal O}(N_p)$ operations. 
Finally, the fluid-solid interaction in equations
(\ref{equ-hybrid-discr-algo-lag-interpol})-(\ref{equ-hybrid-discr-algo-eul-force})
is performed in ${\mathcal O}(N_p\,N_L)$ operations. Since the number
of Lagrangian force points $N_L$ is chosen such that each one controls
a volume corresponding to a grid cell we have
approximately from (\ref{equ-cond-nl-3d}) that $N_L\sim(D/h)^2$. 
Introducing a characteristic macroscopic length scale $L=N\,h$, we
arrive at the following count for the fluid-solid coupling: ${\mathcal
  O}(N_p\,(D/L)^2\,N^2)$. This shows that even for tightly packed
particles (i.e.\ $N_p=(L/D)^3$) the work needed for treating the pure
fluid part of the code asymptotically outweighs the remaining
contributions (since $L/D<N$). 

In order to deal with large-scale problems the algorithm was
implemented for multi-processor machines with
distributed-memory. Classical domain-dis\-tri\-bu\-tion was used for the
fluid part and a master-slave technique was employed for the
particle-related operations~\cite{uhlmann:03c}. 
Table~\ref{tab-timing} shows some execution times per time step of
``production size'' cases. For
the largest case involving $512^2\times 1024$ grid nodes and 128
processors, it can be seen that increasing the number of particles
from 1000 to 2000 increases the execution time by less than 2\%.

\section{Conclusion}\label{sec-conclusions}
We have presented  
an improved 
immersed-boundary method with a 
direct formulation of the fluid-solid interaction force. 
The regularized delta function of Peskin~\cite{peskin:02} is used for
the association between arbitrary Lagrangian and discrete Eulerian 
positions. Thereby, the hydrodynamic forces acting upon the particles
are free from significant oscillations, allowing for smooth motion of
the particles. On the other hand, the direct (not feed-back) character
of the forcing scheme avoids additional restrictions of the
time-step. 

The current method was implemented in a finite-difference and
fractional-step context. The fluid equations are weakly coupled to the
Newton equations for the rigid-body motion of the particles which
imposes a lower limit for the density ratio between particles and
fluid of approximately $1.2$ for stable integration. 

The new scheme was applied to Taylor-Green flow, flow around fixed and
oscillating cylinders as well as sedimentation problems in two and
three space dimensions. By comparison of our present results with
reference values from experiments and independent numerical
simulations we have demonstrated the high accuracy of the
method. Furthermore, our simulations of many-particle systems in truly
three-dimensional domains using multi-processor machines show that
the study of large-scale configurations is feasible with the current
approach.

\vspace*{1ex}
This work was supported by the Spanish Ministry of Education and
Science under the Ram\'on y Cajal program (contract
DPI-2002-040550-C07-04) and through grant
DPI-2002-1314-C07-04.  
Part of the work was done while the author was visiting the
Department of Aeronautics and Astronautics at the University of Kyoto,
Japan, with the aid of a grant from the Japanese Society for the
Promotion of Science.  
Additional computing time was provided by the Potsdam 
Institute for Climate Impact Research, Germany.
\bibliographystyle{model2-names}

%
\appendix
  \section{Distribution of Lagrangian force points and associated
volumes}\label{app-marker} 
In practice we want each force point to control a volume which is
equivalent to a finite volume of the Eulerian grid, i.e.\ $\Delta
V_l\approx h^n$, where $n$ is the number of space-dimensions. We have
verified that further increasing the number of force points $N_L$ does not
significantly improve the solution. Therefore, in all present
simulations $N_L$ was determined from equations
(\ref{equ-cond-nl-2d}) and (\ref{equ-cond-nl-3d}) below. 
\subsection{Circular particles}\label{app-marker-2d}
\begin{figure}
  \begin{center}
    \begin{minipage}{.48\linewidth}
      \includegraphics*[width=\linewidth]{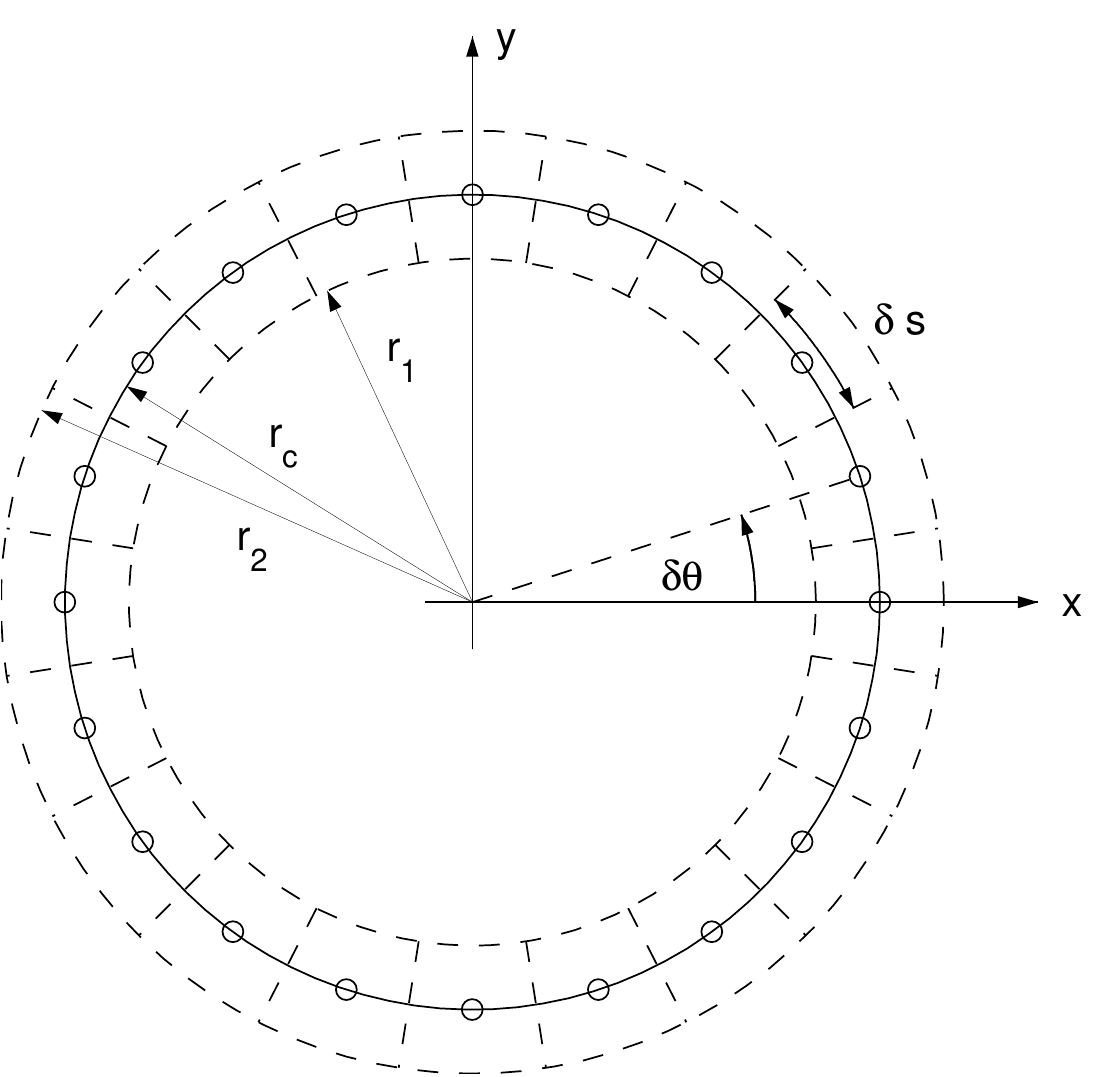}
    \end{minipage}
  \end{center}
  \caption{Definitions for a circular particle. %
    $r_c$ is the actual particle radius and small circular symbols
    indicate the (equidistant) locations of Lagrangian force points;
    the dashed lines show the extent of the volumes associated with
    each force point.  
  }
  \label{fig-grid1}
\end{figure}

We define a number of $N_L$ elements around the circumference
of a circular solid object, as shown in figure~\ref{fig-grid1}. The
elements are equi-partitioned sectors of an annulus 
with inner and outer radii $r_1$, $r_2$, respectively. The actual
particle radius $r_c$ is located at the midpoint of these two
radii $r_c=(r_2+r_1)/{2}$. 
Furthermore, we take the radial width of an element to be equal to
the mesh size $h$, $h=r_2-r_1$. 
The arc-length, measured at radius $r_c$, is given by
$\delta s=2\pi r_c/N_L$. 
It follows then that  
\begin{equation}\label{equ-hybrid-radii-solved}
  r_1=r_c-\frac{h}{2}\,,\qquad 
  r_2=r_c+\frac{h}{2}\,,
\end{equation}
which gives for the surface of an element $\Delta V_l$:
\begin{equation}\label{equ-hybrid-marker-volume}
  \Delta V_l=\frac{2\pi r_c h}{N_L}\,.
\end{equation}
We associate a Lagrangian force point to each of the above elements
and locate it equidistantly on the actual circumference of the
particle (i.e.\ in the center of an element).

Requiring that $\Delta V_l\approx h^2$ leads to the following
condition for the number of force points: 
\begin{equation}\label{equ-cond-nl-2d}
  N_L\approx 2\pi\frac{r_c}{h}
\end{equation}
\subsection{Spherical particles}\label{app-marker-3d}
\subsubsection{Force point distribution}\label{app-marker-3d-vertices}
\begin{figure}
  \begin{minipage}{.48\linewidth}
    \includegraphics*[width=\linewidth]{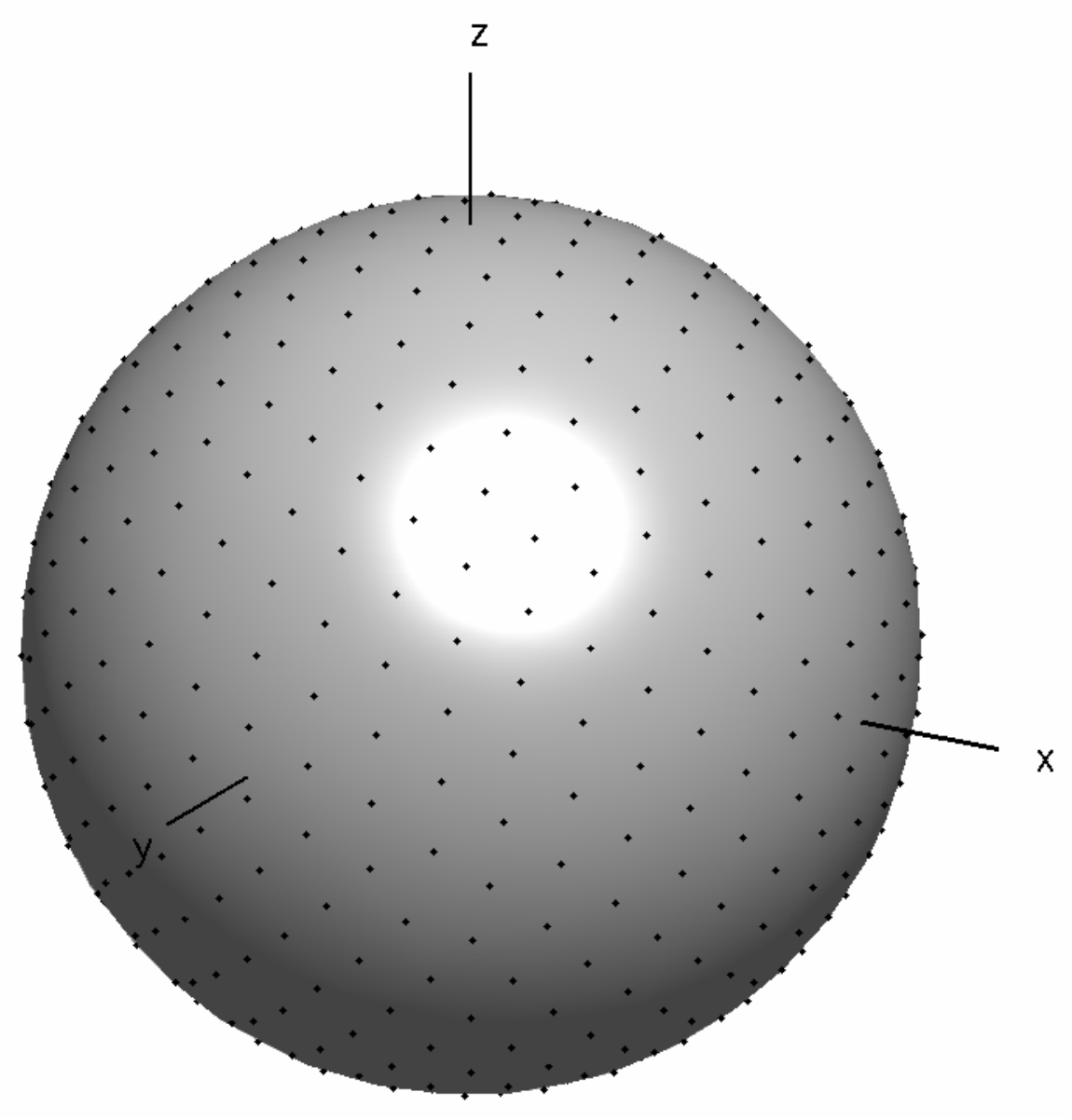}
  \end{minipage}
  \hfill
  \begin{minipage}{.48\linewidth}
    \includegraphics*[width=\linewidth]{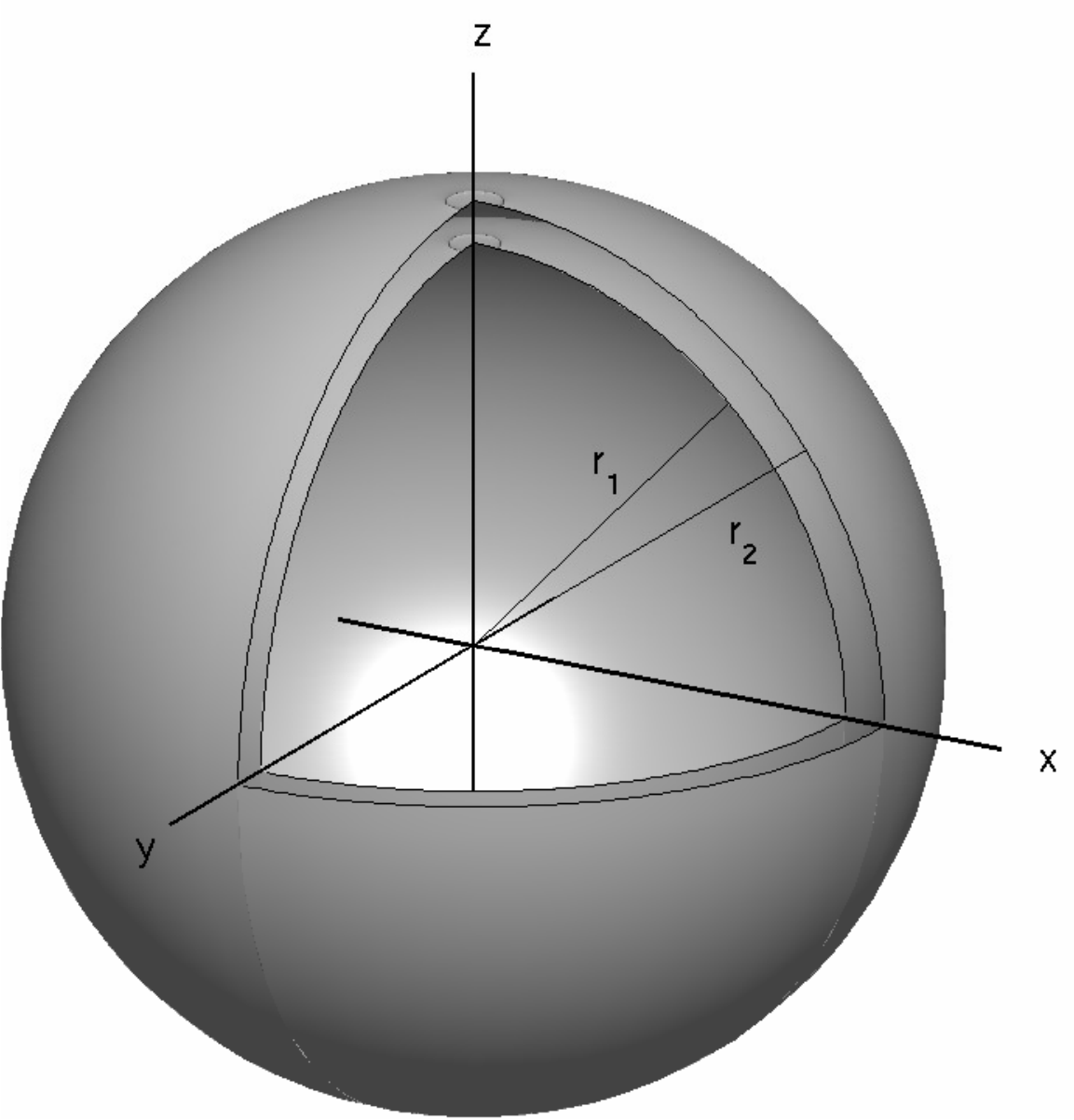}
  \end{minipage}
  \caption{Definitions for a spherical particle. In the graph on the
    left an iteratively obtained distribution of the Lagrangian force
    points for the case $N_L=515$ is shown. The right graph
    shows a cut-away view of the thin shell between radii $r_1$ and
    $r_2$ where forcing is applied. 
  }
  \label{fig-grid2}
\end{figure}

The even distribution of an arbitrary number of points on the surface
of a sphere is an unsolved problem in geometry~\cite{saff:97}. In
fact the very definition of ``even'' is not evident. In practice two
methods appear feasible for our case: 
\begin{enumerate}
\item Start with one of the three triangular-faced regular polyhedra
  (tetrahedron, octahedron, icosahedron) whose vertices lie on the
  surface of a sphere. Among these Platonic
  solids, the icosahedron has the highest number of rotational
  symmetries and therefore leads to the most ``even'' distribution of
  points. In each refinement step,
  ``pull-up'' the centroid of each edge to the sphere's surface and
  thereby subdivide each of its previous faces into four new
  triangular faces. The final number of vertices after $m$ such
  refinement steps is $N_L=n_v+\sum_{i=1}^{m}n_e4^{i-1}$, where $n_v$,
  $n_e$ are the number of initial vertices and edges,
  respectively. Obviously the values of $N_L$ which can be obtained in
  this fashion are very sparsely distributed. For values outside this
  set, one needs to resort to the following method. 
\item Define ``even'' as the configuration of points which minimizes
  the total repulsive energy in a system of charged particles. For a
  given value of $N_L$, run a simulation of the motion of
  point-particles confined to the surface of a sphere. From some
  initial state, using a mutual repulsive force which is proportional
  to the inverse of the square of the inter-particle distance, an
  equilibrium configuration can be obtained iteratively. Several runs
  with different initial conditions might be necessary in order to
  find a global energy minimum. An example for $N_L=515$ is shown in
  figure~\ref{fig-grid2}. 
\end{enumerate}
\subsubsection{Definition of forcing volumes}\label{app-marker-3d-vol} 
A spherical shell between the radii $r_1=r_c-h/2$ and $r_2=r_c+h/2$
shall be forced (cf.\ figure~\ref{fig-grid2}). Therefore, we associate
the following partial volume with each force point:
\begin{equation}\label{equ-hybrid-marker-volume-3d}
  \Delta V_l=\frac{\pi h}{3N_L}(12r_c^2+h^2)
\end{equation}

Requiring that $\Delta V_l\approx h^3$ leads to the following
condition for the number of force points:
\begin{equation}\label{equ-cond-nl-3d}
  N_L\approx \frac{\pi}{3}\left(12\frac{r_c^2}{h^2}+1\right)
\end{equation}
\section{Evaluation of the hydrodynamic forces acting upon particles}
\label{app-newton}
Let us write Newton's equations for the motion of a single rigid
particle, viz.
\bse{equ-particles-newton-2}
\bea\label{equ-particles-newton-2-translation}
\rho_pV_c\,\dot{\mathbf{u}}_c&=&\rho_f\oint_{\partial{\mathcal S}}
\boldsymbol{\tau}\cdot\mathbf{n}\,\mbox{d}\sigma
+(\rho_p-\rho_f)V_c\,\mathbf{g}
\\\label{equ-particles-newton-2-rotation}
I_c\,\dot{\boldsymbol{\omega}}_c&=&\rho_f\oint_{\partial{\mathcal S}}
\mathbf{r}\times\left(\boldsymbol{\tau}
\cdot\mathbf{n}\right)\,\mbox{d}\sigma 
\eea
\ese
where $\boldsymbol{\tau}=
-Ip+\nu\left(\nabla\mathbf{u}+\nabla\mathbf{u}^T\right)$ is the
hydrodynamic stress tensor, $\mathbf{n}$ the outward-pointing normal
vector on the fluid-solid interface $\partial{\mathcal S}$ and
$\mathbf{r}=\mathbf{x}-\mathbf{x}_c$. Cauchy's principle
states for the hydrodynamic force and torque
terms~\cite[p.100]{aris:62}: 
\bse{equ-particles-def-stress}
\bea\label{equ-particles-def-stress-linear}
\oint_{\partial{\mathcal S}}
\boldsymbol{\tau}\cdot\mathbf{n}\,\mbox{d}\sigma&=&
-\int_{{\mathcal
    S}}\mathbf{f}\,\mbox{d}\mathbf{x}
+\frac{\mbox{d}}{\mbox{d}t}\int_{\mathcal
  S}\mathbf{u}\,\mbox{d}\mathbf{x}
\\\label{equ-particles-def-stress-angular}
\oint_{\partial{\mathcal S}}
\mathbf{r}\times\left(\boldsymbol{\tau}\cdot\mathbf{n}\right)\,\mbox{d}\sigma&=&  
-\int_{{\mathcal
    S}}\mathbf{r}\times\mathbf{f}\,\mbox{d}\mathbf{x}
+\frac{\mbox{d}}{\mbox{d}t}\int_{\mathcal
  S}\left(\mathbf{r}\times\mathbf{u}\right)\,\mbox{d}\mathbf{x}
\eea
\ese
The first term on the r.h.s.\ of both equations in
(\ref{equ-particles-def-stress}) is simply the negative of the sum of
the fluid-solid coupling force/\-torque defined in
\S~\ref{sec-forcing-force}. Using the equalities 
(\ref{equ-deltah-prop-conseq}) these can be efficiently evaluated as
sums over the Lagrangian force points. Concerning the rate-of-change
term in the force relation (\ref{equ-particles-def-stress-linear}) it
can be shown that the following expression holds for an incompressible
fluid which satisfies a rigid-body motion on the interface
$\partial{\mathcal S}$~\cite{uhlmann:03}: 
\begin{equation}\label{equ-rate-of-change-force}
  \frac{\mbox{d}}{\mbox{d}t}\int_{\mathcal
  S}\mathbf{u}\,\mbox{d}\mathbf{x}=V_c\,\dot{\mathbf{u}}_c
\end{equation}
irrespective of the actual type of motion inside the volume ${\mathcal
  S}$. Conversely, the analogous relation for the torque only holds in
the case of rigid-body motion throughout the volume ${\mathcal
  S}$: 
\begin{equation}\label{equ-rate-of-change-torque}
  \mathbf{u}(\mathbf{x})=\mathbf{u}_c+{\boldsymbol\omega}_c\times\mathbf{r}(\mathbf{x})\quad
  \mathbf{x}\in{\mathcal S}:\qquad  
  \frac{\mbox{d}}{\mbox{d}t}\int_{\mathcal
  S}\mathbf{r}\times\mathbf{u}\,\mbox{d}\mathbf{x}=\frac{I_c}{\rho_p}
\,\dot{\boldsymbol\omega}_c
\end{equation}
In our case, i.e.\ when the inner part of the solid particles is not
forced, no simplification could be found and the rate-of-change of the
integral of the torque must be evaluated numerically. Finally,
collecting all terms yields the form of Newton's equations
(\ref{equ-particles-newton-2-present}) given in the main text.  
%
%
\clearpage
\clearpage
\clearpage
\clearpage
\clearpage
\clearpage
\clearpage
\clearpage
\clearpage
\clearpage
\clearpage
\clearpage
\clearpage
\clearpage
\clearpage
\clearpage
\clearpage
\clearpage
\clearpage
\clearpage
\clearpage
\clearpage
\clearpage
\clearpage
\clearpage
\begin{table}
  \caption{
    Dimensionless coefficients obtained from the simulation of
    the flow around a stationary cylinder at $Re_D=100$, using
    $D/h=38.4$, $\Delta t=0.003$. The enlarged domain $\Omega_2$
    measures $40D\times 40D$.}  
  \label{tab-cylinder-stat}
  \begin{center}
    \begin{tabular}{l*{4}{c}}
      &$\bar{C}_D$&$C_D^\prime$&$C_L^\prime$&$St$\\
      present&$1.501$&$\pm 0.011$&$\pm 0.349$&$0.172$
      \\
      present, enlarged domain $\Omega_2$&$1.453$&$\pm
      0.011$&$\pm 0.339$&$0.169$ 
      \\
      Liu {\it et al.}~\cite{liu:98}&
      $1.350$&$\pm 0.012$&$\pm 0.339$&$0.165$
    \end{tabular}
  \end{center}
\end{table}\clearpage
\begin{table}
  \caption{
    Dimensionless coefficients obtained from the simulation
    of the flow around a cylinder at $Re_D=185$ which oscillates
    near the natural shedding frequency and using $D/h=38.4$ and
    $\Delta t=0.003$. The domain $\Omega_1$ has been used,
    except where otherwise stated.}
  \label{tab-cylinder-osc}
  \begin{center}
    \begin{tabular}{l*{3}{c}}
      &$\bar{C}_D$&$C_D^\prime$&($C_L)_{rms}$\\
      present %
      &$1.380$&$\pm 0.063$&$0.176$ 
      \\
      present, enlarged domain $\Omega_2$%
      &$1.354$&$\pm 0.065$&$0.166$ 
      \\
      present, 4-point $\delta_h$ of \cite{peskin:02}
      &$1.402$&$\pm 0.064$&$0.172$ 
      \\
      Kajishima \& Takiguchi's scheme~\cite{kajishima:02}&$1.282$&$\pm
      0.088$&$0.223$ 
      \\
      Lu and Dalton~\cite{lu:96}&$1.25$&&$0.18$
    \end{tabular}
  \end{center}
\end{table}\clearpage
\begin{table}
  \caption{Terminal particle Reynolds number $Re_D$ in the case of a
    single sedimenting sphere, compared to the experiment of 
    reference~\cite{mordant:00}.}
  \label{tab-mordant-re}
  \begin{center}
    \begin{tabular}{*{4}{c}}
      case&1&2&4\\
      present&$41.12$&$366.69$&$282.45$
      \\
      experiment&$41.17$&$362.70$&$280.42$
    \end{tabular}
  \end{center}
\end{table}\clearpage
\begin{table}
  \caption{Definitions for the three different configurations used in
    triply-periodic many-particle simulations in
    \S~\ref{sec-results-3d-many}.} 
  \label{tab-many-p-config}
  \begin{center}
    \begin{tabular}{*{5}{c}}
      case&&1&2&3\\
      no.\ of particles&$N_p$&1&63&1000\\
      initial locations&&(irrelevant)&
      $3\times 3\times 7$ array&
      $10\times 10\times 10$ array\\

      domain size&$\Omega$&$[0,1.\bar{6}]^2\times[0,5]$&
      $[0,1.\bar{6}]^2\times[0,5]$&
      $[0,6.\bar{6}]^3$\\
      volume fraction of solids&
      $\epsilon_p$&$0.000175$&$0.011$&$0.0082$\\
      mass loading ratio of solids&
      $\phi_p$&$0.000447$&$0.0285$&$0.0211$
    \end{tabular}
  \end{center}
\end{table}\clearpage
\begin{table}
  \caption{Execution time (per full time step), $t_{exec}$, of the
    present scheme 
    on an IBM p655 cluster with Power 4, 1.1 GHz CPUs and Colony
    switch, 64 bit arithmetic, performing 6-7 multi-grid iterations for
    each Poisson problem. 
    The parameters are: grid size $N_x,N_y,N_z$, number of
    particles $N_p$, number of Lagrangian force points per particle
    $N_L$ and number of processors {\tt nproc}.}  
  \label{tab-timing}
  \begin{center}
    \begin{tabular}{*{5}{c}}
      $N_x\times N_y\times N_z$&$N_p$&$N_{L}$&{\tt nproc}&
      $t_{exec} [s]$\\[.5ex]
      $512\times512\times512$&$1000$&$515$&64&$115.0$\\
      $512\times512\times1024$&$1000$&$515$&128&$144.9$\\
      $512\times512\times1024$&$2000$&$515$&128&$147.4$\\
    \end{tabular}
  \end{center}
\end{table}\clearpage
\end{document}